\documentclass[twoside,twocolumn,9pt]{article}
\usepackage{extsizes}
\usepackage[super,sort&compress,comma]{natbib} 
\usepackage[version=3]{mhchem}
\usepackage[left=1.5cm, right=1.5cm, top=1.785cm, bottom=2.0cm]{geometry}
\usepackage{balance}
\usepackage{mathptmx}
\usepackage{sectsty}
\usepackage{graphicx} 
\usepackage{lastpage}
\usepackage[format=plain,justification=justified,singlelinecheck=false,font={stretch=1.125,small,sf},labelfont=bf,labelsep=space]{caption}
\usepackage{float}
\usepackage{fancyhdr}
\usepackage{fnpos}
\usepackage[english]{babel}
\addto{\captionsenglish}{%
  
}
\usepackage{array}
\usepackage{droidsans}
\usepackage{charter}
\usepackage[T1]{fontenc}
\usepackage[usenames,dvipsnames]{xcolor}
\usepackage{setspace}
\usepackage[compact]{titlesec}
\usepackage{hyperref}
\usepackage{amsfonts}
\usepackage{bigints}
\usepackage{amssymb}
\usepackage{subfigure}
\usepackage{epstopdf}%This line makes .eps figures into .pdf - please comment out if not required.

\definecolor{cream}{RGB}{222,217,201}

\begin{document}

\pagestyle{fancy}
\thispagestyle{plain}
\fancypagestyle{plain}{
%%%HEADER%%%
\renewcommand{\headrulewidth}{0pt}
}
%%%END OF HEADER%%%

%%%PAGE SETUP - Please do not change any commands within this section%%%
\makeFNbottom
\makeatletter
\renewcommand\LARGE{\@setfontsize\LARGE{15pt}{17}}
\renewcommand\Large{\@setfontsize\Large{12pt}{14}}
\renewcommand\large{\@setfontsize\large{10pt}{12}}
\renewcommand\footnotesize{\@setfontsize\footnotesize{7pt}{10}}
\makeatother

\renewcommand{\thefootnote}{\fnsymbol{footnote}}
\renewcommand\footnoterule{\vspace*{1pt}% 
\color{cream}\hrule width 3.5in height 0.4pt \color{black}\vspace*{5pt}} 
\setcounter{secnumdepth}{5}

\makeatletter 
\renewcommand\@biblabel[1]{#1}            
\renewcommand\@makefntext[1]% 
{\noindent\makebox[0pt][r]{\@thefnmark\,}#1}
\makeatother 
\renewcommand{\figurename}{\small{Fig.}~}
\sectionfont{\sffamily\Large}
\subsectionfont{\normalsize}
\subsubsectionfont{\bf}
\setstretch{1.125} %In particular, please do not alter this line.
\setlength{\skip\footins}{0.8cm}
\setlength{\footnotesep}{0.25cm}
\setlength{\jot}{10pt}
\titlespacing*{\section}{0pt}{4pt}{4pt}
\titlespacing*{\subsection}{0pt}{15pt}{1pt}
%%%END OF PAGE SETUP%%%

%%%FOOTER%%%
\fancyfoot{}
\fancyfoot[LO,RE]{\vspace{-7.1pt}\includegraphics[height=9pt]{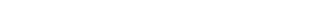}}
\fancyfoot[CO]{\vspace{-7.1pt}\hspace{13.2cm}\includegraphics{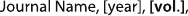}}
\fancyfoot[CE]{\vspace{-7.2pt}\hspace{-14.2cm}\includegraphics{head_foot/RF}}
\fancyfoot[RO]{\footnotesize{\sffamily{1--\pageref{LastPage} ~\textbar  \hspace{2pt}\thepage}}}
\fancyfoot[LE]{\footnotesize{\sffamily{\thepage~\textbar\hspace{3.45cm} 1--\pageref{LastPage}}}}
\fancyhead{}
\renewcommand{\headrulewidth}{0pt} 
\renewcommand{\footrulewidth}{0pt}
\setlength{\arrayrulewidth}{1pt}
\setlength{\columnsep}{6.5mm}
\setlength\bibsep{1pt}
%%%END OF FOOTER%%%

%%%FIGURE SETUP - please do not change any commands within this section%%%
\makeatletter 
\newlength{\figrulesep} 
\setlength{\figrulesep}{0.5\textfloatsep} 

\newcommand{\topfigrule}{\vspace*{-1pt}% 
\noindent{\color{cream}\rule[-\figrulesep]{\columnwidth}{1.5pt}} }

\newcommand{\botfigrule}{\vspace*{-2pt}% 
\noindent{\color{cream}\rule[\figrulesep]{\columnwidth}{1.5pt}} }

\newcommand{\dblfigrule}{\vspace*{-1pt}% 
\noindent{\color{cream}\rule[-\figrulesep]{\textwidth}{1.5pt}} }

\makeatother
%%%END OF FIGURE SETUP%%%

%%%TITLE, AUTHORS AND ABSTRACT%%%
\twocolumn[
  \begin{@twocolumnfalse}
{\includegraphics[height=30pt]{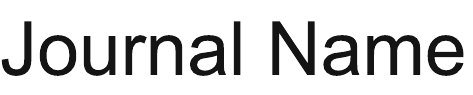}\hfill\raisebox{0pt}[0pt][0pt]{\includegraphics[height=55pt]{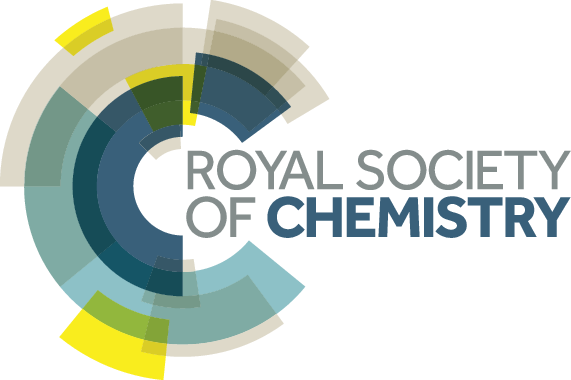}}\\[1ex]
\includegraphics[width=18.5cm]{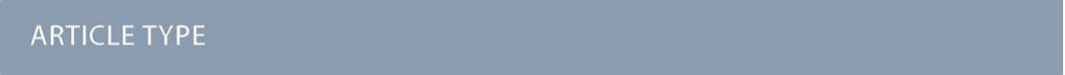}}\par
\vspace{1em}
\sffamily
\begin{tabular}{m{4.5cm} p{13.5cm} }

\includegraphics{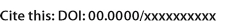} & \noindent\LARGE{\textbf{Supercoiled ring polymers under shear flow}} \\%Article title goes here instead of the text "This is the title"
\vspace{0.3cm} & \vspace{0.3cm} \\

& \noindent\large{
 Christoph Schneck,\textit{$^{{a,b}}$} Jan Smrek,\textit{$^{{a}\ddag}$} Christos N.\ Likos,$^{\ast}$\textit{$^{{a}}$} and Andreas Z{\"o}ttl\textit{$^{{a}\ddag}$}
 } \\%Author names go here instead of "Full name", etc.

\includegraphics{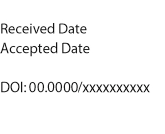} & \noindent\normalsize{We apply monomer-resolved computer simulations of supercoiled ring polymers under shear, taking full account of the hydrodynamic interactions, accompanied, in parallel, by simulations in which these are switched off. The combination of bending and torsional rigidities inherent in these polymers, in conjunction with hydrodynamics, has a profound impact on their flow properties. In contrast to their flexible counterparts, which dramatcially deform and inflate 
under shear [Liebetreu \textit{et al.}, \textit{Commun.\ Mater.} 2020, {\bf 1}, 4], supercoiled rings undergo only weak changes
in their overall shape and they display both a reduced propensity
to tumbling (at fixed Weissenberg number) and a much stronger orientational resistance with respect to their flexible counterparts. In the presence of hydrodynamic interactions, the coupling of the polymer to solvent flow is capable of bringing about a topological transformation of writhe to twist at strong shear upon conservation of the overall linking number.} \\%The abstract goes here instead of the text "The abstract should be..."

\end{tabular}

 \end{@twocolumnfalse} \vspace{0.6cm}

  ]
%%%END OF TITLE, AUTHORS AND ABSTRACT%%%

%%%FONT SETUP - please do not change any commands within this section
\renewcommand*\rmdefault{bch}\normalfont\upshape
\rmfamily
\section*{}
\vspace{-1cm}

%%%FOOTNOTES%%%

\footnotetext{\textit{$^{a}$~Faculty of Physics, University of Vienna, Boltzmanngasse 5, 1090 Vienna, Austria. E-mail: christoph.schneck@ehu.eus, jan.smrek@univie.ac.at, christos.likos@univie.ac.at, andreas.zoettl@univie.ac.at}}
\footnotetext{\textit{$^{b}$~Centro de Física de Materiales (CSIC, UPV/EHU) and Materials Physics Center MPC, Paseo Manuel de Lardizabal 5, 20018 San Sebastián, Spain.}}
%\footnotetext{\textit{$^{b}$~Address, Address, Town, Country. }}

%Please use \dag to cite the ESI in the main text of the article.
%If you article does not have ESI please remove the the \dag symbol from the title and the footnotetext below.
%\footnotetext{\dag~Electronic Supplementary Information (ESI) available: [details of any supplementary information available should be included here]. See DOI: 00.0000/00000000.}
%additional addresses can be cited as above using the lower-case letters, c, d, e... If all authors are from the same address, no letter is required

\footnotetext{\ddag~These authors contributed equally to this work.}

%%%END OF FOOTNOTES%%%

%%%MAIN TEXT%%%%

\section{Introduction}

%General introduction about ring polymers and supercoiled ring polymers, and importance in biology - please Jan and Christos can you add something here? Include modeling approaches used so far for such polymers and summary of most important previous results? Explain why it is useful to study them under flow?
DNA molecules have been largely adopted in nanotechnology thanks to the designability and specificity of the complementary sequences, allowing for construction of highly complex meso-structures from DNA nano-components. \cite{Pinheiro2011}  It is being progressively recognized that also topology \cite{Seeman1993, Chen1991} and geometry of DNA molecules are highly versatile control parameters for future material applications. \cite{Stiakakis2021,janspaper} While it is well known that any particular chain topology (linear, branched, circular...) affects the effective interactions with other chains and the solvent, \cite{Likos2001,Stano2023a} the DNA with ring topology provides extra  tuning parameters. Besides the knottedness, \cite{Narros2010} joining of the ends of a linear double-stranded DNA into the circular unknotted form allows to couple the topological and the geometrical features of the molecule by locking in any pre-existing excess linking between the two strands. Such ring molecules can then exhibit supercoiled conformations as a result of the competition between the bending and the torsional stiffness. % in the presence of other forces in the system. 
While much of the interest has been initially devoted to equilibrium structures, \cite{Vologodskii1979,janspaper} many of the challenges for technological applications of DNA materials lie out of equilibrium. The material response functions, crucial in material characterization and processing, are strongly impacted out of the linear regime as the polymeric degrees of freedom have to adapt to the external stresses in the presence of hydrodynamic interactions and other chains. 

Besides the materials perspective, the properties of non-equilibrium supercoiled DNA polymers touch upon two other highly active research areas. First, the circular DNA is found naturally in the form of (typically supercoiled) bacterial plasmids, \cite{Phillips2004}
% transcription induces supercoiling...
extrachromosomal DNA of eukaryotes,\cite{Koche2020} kinetoplast DNA \cite{Klotz2020a, Chen1995a} of trypanosoma and supercoiled segments have been suspected to form distinct architectural and functional domains also on chromosomal DNA. \cite{Racko2018} Supercoiling itself can affect gene expression \cite{Ding2014} or DNA metabolism \cite{Irobalieva2015} by modulating access to distinct regions. In the biological context the DNA molecules are also typically out of equilibrium, subject to flows and stresses arising through the action of molecular motors and inducing non-equilibrium conformations \cite{Fudenberg2016} and dynamics \cite{Zidovska2013} that in turn can impact the biological function.\cite{Arnould2021} 
 
Second, the properties of unknotted ring polymers under various conditions are one of the largest unsolved problems in polymer physics. The crux lies in the difficulty of capturing the restrictions that the fixed ring topology imposes on the phase space in the presence of other interactions. This general principle is present across different concentration regimes. In dense many-chain systems the chain topology prevents mutual ring linking, resulting in compact, fractal tree-like, conformations,\cite{Halverson2011a,Rosa2014} power-law stress relaxation,\cite{Kapnistos2008a} distinct shear-thinning exponents,\cite{Parisi2021} extensional viscosity thickening,\cite{Huang2019,Oconnor2020}  viscosity of ring-linear blend exceeding that of the two components \cite{Halverson2012,OConnor2022} or vitrification 
upon heterogeneous activation.\cite{Smrek2020a} Some of these effects are due to ring interpenetration (threading) by other chains, although their general impact on the viscoelastic properties is still an open question. \cite{Ge2016,Smrek2015,Rosa2019,Ghobadpour2021} However, the impact of threadings is apparent in simulations and experiments on supercoiled plasmids. There, the supercoiling induces tighter branched ring conformations, reduces mutual ring interpenetration and speeds up the dynamics.\cite{janspaper} Therefore, investigating the systems with torsional degrees of freedom can help to understand dynamical effects also in torsionally relaxed rings, by giving a control on the degrees of the branching and the threading. In semi-dilute and dilute conditions, both, the interchain interactions and the hydrodynamics coupled with the ring topology of the chains affect the chain relaxation dynamics and fluctuations of the metric properties \cite{Schroeder_RL_NATCOM19,Schroeder_Zhou_JR21,Schroeder_Zhou_sims_JR21}.
For dilute solutions under non-equilibrium flow conditions, the combination of topology with hydrodynamic interactions 
has dramatic effects on the conformations and dynamics of the solutes. Vorticity swelling in shear,\cite{liebetreu:acsml:2018} extensional or mixed flows\cite{Hsiao2016,young:pre:2019} has been predicted in simulations and recently confirmed experimentally,\cite{tu:mm:2020} whereas hydrodynamic inflation
under steady shear\cite{liebetreu:commats:2020} results
into a strong suppression of tumbling and of the Brownian nature
of the polymer dynamics for certain domains of shear rate 
for flexible polymers.
To summarize, the supercoiling reduces penetrable area of rings 
which is relevant for viscoelasticity, but hydrodynamic swelling 
acts in the opposite way. Therefore here we aim to study 
systematically in simulation how does the hydrodynamic interaction couple with 
the supercoiled ring structure in dilute (single chain) conditions in shear. To do so, we employ the combination of muliparticle collision dynamics to account for the hydrodynamic interaction in combination with molecular dynamics for the time evolution of the polymer.
  
Under non-equilibrium conditions such as 
shear flow that we focus on in here,\cite{Ripoll2007,liebetreu:commats:2020} non-local hydrodynamic effects play an essential role for the transport and dynamical properties of polymers.
Considering the molecular dynamics of all of the fluid particles interacting with the polymers would be too time-consuming on the typical experimental time-scales (seconds). Hence, several coarse-grained simulation techniques had been developed which capture hydrodynamic interactions in polymers, such as dissipative particle dynamics (DPD),\cite{Pastorino2007} the lattice Boltzmann method (LBM) \cite{Dunweg2009} and multiparticle collision dynamics (MPCD).\cite{Gompper2009}
In particular MPCD had been heavily used in the last two decades to study polymer dynamics under equilibrium and under shear flow conditions. In MPCD the Newtonian background solvent  is modeled by effective point-like fluid particles which perform alternating streaming and collision steps.\cite{Kapral2008,Gompper2009}
While intramolecular forces are advanced through molecular dynamics (MD) purely between monomers, the hydrodynamic effects on the monomers stem from monomer-fluid particle interactions.
In the simplest and most popular approach of such a hybrid MPCD-MD scheme the monomers are assumed to be pointlike but heavier than the fluid particles and exchange momentum with the local fluid particles in the collision step.\cite{Malevanets2000} 
This allows simulations of single/dilute\cite{Winkler2004,Ryder2006,Ripoll2006} and 
semidilute/dense\cite{Ripoll2008,huang:mm:2010} polymer solutions in the presence of an explicit solvent.
Simple unidirectional shear flow is usually realized by applying 
Lees-Edwards boundary conditions,\cite{Lees1972,Winkler2004,Ryder2006,Kikuchi2003}
and the flow can be disturbed by the presence of polymers and their coupling to the fluid motion. 

MPCD had been used to quantify the shear-thinning behavior of linear flexible,\cite{Ryder2006,huang:mm:2010} semiflexible \cite{Nikoubashman2017} and stiff \cite{Winkler2004,Ripoll2008} polymer solutions, as well as the dynamics of single crosslinked polymer chains.\cite{Formanek2019,Novev2020}
Furthermore, the MPCD method can easily be adapted to simulate pressure-driven Poiseuille flow bounded by no-slip channel walls.\cite{Allahyarov2002,Bolintineanu2012,weiss:acsml:2017}
Consequently the dynamics of single linear flexible \cite{Cannavacciuolo2008} and 
semiflexible \cite{Chelakkot2010} polymers, as well as the shape transitions of a tethered 
linear polymer \cite{Webster2005} had been investigated.
The properties of polymers with other architectures under shear flow had also be studied in detail, such as star polymers and dendrimers,\cite{Ripoll2006,Ripoll2007,Nikoubashman2010b,Fedosov2012,Singh2013,Jaramillo2018,Toneian2019,Jaramillo2020} as well as ring polymers.\cite{Chen2013,Chen2013b,liebetreu:acsml:2018,liebetreu:commats:2020} Noteworthy, in MPCD hydrodynamic interactions can be turned off easily without modifying other physical 
properties of the fluid,\cite{Ripoll2007,liebetreu:commats:2020}. 
This allows to directly determine the influence of hydrodynamic interactions in 
polymer simulations, of particular importance in out-of-equilibrium conditions. \cite{Zottl2019,Qi2020,Zottl2023a}

The article is organized as follows: In sec.~\ref{sec:model0}
we introduce the model and in sec.~\ref{sec:polymer} the quantities that characterize the topology and the 
shapes of the supercoiled polymers in equilibrium and under shear. In secs.~\ref{sec:equ} and \ref{sec:shear} we present the extracted equilibrium properties and the 
behavior under shear flow of supercoiled ring polymers, respectively.
We conclude our work in sec.~\ref{sec:conc}, whereas certain technical details of our work are 
delegated to the Appendix.

\section{Model}
\label{sec:model0}

\subsection{Multiparticle collision dynamics}

Multiparticle collision dynamics (MPCD) is a method of simulating a particle-based solvent put forward by Malevanets and Kapral \cite{mpcd1999} that is used to generate hydrodynamic interactions (HI) and mimic thermal fluctuations in 
molecular dynamics (MD) simulations of embedded objects such as colloids and polymers. It models the solvent fluid as $N_{\mathrm{s}}$ individual point particles with mass $m_{\mathrm{s}}$ within a rectangular simulation box with side lengths $L_x,L_y,L_z$. Positions and velocities are described by continuous variables $\vec{r}_{\mathrm{s},i}, \vec{v}_{\mathrm{s},i}$, $i \in \{1,...,N_{\mathrm{s}}\}$. The algorithm consists of two alternating steps. During the streaming step the particles are allowed to move ballistically for a time period $h$ according to their current velocities:
\begin{equation}
    \vec{r}_{\mathrm{s},i}(t+h)=\vec{r}_{\mathrm{s},i}(t)+h\vec{v}_{\mathrm{s},i}(t).
    \label{streaming step}
\end{equation}

Periodic boundary conditions are applied for particles crossing the dimensions of the simulation box. In the subsequent collision step the entire simulation box is coarse-grained into cubic cells of a grid with lattice constant $a_{\mathrm{c}}$ and the solvent particles are sorted according to their current positions. The velocities of all $N_{\mathrm{c}}$ particles in one particular cell are summed up to a center-of-mass velocity, 
$\vec{v}_{\mathrm{cm}}(t)=\frac{1}{N_{\mathrm{c}}}\sum_{i \in \mathrm{cell}}\vec{v}_{\mathrm{s},i}(t)$, and a rotation matrix $\textbf{R}[\alpha]$ is generated and assigned to the cell. The orientation of the rotation axis is picked randomly at each step and varies from cell to cell, whereas the rotation angle $\alpha$ is fixed. This randomness is the ingredient endowing MPCD with thermal fluctuations. The velocities of all particles within the cell are then rotated relative to the center-of-mass velocity of the cell,
\begin{equation}
    \vec{v}_{\mathrm{s},i}(t+h)=\vec{v}_{\mathrm{cm}}(t)+\textbf{R}[\alpha](\vec{v}_{\mathrm{s},i}(t)-\vec{v}_{\mathrm{cm}}(t)),
    \label{collision step}
\end{equation}
which represents the collision of the particles. The procedure is conducted for all $L_xL_yL_za_{\mathrm{c}}^{-3}$ cells contained in the simulation box and allows for local momentum exchange between particles within a cell without changing $\vec{v}_{\mathrm{cm}}$. The internal interaction of the simulated fluid is coarse-grained both in space (the division into cells) and time (collision at discrete time steps). Because of the lack of systematic forces, MPCD solvent particles obey an ideal gas equation of state \cite{Gompper2009} and one can show that the algorithm conserves mass, momentum, and energy locally at cell level. \cite{mpcd1999} MPCD is an effective way of solving the Navier-Stokes equations and reproducing hydrodynamic interactions and thermal fluctuations down to the size of a collision cell.

There are three basic units all set to unity: the mass $m_{\mathrm{s}}$ of the fluid particles, the length of the cubic collision cells $a_{\mathrm{c}}$, and the energy $k_{\mathrm{B}} T$ specifying the average kinetic energy of a solvent particle by 
$3k_{\mathrm{B}} T= m_{\mathrm{s}}\langle\vec{v}_{\mathrm{s},i}^2\rangle$. This corresponds to expressing length, energy, and mass in units of $a_{\mathrm{c}}$, $k_{\mathrm{B}} T$, and $m_{\mathrm{s}}$. A reference unit for time can then be derived as 
$\tau_{\mathrm{MPCD}}=a_{\mathrm{c}}\sqrt{{m_{\mathrm{s}}}/{k_{\mathrm{B}} T}}=1$. The box dimensions are set to $L_x = 80a_{\mathrm{c}}$, $L_y = 50a_{\mathrm{c}}$, $L_z = 50a_{\mathrm{c}}$, which has been shown to be large enough to avoid flow interference through periodic boundary conditions or polymer self-interactions for similar simulations \cite{liebetreu:commats:2020}. The box is rectangular with its
long side in the shear flow direction - see later below the description of the system. The time between collisions is set to $h = 0.1\tau_{\mathrm{MPCD}}$, 
the collision angle $\alpha = 130^\circ$
and the solvent particle number density $\rho_{\mathrm{s}} = \frac{N_{\mathrm{s}}}{L_x L_y L_z} = 10a_{\mathrm{c}}^{-3} \equiv \langle N_{\mathrm{c}} \rangle a_{\mathrm{c}}^{-3}$, where $\langle N_{\mathrm{c}} \rangle = 10$
is the average number of solvent particles in each collision cell.
This choice of parameters corresponds to a dynamic viscosity \cite{GK_resummation} of 
$\eta = 8.7\,\tau_{\mathrm{MPCD}}^{-1} m_{\mathrm{s}} a_{\mathrm{c}}^{-1}$.

For short time steps $h$ the mean free path $\Lambda=h\sqrt{\frac{k_{\mathrm{B}} T}{m_{\mathrm{s}}}}$ between consecutive collisions is small compared to the cell size. The discrete nature of the grid causes particles to stay in the same cell for several collisions before particle exchange with neighbouring cells occurs. This causes particles within one collision cell to build up correlations and breaks the translational symmetry of the system. Galilei invariance can be recovered by performing a random shift of the sorting grid by the shift vector $\delta \vec{r}$ with the components that are sampled uniformly from $[-\frac{a_{\mathrm{c}}}{2},\frac{a_{\mathrm{c}}}{2}]$. Additionally, the shifting procedure accelerates momentum transfer between particles. \cite{gridshift}

 To generate shear flow with a tunable shear rate $\dot{\gamma}$ the periodic boundaries are modified to Lees-Edwards boundary conditions. \cite{Lees1972} 
 This defines three distinct space directions: the shear motion is along the 
 flow direction ($x$-axis), the magnitude of shear flow varies along the gradient direction ($y$-axis), and the remaining one is called the vorticity direction ($z$-axis). Thereby, energy is 
 constantly pumped into the system, and then is converted into heat due to the MPCD-fluid viscosity, which, if left unattended, would result into 
 constant viscous heating of the system. A thermostat is thus necessary in such a non-equilibrium system 
 to maintain the temperature at its prescribed value. We employ here a cell-level thermostat, \cite{Gompper2009,huang2015thermostat} which further ensures that the particle velocities follow a Maxwell-Boltzmann distribution and it 
 restricts the fluctuations of the total energy of the system to those leading to the desired value $k_{\mathrm{B}} T$. 
 
To create coupling with the fluid and thereby enable HI, we include the monomers in the MPCD collision, step\cite{Malevanets2000} but other approaches also exist.\cite{Lee2006} An embedded monomer $i$ with mass $m_{\mathrm{m}}$ and velocity $\vec{v}_{i}$ contributes to the center-of-mass velocity of the 
collision cell. If $N_{\mathrm{m}}$ is the number of monomers in a particular cell, the center-of-mass velocity is
\begin{equation}
    \label{vcompolymer}
    \vec{v}_{\mathrm{cm}}=\frac{m_{\mathrm{s}} \sum_{i \in \mathrm{cell}}\vec{v}_{\mathrm{s},i} + m_{\mathrm{m}} \sum_{i \in \mathrm{cell}}\vec{v}_{i}}{N_{\mathrm{s}} m_{\mathrm{s}} + N_{\mathrm{m}} m_{\mathrm{m}}}.
\end{equation}
A stochastic rotation as in eq.~(\ref{collision step}) is then performed on the relative velocities of both the solvent particles and the embedded monomers. This results in an exchange of momentum and energy between all particles within a collision cell. The updated monomer velocities are then used as initial conditions for the subsequent MD time step. 100 MD time steps are performed between consecutive MPCD collisions. 

Gompper \textit{et al.}~\cite{Gompper2009} have noted that the average number of monomers per collision cell should be smaller than unity in order to properly resolve HI between them. The bond length of polymer chains should be of order of the cell size $a_{\mathrm{c}}$ so that monomers are close enough to display effects due to HI but far enough apart to avoid multiple monomers within a cell. Furthermore, 
Ripoll \textit{et al.}~\cite{Ripoll2005} have demonstrated that the mass ratio ${m_{\mathrm{m}}}/{m_{\mathrm{s}}}$ should equal the average number of solvent particles per cell $\langle N_{\mathrm{c}} \rangle$ (assuming there is only one monomer per cell). Under this condition the monomer collides with a piece of solvent with equal mass so that the mutual momentum exchange is balanced. For $m_{\mathrm{m}} \gg \langle N_{\mathrm{c}} \rangle m_{\mathrm{s}}$ the polymer is barely affected by the solvent, whereas for $m_{\mathrm{m}} \ll \langle N_{\mathrm{c}} \rangle m_{\mathrm{s}}$ the monomers are kicked around too violently. Hence we chose $m_{\mathrm{m}}=\langle N_{\mathrm{c}} \rangle m_{\mathrm{s}} = 10m_{\mathrm{s}}$.

In order to study the specific effects of the hydrodynamic interactions on the dynamics of the system the same simulations have to be performed with and without HI. An alternative concept of simulating the solvent and the interaction with the monomer has to be found, that differs as little as possible from the original conditions except for the presence of HI. 
Ripoll \textit{et al.} invented an efficient method to switch off HI in a MPCD algorithm.\cite{random_mpcd} They suggest that the positions of the solvent particles do not contribute in the collision step and therefore, no explicit particles have to be considered at all. Instead, every monomer is coupled with an effective solvent momentum $\vec{P}$ that is chosen randomly from a Maxwell-Boltzmann distribution of variance $m_{\mathrm{s}}\langle N_{\mathrm{c}} \rangle k_{\mathrm{B}} T$ and zero mean. A center-of-mass velocity is assigned to every monomer velocity $\vec{v}_{i}$ given by:
\begin{equation}
    \label{random_collision}
    \vec{v}_{\mathrm{cm},i} = \frac{m_{\mathrm{m}} \vec{v}_{i} + \vec{P}}{m_{\mathrm{m}} + m_{\mathrm{s}} \langle N_{\mathrm{c}} \rangle},
\end{equation}
with which the collision step is performed. Then the same rotation angle is used as in standard MPCD simulations. The algorithm has similar properties to MPCD, but it does not include HI because there are no particles carrying momentum from one monomer to another. Shear flow with the shear rate $\dot{\gamma}$ is added to the random solvent by setting the mean of $\vec{P}$ from zero to $\left(m_{\mathrm{s}} N_{\mathrm{c}}^P \dot{\gamma} r_{iy}, 0, 0\right)$, where the monomer's position in gradient direction $r_{iy}$ determines the local shear velocity in flow direction. The number of collision partners $N_{\mathrm{c}}^P$ is picked randomly from a Poisson distribution with expectation value $\langle N_{\mathrm{c}} \rangle$ which corresponds to the distribution of the number density in a particle-based MPCD cell. In the following this heat bath will be referred to as random MPCD solvent or `$-$HI', meaning there are no hydrodynamic interactions, whereas the conventional particle-based procedure, with hydrodynamic interactions, will be referenced to as `$+$HI'.

\subsection{Polymer model}

%%%%%%%%%%%%%%%%%%%%%%%%%%%%%%%%%%%%%%%%%%%%%%%%%%%%%%%%%%%%%%%%%%%%%%%%%%%%%%
\begin{figure*}[htb]
\centering
  \includegraphics[width=0.6\textwidth]{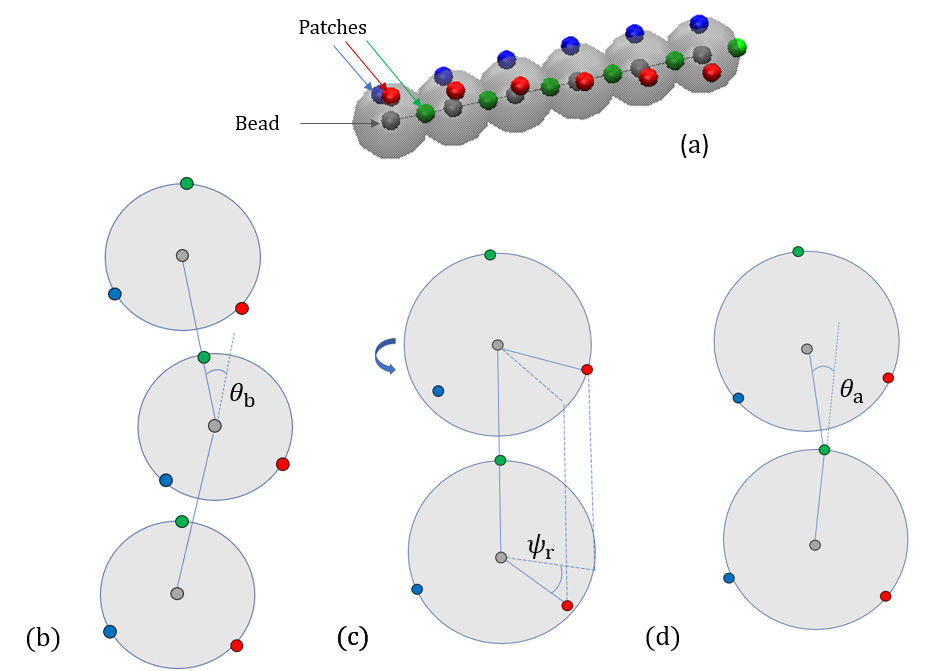}
  \caption{Illustration of monomers endowed with patches, [(a)], and of potentials concerning bending and 
torsional degrees of freedom, [(b)-(d)]. (a) The grey surface indicates the extension of the
monomer’s rigid body with radius $l_{\mathrm{pat}}$. The bead positions are the grey centers and the three
patches are fixed on the surface. They are used to define: (b) bending between consecutive triplets forming the angle $\theta_{\mathrm{b}}$, (c) torsion between neighbouring monomers (the red torsional angle $\psi_{\mathrm{r}}$ is depicted as an example, the blue patches form a dihedral system as well which is omitted here), and (d) alignment of the green patches with the connection vector $\vec{r}_{i+1}-\vec{r}_{i}$.}
  \label{Fig:patches}
\end{figure*}
%%%%%%%%%%%%%%%%%%%%%%%%%%%%%%%%%%%%%%%%%%%%%%%%%%%%%%%%%%%%%%%%%%%%%%%%%%%%%%

We use a coarse-grained polymer model used previously\cite{janspaper} with parameters tuned to represent typical properties of double-stranded DNA (see below).
The polymer rings consist of $N = 100$ beads with $\vec{r}_{i}$, $\vec{v}_{i}$ representing their positions and velocities respectively. Their time evolution is governed by a velocity Verlet algorithm with time step length 
$\delta t = 10^{-2} h$. The beads are subject to a pair-wise Weeks-Chandler-Andersen (WCA) potential \cite{liebetreu:commats:2020, janspaper} with a cutoff distance at $r_{\mathrm{WCA}} = \sqrt[6]{2} \sigma$,
\begin{equation}
    \label{WCA}
    U_{\mathrm{WCA}}(r_{ij}) = 4\epsilon\left[\left(\frac{\sigma}{r_{ij}}\right)^{12} - \left(\frac{\sigma}{r_{ij}}\right)^6 + \frac{1}{4} \right] \Theta\left(r_{\mathrm{WCA}}-r_{ij}\right),
\end{equation}
which 
simulates an excluded volume and a soft repulsion between any two beads $i$ and $j$ at distance $r_{ij} = \left|\vec{r}_i - \vec{r}_j \right|$. 
The parameters $\epsilon$ and $\sigma$ define energy and length scales of the polymer and are set to $\epsilon = k_{\mathrm{B}} T = 1$ and $\sigma = a_{\mathrm{c}} = 1$.

Neighbouring beads are attached to each other along the backbone by applying a pair-wise attractive FENE (finitely extensible nonlinear elastic) potential given by
\begin{equation}
    \label{FENE}
    U_{\mathrm{FENE}}(r_{ij}) = -\frac{1}{2}k_{\mathrm{FENE}}R_0^2\, \mathrm{ln}\left[ 1 - \left(\ \frac{r_{ij}}{R_0} \right)^2\right],
\end{equation}
where $k_{\mathrm{FENE}} = 40 k_{\mathrm{B}} T$ determines the strength of attraction and $R_0 = 1.6\sigma$ sets the maximal extent of the spring. \cite{liebetreu:commats:2020,janspaper} The combination of potentials (\ref{WCA}) and (\ref{FENE}) acting on a pair of beads results in an effective bonding potential with a minimum at $r_{min}=0.94\sigma$. The monomer point particle $i$ interacts with its respective neighbours $i-1$ and $i+1$ via FENE. In what follows when talking about neighbors $i-1$ or $i+1$ of bead $i$ we always mean neighbors in respecting the ring topology, i.e.~if $i=N$, its neighbors are $N-1$ and $1$, and if $i=1$, its neighbors have indices $N$ and $2$. The WCA potential acts on all pairs $ij$ keeping two beads from overlapping. All rings are constructed as the trivial knot, i.e., they are unknotted.

We further include bending and torsional stiffness as in Smrek \textit{et al.},~\cite{janspaper} to mimic the coarse-grained properties of the DNA. We show a sketch of the polymer in Fig.~\ref{Fig:patches}(a),
and we describe the model in more 
detail in what follows. The bending energy is 
introduced between every consecutive triplets of chain neighbours $\vec{r}_{i-1}$, $\vec{r}_{i}$, and 
$\vec{r}_{i+1}$ (again respecting the ring topology):
\begin{equation}
    \centering
    U_{\mathrm{bend}}(\theta_{\mathrm{b}}) = k_{\mathrm{bend}}\left( 1 - \cos\theta_{\mathrm{b}} \right),
    \label{bending_potential}
\end{equation}
where $\theta_{\mathrm{b}}$ is the angle formed by two adjacent bonds $\vec{r}_{i-1,i}$ and $\vec{r}_{i,i+1}$, 
see Fig. \ref{Fig:patches}(b), so that
\begin{equation}
\cos\theta_{\mathrm{b}} = \frac{\vec{r}_{i-1,i} \cdot \vec{r}_{i,i+1}} {|\vec{r}_{i-1,i}| |\vec{r}_{i,i+1}|}.
\label{eq:costhb}
\end{equation}
 The bending constant is chosen to be $k_{\mathrm{bend}} = 20 k_{\mathrm{B}} T$ leading to a persistence length $l_{\mathrm{p}} = 20 \sigma$. The scale of the monomer diameter should be mapped\cite{janspaper} to DNA's thickness in real units of $\sigma = 2.5\,\mathrm{nm}$, i.e., 
 $2.5/0.34=7.35\,{\mathrm{bp}}$ per bead and $l_{\mathrm{p}} = 150\,{\mathrm{bp}} = 50\,{\mathrm{nm}}$.

To model the torsional degrees of freedom, every bead is equipped with three patches which remain at a fixed distance $l_{\mathrm{pat}} = r_{\mathrm{WCA}}/2 = {\sqrt[6]{2}\sigma}/{2}$. The vectors $\vec{r}_{\mathrm{pat},i,k}$ are defined as the vectors going from the position of bead $i$ to its $k$-th patch where $k = 1,2,3$ so that each triplet $\{\vec{r}_{\mathrm{pat},i,k}\}_{k=1,2,3}$ forms an orthogonal basis with the respective bead position $\vec{r}_i$ as the origin. 
In Fig.~\ref{Fig:patches}(a), we show a short segment of a polymer chain with patches where colours blue, red, and green correspond to the numeration $k = 1, 2, 3$, respectively. 
The blue and red patches are used to introduce two dihedral springs that constrain the torsional angle $\psi_{\mathrm{b/r}}$ between consecutive beads close to a fixed equilibrium value $\psi_0$. 
Here, $\psi_{\mathrm{b/r}}$ is defined as the angle between the planes with normal vectors $\vec{r}_{i,i+1} \times \vec{r}_{\mathrm{pat},i,\mathrm{b/r}}$ and $\vec{r}_{i,i+1} \times \vec{r}_{\mathrm{pat},i+1,\mathrm{b/r}}$, 
see Fig.~\ref{Fig:patches}(c), and 
it is evaluated for the blue and red patches, respectively. The corresponding potential is
given by the expression:
\begin{equation}
    \centering
    \begin{aligned}
        U^{\mathrm{b/r}}_{\mathrm{torsion}}(\psi_{\mathrm{b/r}}) & = k_{\mathrm{torsion}}\left[ 1 - \cos\left(\psi_{\mathrm{b/r}}-\psi_0\right) \right]
        \\
        & = k_{\mathrm{torsion}}\left( 1 - \cos{\psi_0}\cos\psi_{\mathrm{b/r}} - \sin{\psi_0}\sin\psi_{\mathrm{b/r}} \right), 
    \end{aligned}
    \label{torsion_potential}
\end{equation}
with the torsion constant $k_{\mathrm{torsion}} = 50 k_{\mathrm{B}} T$. The equilibrium angle $\psi_{0}$ determines the polymer  degree of supercoiling $\sigma_{\mathrm{sc}} = \psi_0/2\pi$ of the chain. To see the impact of the torsional constraint the simulations are also performed for torsionally unconstrained rings for which $k_{\mathrm{torsion}}$ is set to zero. These are referred to as 'relaxed' rings in the following and they are different to rings with $\sigma_{\mathrm{sc}}=0$ and nonzero $k_{\mathrm{torsion}}$ which are torsionally constrained to a supercoiling of zero. A third potential aligns the patch reference systems so that the blue and red patches' positions relative to their bead is perpendicular to the segments $\vec{r}_{i-1,i}$ and $\vec{r}_{i,i+1}$. The green patch $\vec{r}_{\mathrm{pat},i,\mathrm{g}}$ is forced onto the segment line $\vec{r}_{i,i+1}$ by the alignment potential:
\begin{equation}
        U_{\mathrm{align}}(\theta_{\mathrm{a}}) = k_{\mathrm{align}}\left( 1 - \cos\theta_{\mathrm{a}}\right),
\label{align_potential}
\end{equation}
where
\begin{equation}
    \cos(\theta_{\mathrm{a}})  =  \frac{\vec{r}_{\mathrm{pat},i,\mathrm{g}}} {l_{\mathrm{pat}}} \cdot \frac{\vec{r}_{i,i+1} - \vec{r}_{\mathrm{pat},i,\mathrm{g}}}{|\vec{r}_{i,i+1} - \vec{r}_{\mathrm{pat},i,\mathrm{g}}|}.
    \label{costheta}
\end{equation}
In this way, blue and red patches are aligned orthogonal to the polymer backbone consisting of the beads so that dihedral angles between consecutive beads can be formed properly, see Fig.~\ref{Fig:patches}(d). 
The constant is chosen to be as high as $k_{\mathrm{align}} = 200 k_{\mathrm{B}} T$ to minimize the fluctuations of the alignment.

 Forces derived from all potentials presented act on the beads $i$ whereas torsion, eq.~(\ref{torsion_potential}), and alignment, 
 eq.~(\ref{align_potential}), also yield forces upon the respective patches $k$ denoted by $\vec{F}_{\mathrm{pat},i,k}$. Forces acting on the beads are calculated as
 \begin{equation}
     \centering
 \vec{F}_i(t)=-\vec{\nabla}_{\vec{r}_i}U_{\mathrm{total}}(\{\vec{r}_j(t)\},\{\vec{r}_{\mathrm{pat},j,k}(t)\}),
 \end{equation}
 where $U_{\mathrm{total}}$ is total potential. The forces acting on the patch are 
 \begin{equation}
 \begin{aligned}
     \centering
 \vec{F}_{\mathrm{pat},i,k}(t) & =-\vec{\nabla}_{\vec{r}_{\mathrm{pat},i,k}} U_{\mathrm{total}} (\{\vec{r}_j(t)\}, \{\vec{r}_{\mathrm{pat},j,k}(t)\}) 
 \\
 & = -\vec{\nabla}_{\vec{r}_{\mathrm{pat},i,k}} U_{\mathrm{pat}}(\{\vec{r}_j(t)\},\{\vec{r}_{\mathrm{pat},j,k}(t)\})
 \end{aligned}
 \end{equation}
 where  $U_{\mathrm{total}} =  U_{\mathrm{FENE}} + U_{\mathrm{WCA}} + U_{\mathrm{bend}} + U_{\mathrm{torsion}}^{b} + U_{\mathrm{torsion}}^{r} + U_{\mathrm{align}}$ and $U_{\mathrm{pat}} =  U_{\mathrm{torsion}}^{b}\delta_{k,1} + U_{\mathrm{torsion}}^{r}\delta_{k,2} + U_{\mathrm{align}}\delta_{k,3}$. The total force on monomer $i$ is computed as 
\begin{equation}
\centering
\vec{F}_{\mathrm{total},i}(t)=\vec{F}_i(t)+\sum_{k=1}^3 \vec{F}_{\mathrm{pat},i,k}(t),
\end{equation}
 which is used as the force acting on the momenta in the velocity Verlet algorithm. The details about the computation of forces and potentials in dependence of bead and patch positions are described in the Appendix. 

Adding patches to the beads effectively transforms the previous point particles $i$ to rigid bodies with a rotational degree of freedom $\chi_i$, angular velocity $\vec{\omega}_i$, and inertia tensor $I_i^{\alpha\beta}$. The set of three angles $\chi_i$ defines the orientation of the monomer $i$ with respect to the lab frame, and hence define the positions of the patches $\vec{r}_{\mathrm{pat},i,k}$ on the monomer $i$. Their explicit representation with quaternions is discussed in the Appendix. The monomers are modeled as extended 
spheres with radius $l_{\mathrm{pat}}$ and continuous mass density $\varrho_{\mathrm{m}} = {3m_{\mathrm{m}}}/({4 \pi l_{\mathrm{pat}}^3})$, which corresponds to a trivial inertia tensor $I_i^{\alpha\beta} = I_{\mathrm{sphere}}\delta_{\alpha,\beta}$ with the 
moment of inertia $I_{\mathrm{sphere}} = \frac{2}{5} m_{\mathrm{m}} l_{\mathrm{pat}}^2$. The position of the bead $\vec{r}_i$ is the central point of the rigid body and coincides with the center-of-mass of the monomer, whereas the patches reside on the sphere's surface. Therefore, only the forces $\vec{F}_{\mathrm{pat},i,k}(t)$ contribute to the total torque $T_i(t)$ on monomer $i$, expressed as
\begin{equation}
    \centering
    \begin{aligned}
    \vec{T}_i(t) & = \sum_{k=1}^3 \vec{r}_{\mathrm{pat},i,k}(t) \times \vec{F}_{\mathrm{pat},i,k}(t)=\frac{\mathrm{d}}{\mathrm{d}t}\vec{L}_i(t)
    \\
    & =\frac{\mathrm{d}}{\mathrm{d}t}\left(I_i^{\alpha\beta} \vec{\omega}_i (t) \right)=I_{\mathrm{sphere}}\,\dot{\vec{\omega}}_i (t),
    \end{aligned}
\end{equation}
with the angular momentum ${\vec L}_i(t)$ of bead $i$.
Rotations around the center-of-mass only affect the orientations $\chi_i$ but not the bead positions. The rotational dynamics are added to the velocity Verlet algorithm, see Appendix for details. When the polymer is suspended in the MPCD solvent, the collisions only rotate the monomer velocities $\vec{v}_i$ but not the angular velocities $\vec{\omega}_i$ and so the solvent-solute interaction does not affect the rotational degrees directly. This is an approximation originating in our choice of the coarse-grained model. More details on the force calculations as well as on the integration of the
equations of motion for the translations and the rotations of the beads are given in the Appendix.

\section{Polymer topology and shapes}
\label{sec:polymer}

\subsection{Writhe and twist}

The conformations of the supercoiled ring polymers are influenced by an interplay of the applied bending and torsional constraints. Their geometry is analyzed by measuring the topological parameters writhe $\mathrm{Wr}(t;\dot{\gamma})$ and twist $\mathrm{Tw}(t;\dot{\gamma})$, where we explicitly denote the time-dependence of these quantities as well as their parametric dependence on the
applied shear rate $\dot\gamma$.
The former is defined for any instantaneous configuration $C_t$ of a closed curve in three-dimensional space as:
\begin{equation}
    \centering
    {\mathrm{Wr}}(t;\dot{\gamma}) = \frac{1}{4\pi} \bigintsss_{C_t} \bigintsss_{C_t} \frac{\left({\mathrm d}\vec{r}_2 \times {\mathrm d}\vec{r}_1 \right)\cdot \vec{r}_{12}}{r_{12}^3},
    \label{writhe_formula}
\end{equation}
where $\vec{r}_1$ and $\vec{r}_2$ are points on the curve $C_t$ and $\vec{r}_{12} = \vec{r}_2 - \vec{r}_1$ at a given time $t$. The writhe gives an intuitive measure of how many crossings an average 2D projection of the molecular backbone forms with itself. For the actual computation we use a version of 
eq.~(\ref{writhe_formula}) detailed in Klenin and Langowski,\cite{writhe2000} where the bead positions serve as the reference points of the discretized curve $C_t$.

%%%%%%%%%%%%%%%%%%%%%%%%%%%%%%%%%%%%%%%%%%%%%%%%%%%%%%%%%%%%%%%%%%%%
\begin{figure}[htb]
\centering
  \includegraphics[width=\columnwidth]{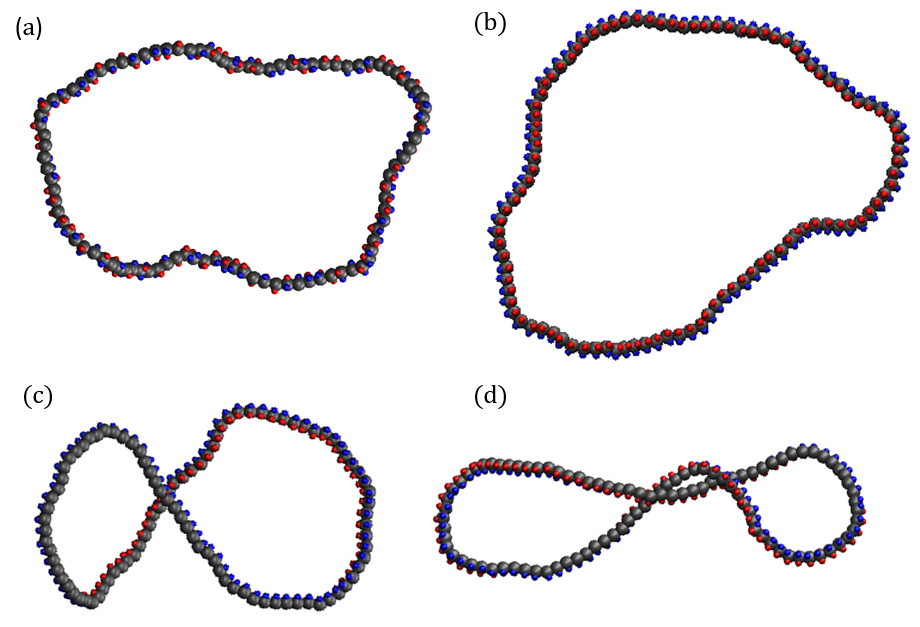}
  \caption{Snapshots of the supercoiled polymers at equilibrium ($\dot\gamma = 0$). 
  (a) Relaxed ring with $k_{\mathrm{torsion}}=0$, (b) $\sigma_{\mathrm{sc}}=0.00$, (c) $\sigma_{\mathrm{sc}}=0.01$, (d) $\sigma_{\mathrm{sc}}=0.02$.}
  \label{Fig:snapshots_zero_shear}
\end{figure}
%%%%%%%%%%%%%%%%%%%%%%%%%%%%%%%%%%%%%%%%%%%%%%%%%%%%%%%%%%%%%%%%%%%%

The twist is computed for the torsionally constrained polymer rings by summing over the difference of all dihedral angles from the respective equilibrium value $\psi_0$:
\begin{equation}
    \centering
    {\mathrm{Tw}}(t;\dot{\gamma}) = \frac{1}{2\pi} \sum_{i=1}^{N} \biggl[\frac{\psi_{\mathrm{b},i}(t;\dot\gamma)+\psi_{\mathrm{r},i}(t;\dot\gamma)}{2} - \psi_0 \biggr].
    \label{twist_formula}
\end{equation}
The blue and red patches each form an individual set of dihedral angles $\{\psi_{\mathrm{b},i}(t;\dot\gamma)\}$, respectively $\{\psi_{\mathrm{r},i}(t;\dot\gamma)\}$ which is why an arithmetic mean over the two is taken in the evaluation of $\mathrm{Tw}(t;\dot{\gamma})$ in eq.~(\ref{twist_formula}).
The alignment potential forces the green patches to be orientated nearly perfectly with the polymer backbone so that the blue and red patches are kept in a perpendicular orientation to it. A high value is picked for the constant $k_{\mathrm{align}}=200 k_{\mathrm{B}} T$ which limits the fluctuations from this structure. For this reason, the vectors $\vec{r}_{\mathrm{pat},i,b/r}$ serve as (unnormalized) normal vectors to the space curve approximated by the bead positions $\vec{r}_i$. With the brackets $\langle ... \rangle_t$ denoting averaging over time, we also define the quantities:
\begin{equation}
{\mathrm{Wr}}(\dot{\gamma}) = \langle{\mathrm{Wr}}(t;\dot{\gamma})\rangle_t;\,\, {\mathrm{Tw}}(\dot{\gamma}) = \langle{\mathrm{Tw}}(t;\dot{\gamma})\rangle_t,
\label{eq:wrtwav}
\end{equation}
expressing expectation values as functions of the applied shear rate $\dot\gamma$.

The patches endow the simulated ring polymer with the structure of a closed ribbon, which therefore obeys Călugăreanu's theorem,
stating that the sum over its writhe and twist, i.e., its
linking number $\mathrm{Lk}$ is a conserved quantity: \cite{Calugareanu1961}
\begin{equation}
    \centering
    \mathrm{Lk}(t;\dot{\gamma}) = \mathrm{Wr}(t;\dot{\gamma}) + \mathrm{Tw}(t;\dot{\gamma}) = N\sigma_{\mathrm{sc}} = \mathrm{const.}
    \label{Calugareanu}
\end{equation}
The linking number is thus a topological invariant of the ribbon, independently of time or applied shear rate, akin to the knotedness of single rings or the catenation connectivities 
of several ones in composite macromolecular entities, such as catenanes. Note that the model is torsionally symmetric, i.e., it does not distinguish between left-handed and right-handed torsion and therefore we have chosen one of them and we do not specify the sign of the supercoiling or the linking number.
  
Supercoiled rings must be initialized in a state with minimal torsional energy to avoid large initial torsional forces that could cause a simulation crash. The initial conformations of rings with $\sigma_{\mathrm{sc}}>0$ are generated by starting from a flat open ribbon with a linking number of zero and performing MD steps while gradually ramping up the offset of the torsional potential $\psi_0$ from 
zero to the desired angle $2\pi \sigma_{\mathrm{sc}}$ in small steps. During this procedure monomers are coupled to the random MPCD solvent. The offset is increased in 
steps of $10^{-3} \cdot 2\pi \sigma_{\mathrm{sc}}$ after every 2000th collision with the solvent until the final angle is reached. The resulting conformations are supercoiled rings
for which $\mathrm{Lk}(t;\dot{\gamma})=N\sigma_{\mathrm{sc}}$ is indeed conserved up to fluctuations around the mean. The initialization process is performed for the supercoiling values 
$\sigma_{\mathrm{sc}}=0.01$ and $\sigma_{\mathrm{sc}}=0.02$. 
In Fig.~\ref{Fig:snapshots_zero_shear}, we show 
snapshots of all four types of ring polymers investigated. 

\subsection{The gyration tensor and its rotational invariants}
The conformations of polymers are investigated and quantified by measuring the ring's gyration tensor: \cite{gyration_tensor}
\begin{equation}
\label{gyration_tensor}
    G_{\alpha\beta}(t;\dot\gamma) = 
    \frac{1}{N}\sum_{i=1}^{N} 
    \overline{r}_{i,\alpha}(t;\dot\gamma) 
    \overline{r}_{i,\beta}(t;\dot\gamma),
\end{equation}
and additional quantities derived from it, to be specified in what follows. Here, $\overline{r}_{i,\alpha}(t;\dot{\gamma}) = (\vec{r}_i(t;\dot{\gamma}) - \vec{r}_{\mathrm{cm}}(t;\dot{\gamma}))_{\alpha}$ is the $\alpha$-coordinate of $i$-th monomer's position relative to the center-of-mass $\vec{r}_{\mathrm{cm}}(t;\dot{\gamma})$ of the polymer at a given time $t$ and shear rate $\dot{\gamma}$. 
The diagonal element $G_{\alpha\alpha}(t;\dot{\gamma})$ reflects the elongation of the polymer along the $\alpha$-coordinate.
Of particular importance are the instantaneous eigenvalues 
$\lambda_1(t;\dot\gamma) > \lambda_2(t;\dot\gamma) > \lambda_3(t;\dot\gamma)$, out of which several useful rotational invariants can be constructed.
The instantaneous radius of gyration squared, $R_\mathrm{g}^2(t;\dot{\gamma})$, is defined as the trace of the gyration tensor:
\begin{equation}
R^2_\mathrm{g}(t;\dot{\gamma}) 
= \sum_{\alpha = x,y,z}G_{\alpha\alpha}(t;\dot\gamma)
= \sum_{i = 1}^3\lambda_i(t;\dot\gamma), 
\label{eq:rg}
\end{equation}
and it is a measure of the overall spatial extension of the polymer.
An alternative way of expressing the extent of the polymer in the flow direction, which can be
particularly useful for microfluidics experiments, is the extent $\Delta x(t;\dot\gamma)$ defined as:
\begin{equation}
\Delta x(t;\dot\gamma) = \max_i\{{\vec r}_{i,x}(t;\dot\gamma)\} - \min_i\{{\vec r}_{i,x}(t;\dot\gamma)\},
\label{eq:dx}
\end{equation}
where ${\vec r}_{i,x}(t;\dot\gamma)$ is the $x$-component of the position vector ${\vec r}_{i}(t;\dot\gamma)$
of the $i$-th monomer.

The shape of the conformation can be characterized by the prolateness $S^*(t;\dot{\gamma}) \in [-0.25,2]$, defined as:\cite{liebetreu:commats:2020}
\begin{equation}
    S^*(t;\dot{\gamma}) = \frac
    {\prod_{i = 1}^3\left(3\lambda_i(t;\dot{\gamma}\right) - R_\mathrm{g}^2(t;\dot{\gamma}))}
    {R_\mathrm{g}^6(t;\dot{\gamma})},
    \label{prolateness}
\end{equation}
which attains negative values for oblate objects and positive values for prolate ones. Further, we consider the relative shape anisotropy $\delta^*(t;\dot{\gamma}) \in [0,1]$,
which is defined as: \cite{liebetreu:commats:2020}
\begin{equation}
    \delta^*(t;\dot{\gamma}) = 1 - 3 \frac{I_2(t;\dot{\gamma})}{R_\mathrm{g}^4(t;\dot{\gamma})},
    \label{anisotropy}
\end{equation}
where $I_2(t;\dot{\gamma}) = \lambda_1 (t;\dot{\gamma})\lambda_2 (t;\dot{\gamma})+ \lambda_2 (t;\dot{\gamma})\lambda_3(t;\dot{\gamma}) + \lambda_3 (t;\dot{\gamma})\lambda_1(t;\dot{\gamma})$.
For this quantity, $\delta^*(t;\dot{\gamma})=0$ occurs if all eigenvalues are identical (e.g. spherical or polyhedral group symmetry)  while $\delta^*(t;\dot{\gamma})=1$ indicates that two eigenvalues are zero (e.g. all monomers on a line). 
The (squared) gyration radius, the prolateness and the relative shape anisotropy are all independent of the  orientation of the polymer in space.
Similar to the writhe and twist, eq.~(\ref{eq:wrtwav}), we define time-averages as
\begin{equation}
\label{gyration_tensor1}
    {\mathcal O}(\dot\gamma) = 
    \langle {\mathcal O}(t;\dot\gamma) \rangle_t,
\end{equation}
where ${\mathcal O} = G_{\alpha\beta}$, $R_\mathrm{g}^2$, $\Delta x$, $S^*$, or $\delta^*$, and we report on their values in what follows.

\section{Equilibrium properties}
\label{sec:equ}

%%%%%%%%%%%%%%%%%%%%%%%%%%%%%%%%%%%%%%%%%%%%%%%%%%%%%%%%%%%%%%%%%%%%%%%%
 \begin{figure*}[htb]
\centering
  \includegraphics[width=0.4\textwidth]{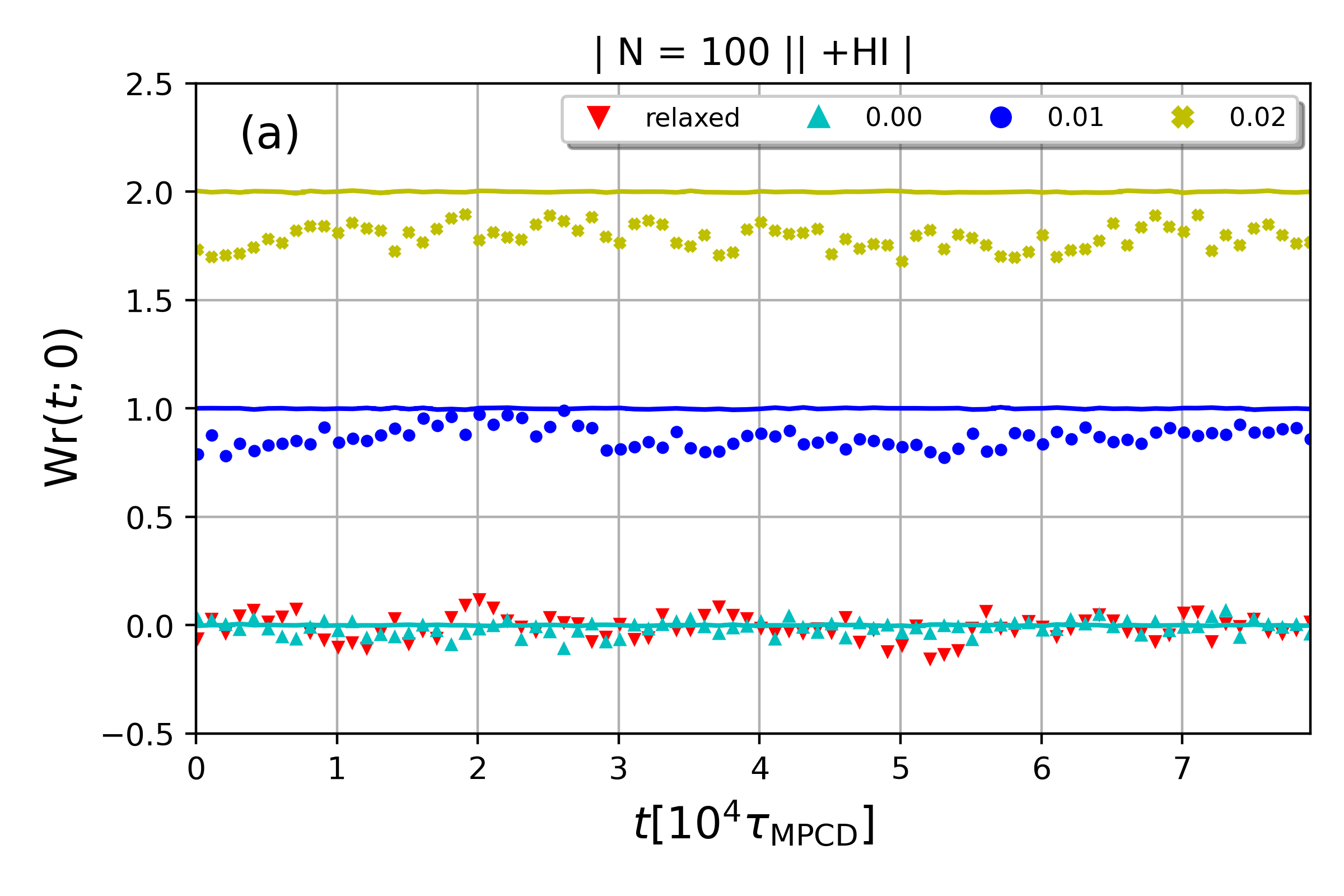}
   \includegraphics[width=0.4\textwidth]{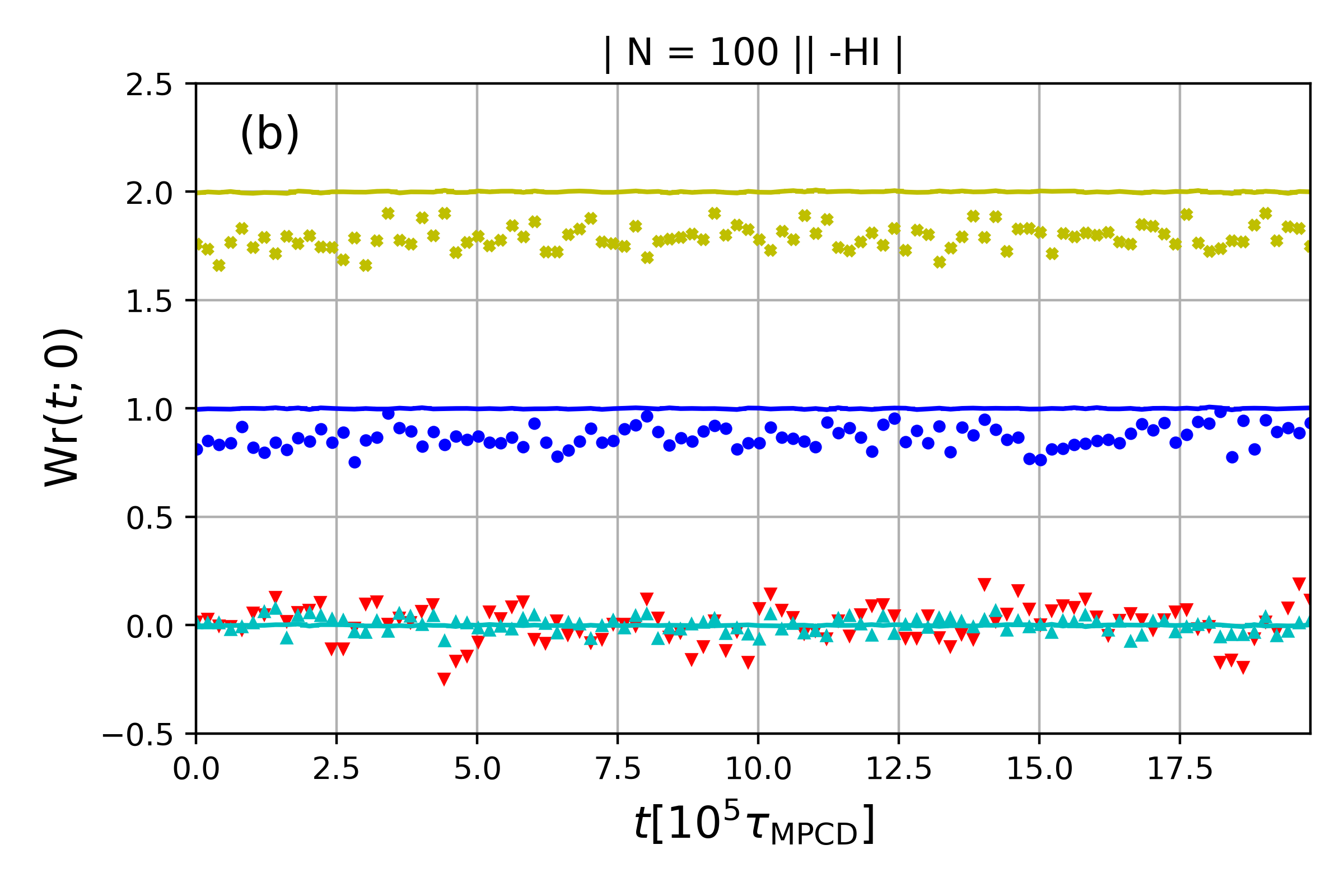}
  \\
\includegraphics[width=0.4\textwidth]{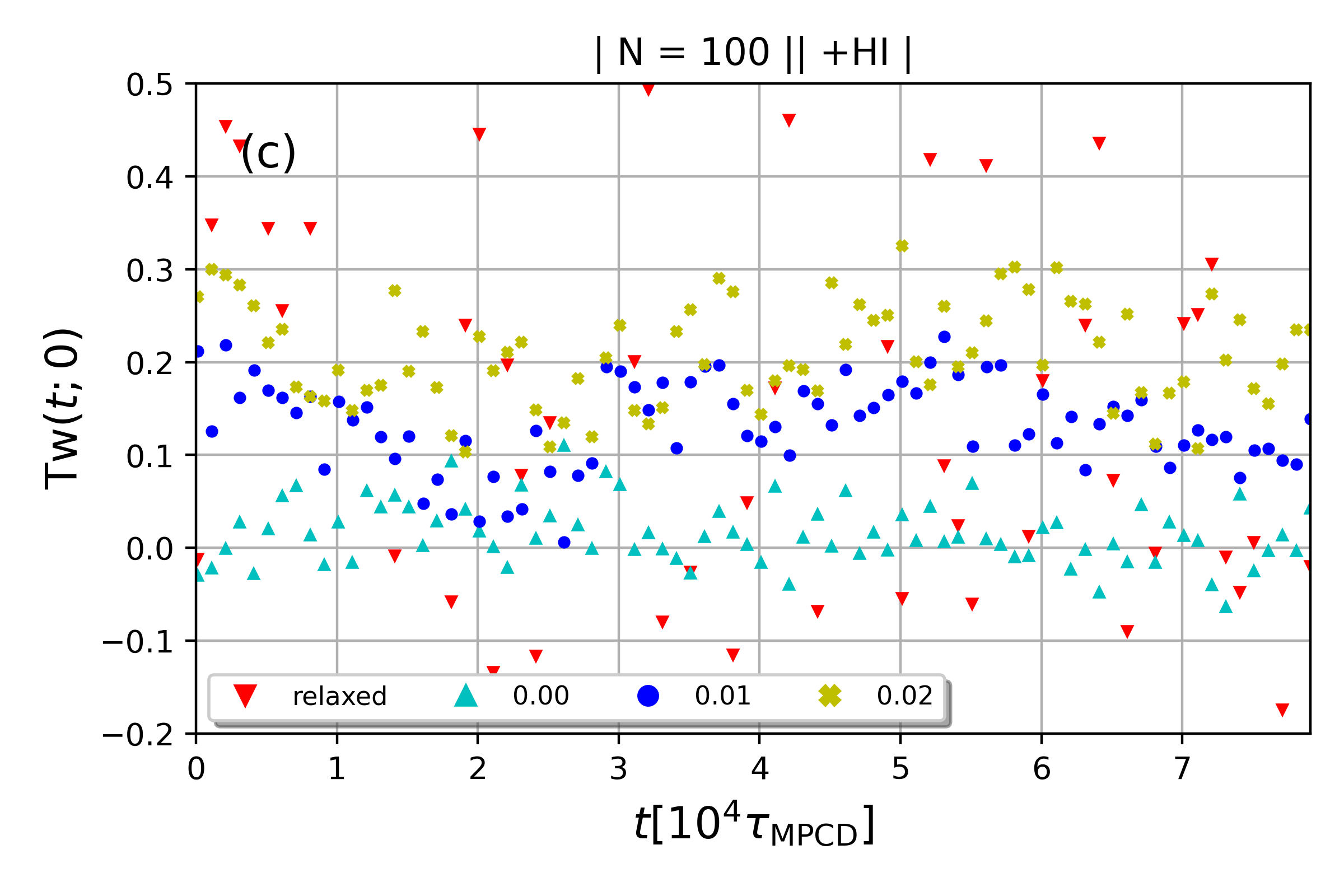}
  \includegraphics[width=0.4\textwidth]{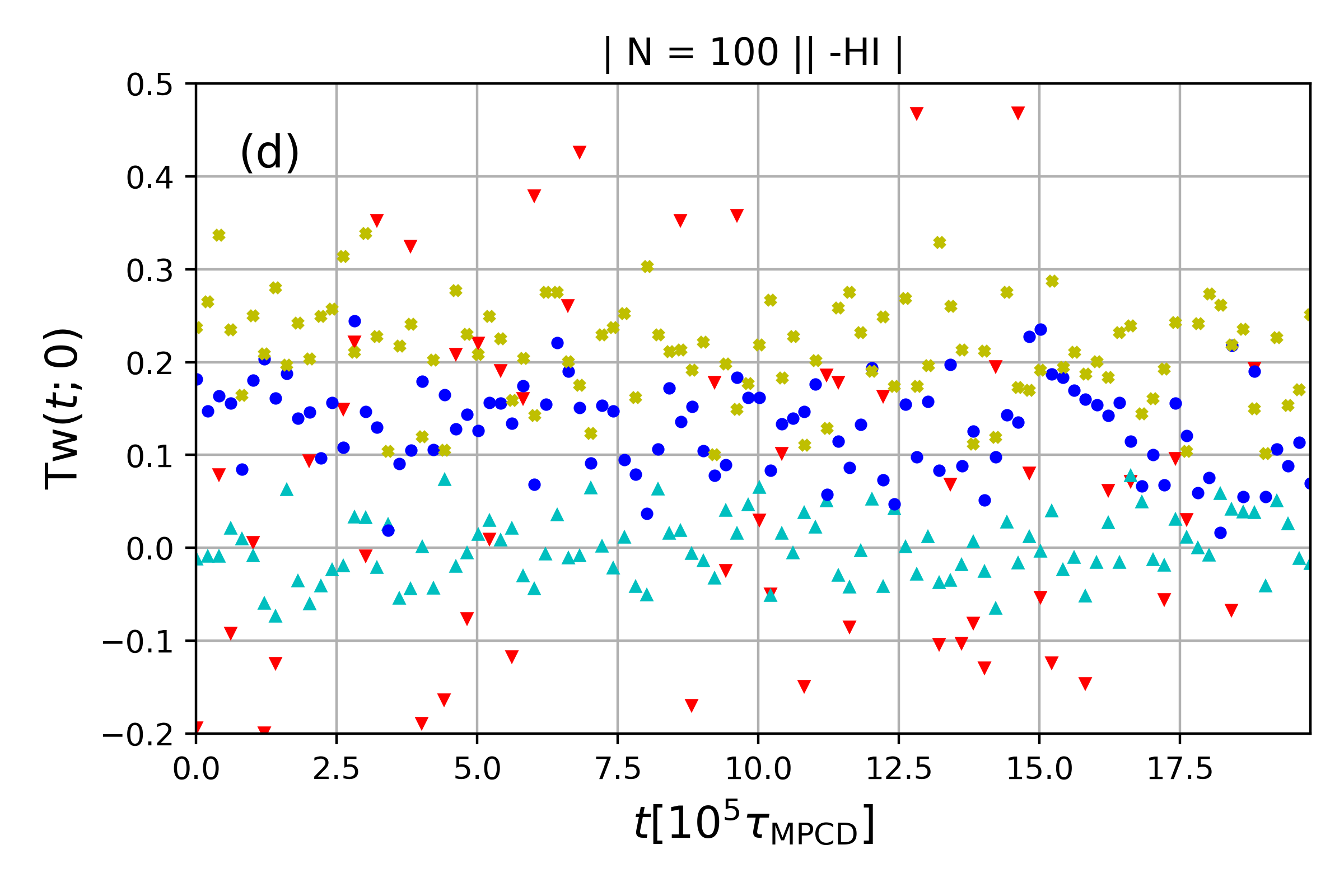}
  \\
    \includegraphics[width=0.4\textwidth]{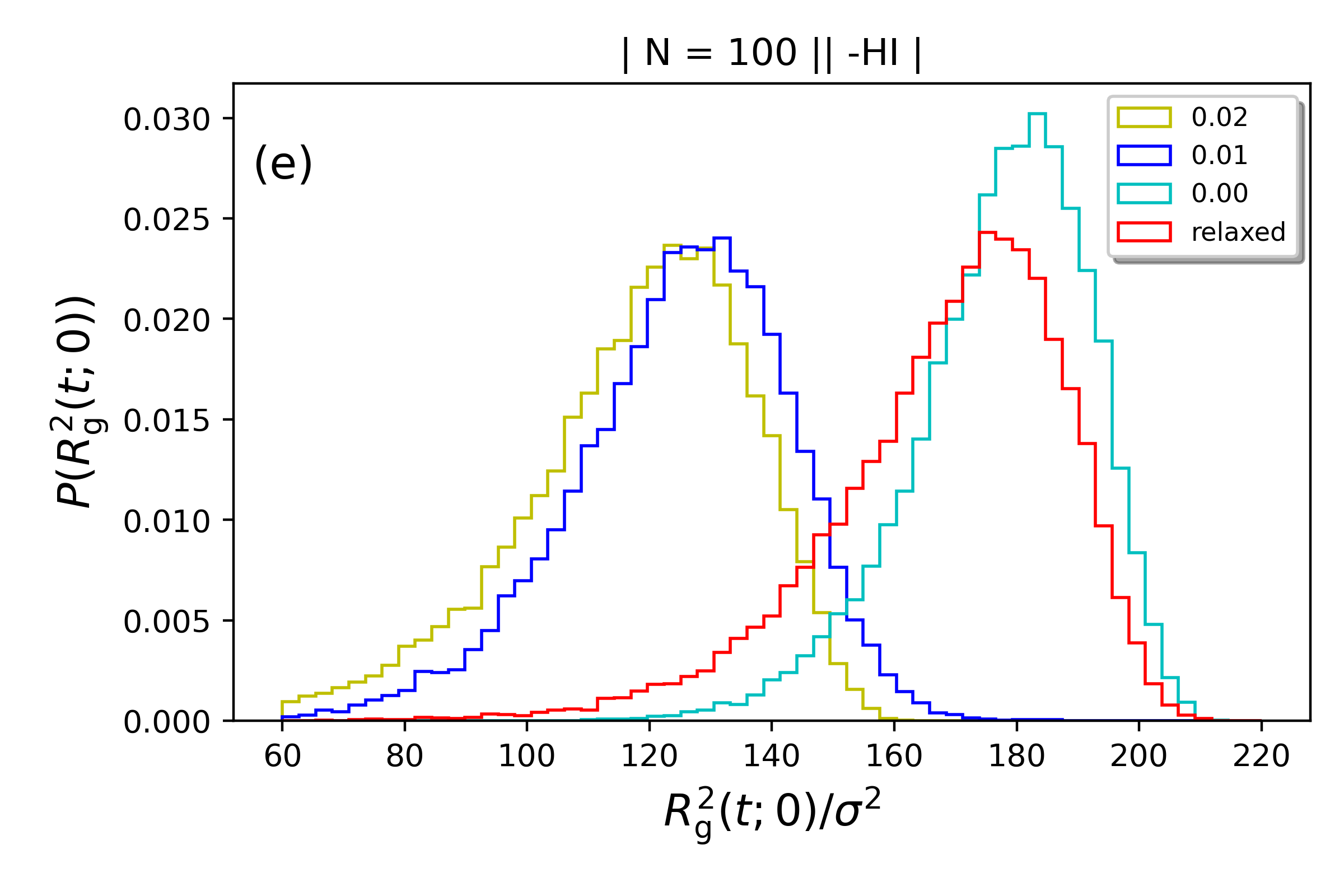}
   \includegraphics[width=0.4\textwidth]{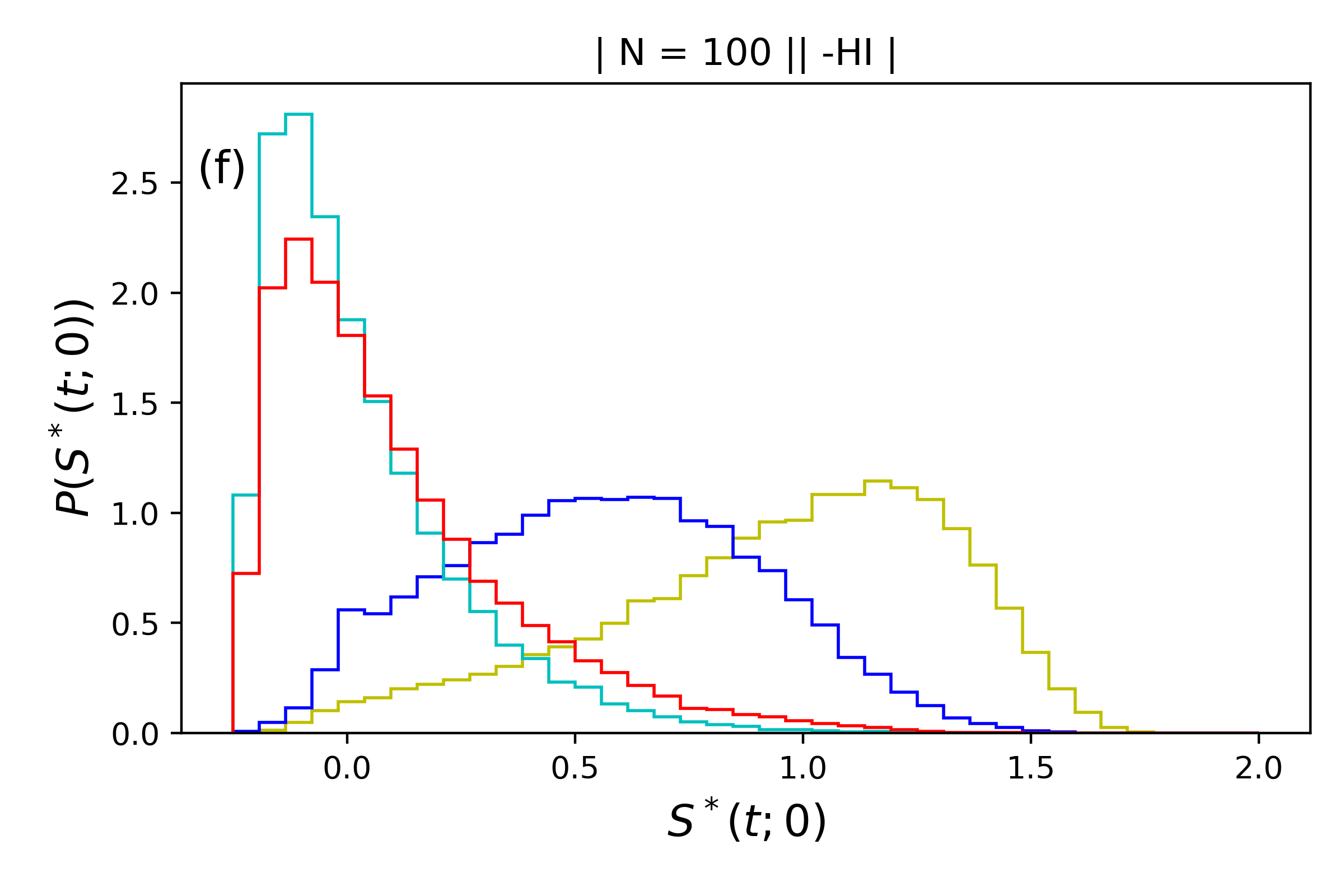}
  \caption{Time series of the topological quantities writhe (${\mathrm{Wr}}(t;0)$) 
  and twist (${\mathrm{Tw}}(t;0)$) at equilibrium for the cases $+$HI [(a),(c)] and $-$HI [(b),(d)].
  The solid lines in the $\mathrm{Wr}(t;0)$-plots, panels (a) and (b), 
  indicate the linking number as the sum $\mathrm{Lk}(t;0) = \mathrm{Wr}(t;0) + \mathrm{Tw}(t;0)$. 
  The equilibrium probability distributions of 
  the squared radius of gyration and the prolateness are shown in panels (e) and (f), respectively.}
  \label{Fig:topological_zero_shear}
\end{figure*}
%%%%%%%%%%%%%%%%%%%%%%%%%%%%%%%%%%%%%%%%%%%%%%%%%%%%%%%%%%%%%%%%%%%%%%%%
Supercoiled rings with $\sigma_{\mathrm{sc}}=0.00,0.01,0.02$ as well as torsionally relaxed ones with $N=100$ monomers
were allowed to evolve freely in time under equilibrium conditions ($\dot{\gamma}=0$) while we monitor the quantities of interest to establish when the equilibrium was reached.
In that process, the monomers were coupled to the MPCD solvent both with and without HI in separate runs to confirm the independence of the equilibrium state on the HI. After the equilibration run of $2\cdot10^7$ MD steps for $+$HI and $5\cdot10^8$ steps for $-$HI followed a production run of $6\cdot10^7$ and $1.5\cdot10^9$ steps, respectively for $+$/$-$HI. All quantities were sampled at least every $100\tau_{\textrm{MPCD}}$ for $+$HI and $1000\tau_{\textrm{MPCD}}$ for $-$HI, while quickly fluctuating quantities, the intrinsic viscosity and the tumbling cross correlation function, were sampled every $10\tau_{\textrm{MPCD}}$. All the quantities measured contributed to their equilibrium averages and distributions. To get more precise results, 38 independent simulation runs were conducted for $\dot{\gamma}=0$ and observed quantities were averaged among them. The error bars were computed as the error of the mean from the 38 independent runs. We simulate longer than the characteristic time as estimated in the sheared case, see sec. \ref{sec:shear}.

For the supercoiled rings with nonzero Linking number, $\sigma_{\mathrm{sc}}>0$, the simulation occasionally 
crashed, due to the relatively large timestep and the buildup of numerical instabilities leading to unphysically large forces. In such cases, we discarded the data. Higher degrees of writhe constrain the rings to take on more contracted conformations and the resulting contorted macrostructure of the polymer forces some beads to be pressed together increasing the risk of building up large repulsive forces. High shear rates amplify this behaviour as the solvent flow stretches out the polymer rings. More specifically, two different kinds of undesired behaviour have been observed. The first is rupture of the polymer, where a small part (usually up to 5 monomers) leaves its place within the chain. The second type of undesired behaviour amounts to $\mathrm{Lk}$ spontaneously being changed by an integer value $\pm 1$, violating the topological constraints that force the ring to have a constant Linking number (apart from small deviations due to the fact that the patches are not necessarily aligned in a perfect right angle relative to the backbone). These crashes happened mostly when using the $-$HI-solvent at high shear rates, since the total number of MD steps was much higher compared to the costly particle-based solvent with $+$HI and the chances of a crashing situation increased with the length of the simulation runs which could not be decreased significantly. Both types of crashes can presumably be avoided by using a smaller $\Delta t_{\mathrm{MD}}$.

In Figs.~\ref{Fig:topological_zero_shear}(a)-\ref{Fig:topological_zero_shear}(d), 
we show time series of the topological quantities, writhe and twist, for all systems simulated,
whereby the curves of the single runs are averaged to a single curve. Whereas both the relaxed rings and the rings without supercoiling
assume conformations that are essentially writhe-free, the rings with $\sigma_{\mathrm{sc}} = 0.01$ and $\sigma_{\mathrm{sc}} = 0.02$ display conformations that carry finite writhe.
Connected to this, and rather insensitively of the degree of supercoiling, the twist fluctuates closely around zero in case of $\sigma_{\mathrm{sc}}=0.00$ whereas for $\sigma_{\mathrm{sc}}>0$, $\mathrm{Tw}(t;\dot{\gamma})$ takes on 
small but nonzero values. Up to this value, the writhe equals the linking number for the entire duration of the simulation and Călugăreanu's theorem \eqref{Calugareanu} is indeed fulfilled for all systems. 
The relaxed chains remain in a state with only few writhe because of the high bending stiffness-cost that a high writhe would incur, and the twist fluctuates arbitrarily for the same chains because of the absence of torsional springs. 
The overall behaviour of the topological parameters does not depend on the presence of HI, as it should be the case for polymers in equilibrium.

\begin{table}[h]
\small
  \caption{\ The expectation values of the squared radii of gyration in equilibrium, $R_\mathrm{g}^2(\dot\gamma = 0)$, 
  for the various types of ring polymers with $N = 100$ monomers considered in this work, along with 
  the value of the same quantity for a fully flexible polymer (last row). The error values indicated are computed as the standard deviation of the equilibrium distribution.}
  \label{tbl:rg}
  \begin{tabular*}{0.48\textwidth}{@{\extracolsep{\fill}}lll}
    \hline
    Polymer type     & Supercoiling $\sigma_{\rm sc}$ & $R_\mathrm{g}^2(\dot\gamma = 0)/\sigma^2$ \\
    \hline
    Semiflexible      & Relaxed                        %& $165.54 \pm 6.12$ \\
    & $171.67 \pm 19.28$ \\
    Supercoiled       & 0.00                           %& $177.69 \pm 4.33$ \\
    & $179.21 \pm 14.47$ \\
    Supercoiled      & 0.01                           %& $121.08 \pm 5.20$ \\
    & $126.52 \pm 17.71$ \\
    Supercoiled      & 0.02                           %& $125.54 \pm 4.94$ \\
    & $120.53 \pm 18.56$ \\
    Fully flexible   & --                             & $ 28.95 \pm 4.00$ \\
    \hline
  \end{tabular*}
\end{table}

The equilibrium distributions of the squared radius of gyration are presented in 
Fig.~\ref{Fig:topological_zero_shear}(e) and the expectation values of this quantity are summarized in 
Table~\ref{tbl:rg}, which also provides a comparison with the gyration radius of a fully flexible ring. As a first remark, we note that the presence of the bending stiffness, independently
of the supercoiling, causes the rings to swell with respect to their flexible counterparts, as expected. 
For $\sigma_{\mathrm{sc}}=0.00$ both the bending and the torsional stiffness force the ring to remain in an open circular 
shape which leads to a sharp peak in the $R_\mathrm{g}^2(t;0)$-distributions. The absence of torsional stiffness softens considerably the relaxed rings, 
leading to slightly smaller sizes as well as a broader distribution of the gyration radius values. 
The $\sigma_{\mathrm{sc}}=0.00$ rings are locked in a rather open and flat conformation and $R_\mathrm{g}^2(t;0)$ fluctuates more weakly around a larger mean 
compared to the relaxed counterparts.
The values of the gyration radius agree with Smrek {\it et al.}, \cite{janspaper} 
which confirming 
the correct implementation of the model. Moreover the $R_\mathrm{g}$-distribution of the relaxed case shows a long tail at low values, indicating a weakly bimodal character of the distribution in agreement with the findings of Smrek \textit{et al.}\cite{janspaper}
Nonzero supercoiling forces the polymers to take on more complex supercoiled shapes, characterized by a smaller $R_\mathrm{g}^2(t;0)$ for $\sigma_{\mathrm{sc}}>0$,
as can be also confirmed by the prolateness data below. The values of $R_\mathrm{g}^2(t;0)$ coincide for the solvents with and without 
HI, again as expected, and confirming that equilibrium has indeed been achieved for both simulation variants.

The equilibrium distributions of the prolateness parameter $S^*$ are depicted in 
Fig.~\ref{Fig:topological_zero_shear}(f). 
As intuitively expected, the $\sigma_{\mathrm{sc}}=0.00$-ring and the 
relaxed one are mainly found in oblate shapes as they tend to be open rings, a property manifested in the fact that their $S^*(t;0)$-distributions peak at values below zero. Nevertheless,
even for these rings the probability distributions have sufficiently long tails into the prolate domain, resulting into expectation values $S^*(0) \cong 0$ overall,
see Fig.~\ref{Fig:shape_parameters_vs_shear}(d) (at $\tau_{\textrm{MPCD}} \dot{\gamma} < 10^{-5}$): to obtain rings that are oblate on average, one would need a higher value of $k_{\mathrm{bend}}$ and/or
a smaller value of $N$, so as to obtain a larger ratio of persistence to contour lengths. On the other hand, 
supercoiling increases the prolateness markedly, as it constrains 
the rings in a more elongated conformation, consistent with the increasingly writhed conformations they assume. Accordingly,
the prolateness distribution of the $\sigma_{\mathrm{sc}}=0.01$-supercoiled ring has a peak at $S^*(t,0) \cong 0.5$ and 
that of the $\sigma_{\mathrm{sc}}=0.02$-ring at $S^*(t,0) \cong 1.25$. Despite the increase in prolateness, the gyration radius
of the $\sigma_{\mathrm{sc}}=0.02$-ring remains unchanged with respect to that of its $\sigma_{\mathrm{sc}}=0.01$-counterpart,
as the elongation in one direction is compensated by shrinking in the two orthogonal ones.

\section{Properties under shear flow}
\label{sec:shear}

 The supercoiled and relaxed ring polymers discussed so far are simulated under shear flow for a wide range of shear rates $\dot{\gamma}$ spanning multiple orders of magnitude. They are coupled both to the particle-based MPCD solvent 
($+$HI) and to the random MPCD solvent ($-$HI) in 10 separate runs to determine any effects of HI on the supercoiled ring conformations. The systems without HI are run for longer times ($5\cdot10^8$ MD steps compared to 
$2\cdot10^7$ for $+$HI) since the particle-based MPCD took up most of the numerical effort when HI is present. On the flip side the polymers with $\sigma_{\mathrm{sc}}>0$ are more prone to crashes (as explained in sec. \ref{sec:equ}) for $-$HI so it is not possible to measure 
their properties for very high shear rates. After the ``equilibration" time of 20\% of the respective simulation run time  we considered the systems to be in a stationary state since all key observed quantities had reached stationary values.

\begin{figure*}[htb]
\centering
  \includegraphics[width=0.33\textwidth]{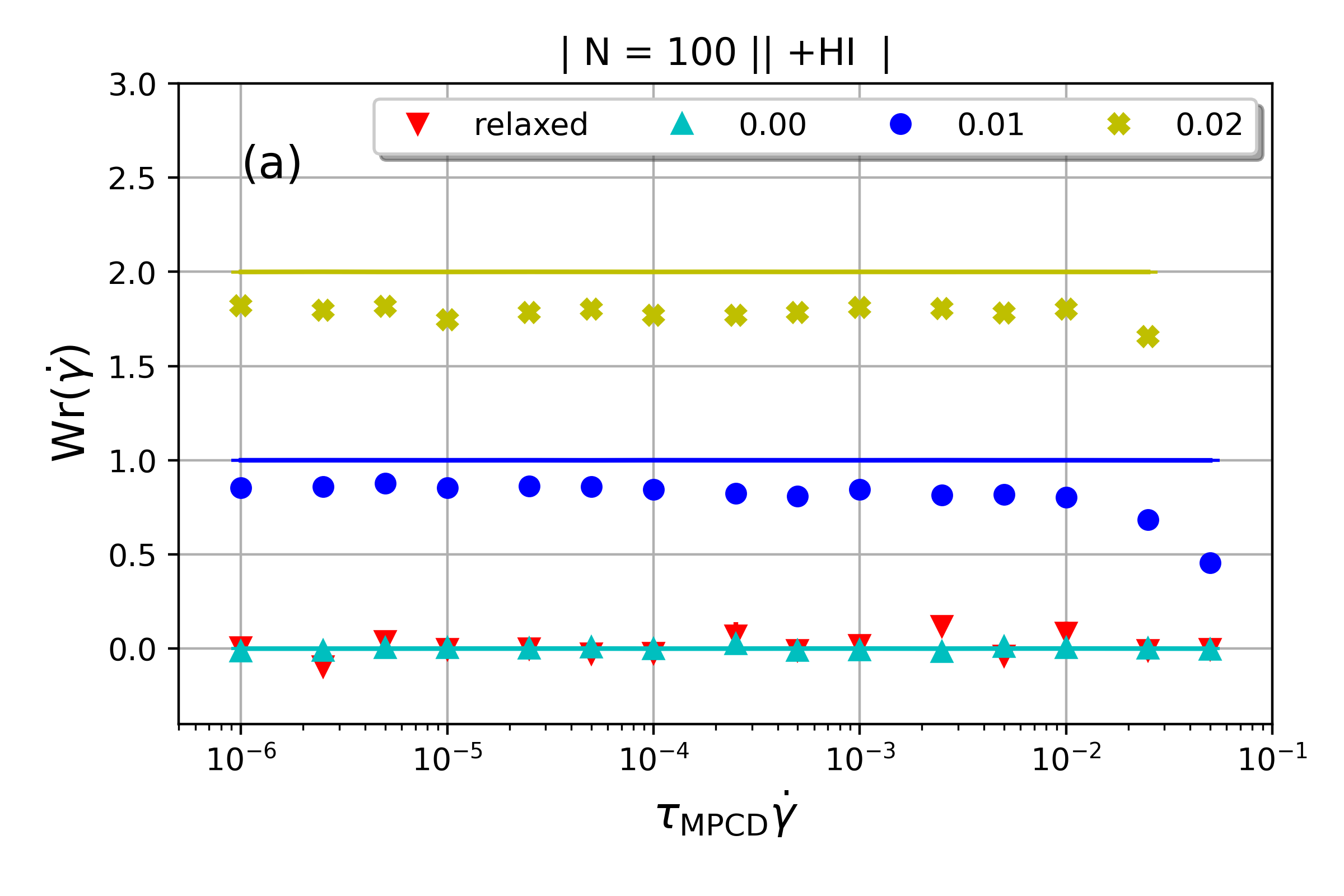}
\includegraphics[width=0.33\textwidth]{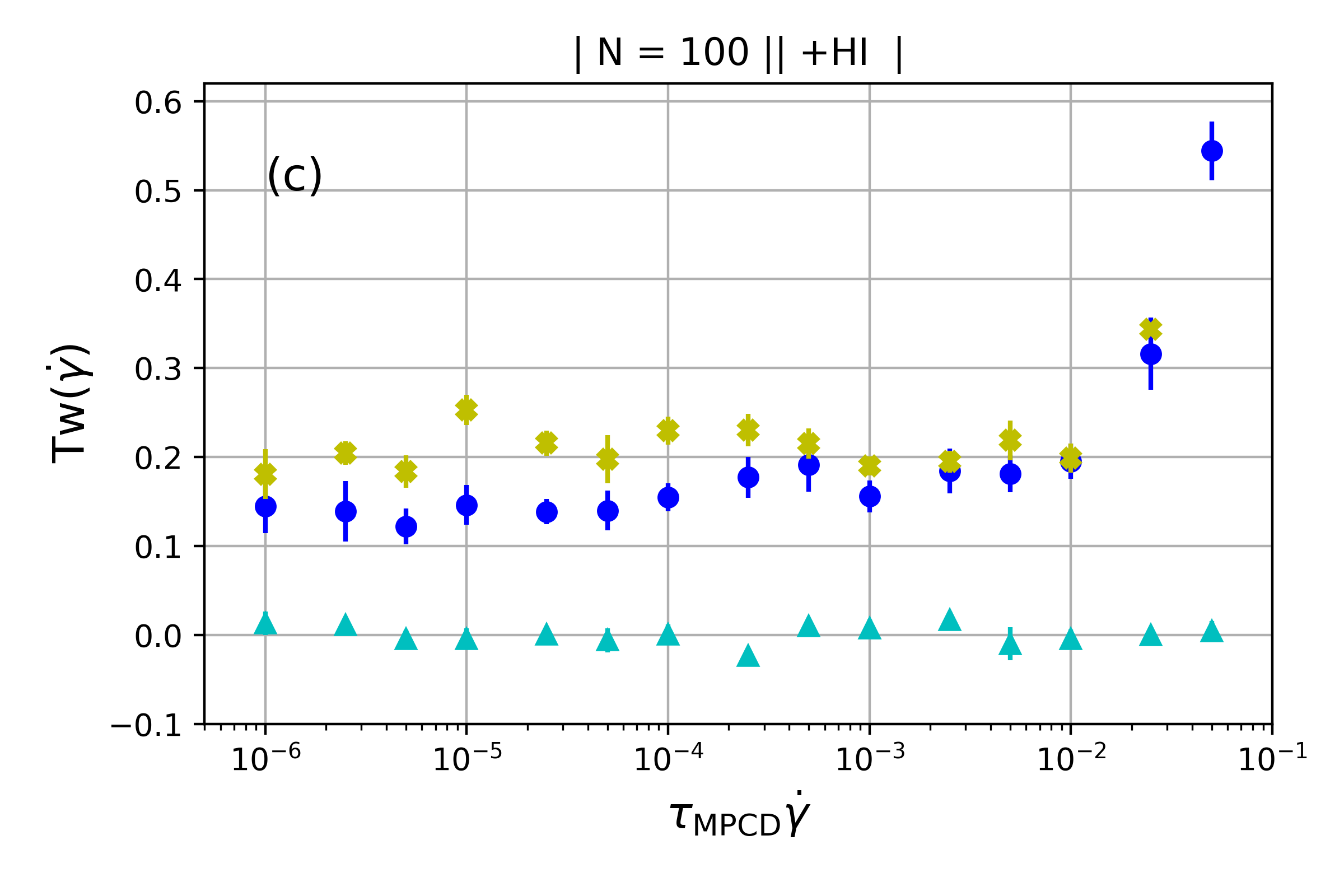}
\includegraphics[width=0.33\textwidth]{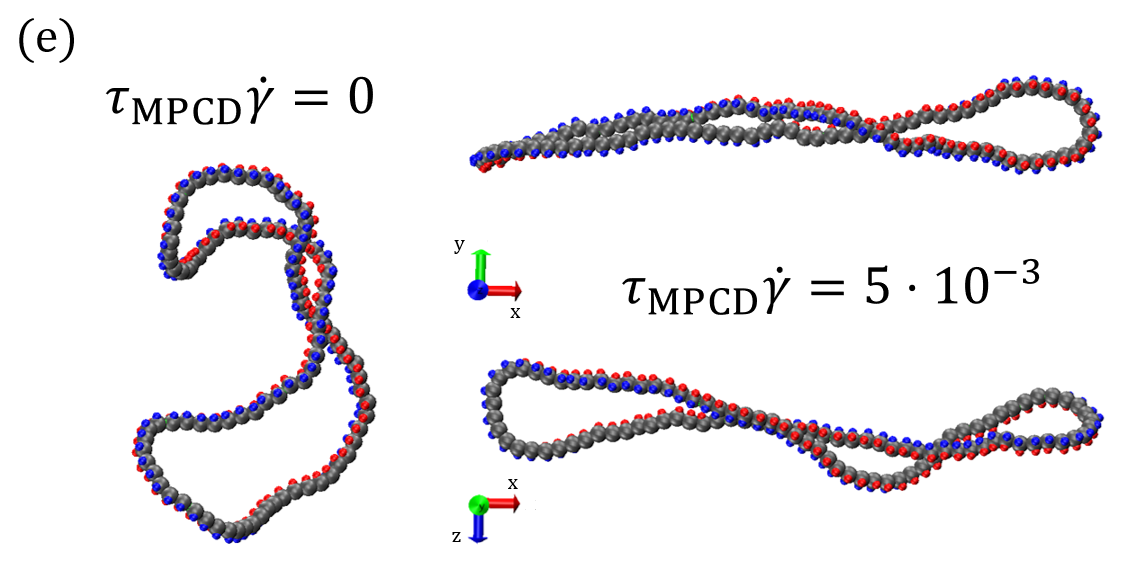}
\includegraphics[width=0.33\textwidth]{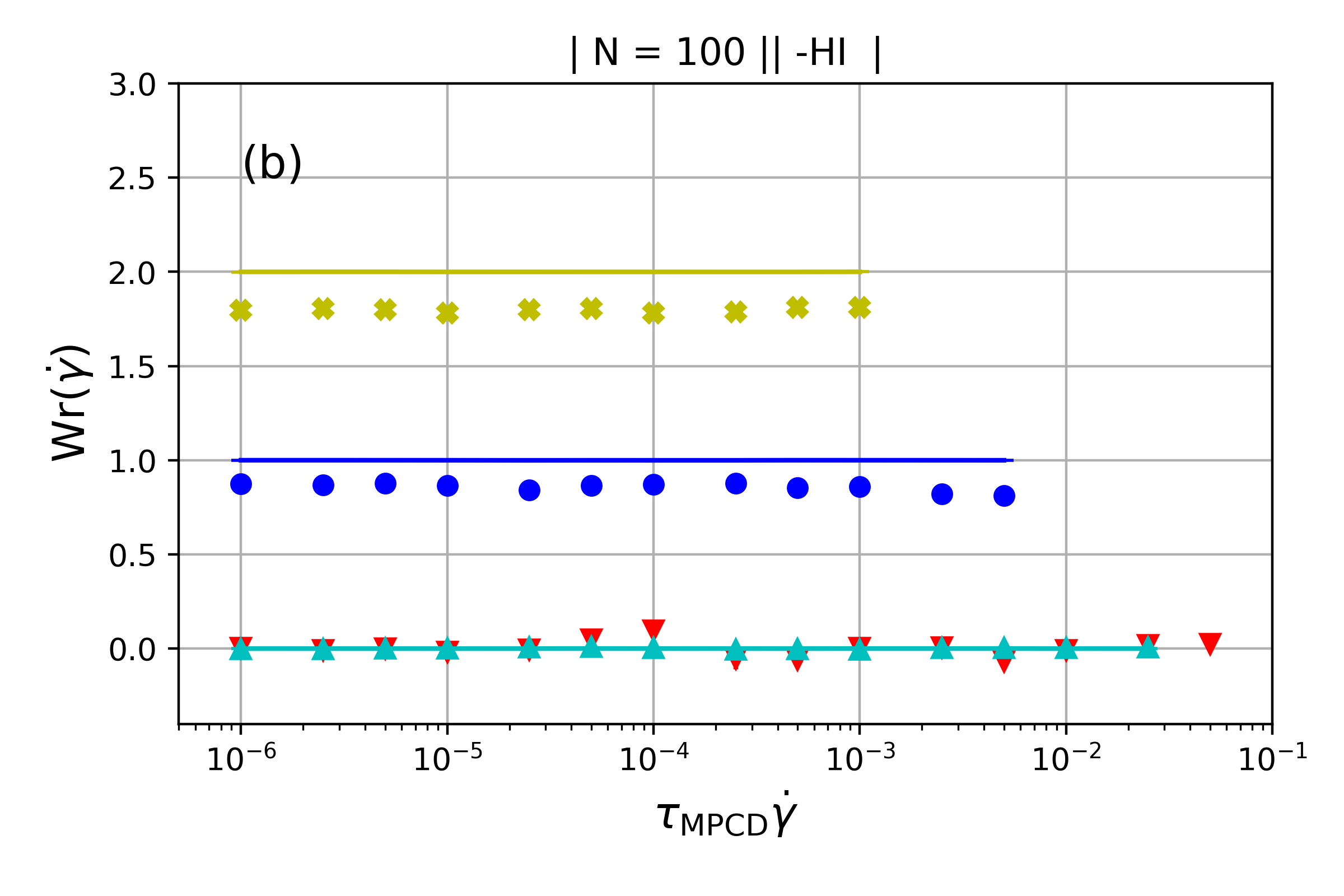}
\includegraphics[width=0.33\textwidth]{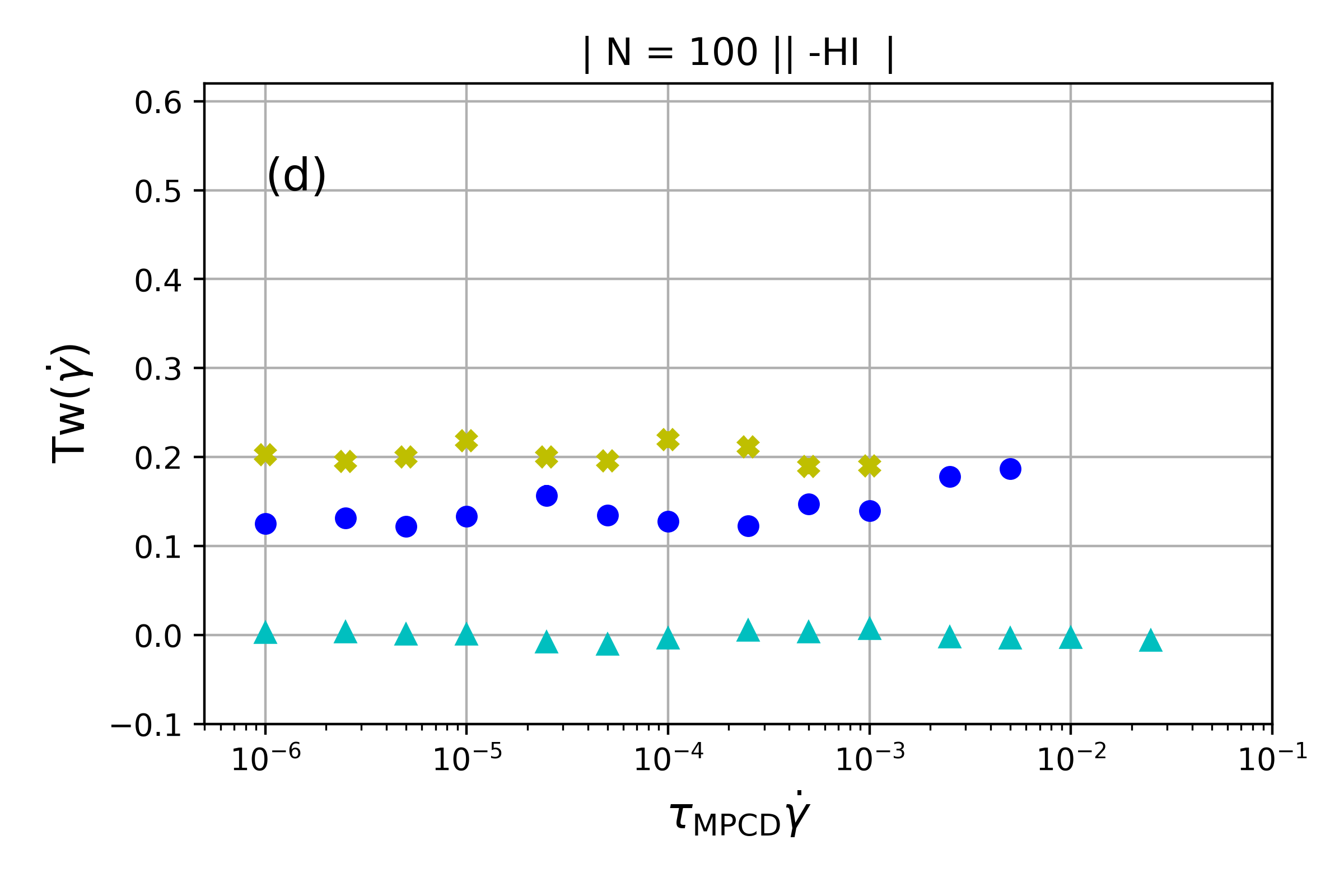}
\includegraphics[width=0.33\textwidth]{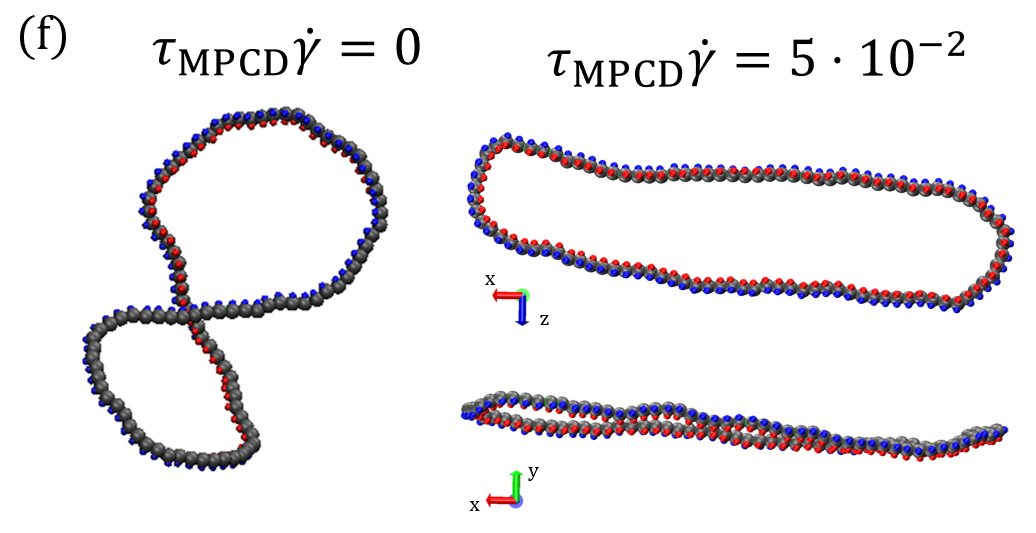}
  \caption{The dependence of the time-averages of writhe, ${\mathrm{Wr}}(\dot\gamma)$,
  and twist, ${\mathrm{Tw}}(\dot\gamma)$ on shear rate, in panels [(a),(b)] and [(c),(d)] respectively.
  Panels (a) and (c) are for the $+$HI-case and panels (b) and (d) for the $-$HI-case.
 The snapshots in panels (e) and (f) compare conformations in equilibrium and under shear ($+$HI). Error bars indicate the standard error of the mean. The solid lines in the $\mathrm{Wr}(t;0)$-plots, panels (a) and (b), 
  indicate the linking number as the sum $\mathrm{Lk}(t;0) = \mathrm{Wr}(t;0) + \mathrm{Tw}(t;0)$. (e): supercoiling $\sigma_{\mathrm{sc}}=0.02$-ring, at shear rate 
  $\tau_{\mathrm{MPCD}}\dot\gamma = 5\cdot 10^{-3}$,
  where $\mathrm{Wr}(\dot{\gamma})$ is not altered; (f)
  supercoiling $\sigma_{\mathrm{sc}}=0.01$ at shear rate 
  $\tau_{\mathrm{MPCD}}\dot\gamma = 5\cdot 10^{-2}$,
  where writhe is converted into twist.}
  \label{Fig:topological_vs_shear}
\end{figure*}

We begin our analysis by looking at the topological characteristics of the supercoiled rings, and in particular at the writhe ${\mathrm{Wr}}(\dot\gamma)$ and the twist ${\mathrm{Tw}}(\dot\gamma)$, reported in 
figs.~\ref{Fig:topological_vs_shear}(a)-\ref{Fig:topological_vs_shear}(d). 
The linking number is also depicted as their sum, which is constant for all $\dot{\gamma}$, serving as a test of the validity of the simulations. 
In all cases and for most shear rates considered, writhe and twist remain constant at their preferred values as in equilibrium according to their degree of supercoiling $\sigma_{\mathrm{sc}}$; hydrodynamic forces
are not capable of ``unwrithing'' the molecules for most of the range of shear rates  considered.
However, for the highest values of $\dot{\gamma}$ tested and in the $+$HI-variant, where simulations at these shear rates could
be performed without crashes,
considerable transformation of $\mathrm{Wr}(\dot{\gamma})$ into $\mathrm{Tw}(\dot{\gamma})$ takes place (see Fig.~\ref{Fig:topological_vs_shear}(a) and (c) at  $\tau_{\textrm{MPCD}}\dot{\gamma}>10^{-2}$). 
This is interpreted as a forced opening of the writhed structure since the process is connected with a high torsional energy penalty. 
The snapshots of supercoiled rings with $\sigma_{\mathrm{sc}}=0.02$ and $\sigma_{\mathrm{sc}}=0.01$ in figs.~\ref{Fig:topological_vs_shear}(e) 
and \ref{Fig:topological_vs_shear}(f), respectively (both $+$HI), are representatives of the polymers' behaviour under shear flow in comparison to their shapes in equilibrium $\dot{\gamma}=0$. 
At zero shear rate the $\sigma_{\mathrm{sc}}=0.02$-ring rotates freely in space and takes on more compact, prolate shapes. Although bending and torsion constrains it 
to have a writhe of nearly $\mathrm{Wr}(t;\dot{\gamma})\approx\sigma_{\mathrm{sc}}N$ it is not elongated in any direction. In contrast, the polymer is almost 
completely stretched out and aligned along flow direction at high shear rate of $\tau_{\mathrm{MPCD}}\dot{\gamma}=10^{-3}$. 
Above $\tau_{\mathrm{MPCD}}\dot{\gamma}>10^{-2}$ the shear flow is even able to counter the torsional forces so that writhe and twist experience large fluctuations and the conformation 
unwrithes, see figs.~\ref{Fig:topological_vs_shear}(e) and \ref{Fig:topological_vs_shear}(f).
%%%%%%%%%%%%%%%%%%%%%%%%%%%%%%%%%%%%%%%%%%%%%%%%%%%%%%%%%%%%%%%%%%%%%%%%
%%%%%%%%%%%%%%%%%%%%%%%%%%%%%%%%%%%%%%%%%%%%%%%%%%%%%%%%%%%%%%%%%%%%%%%%

\begin{figure*}[htb]
\centering
  \includegraphics[width=0.4\textwidth]{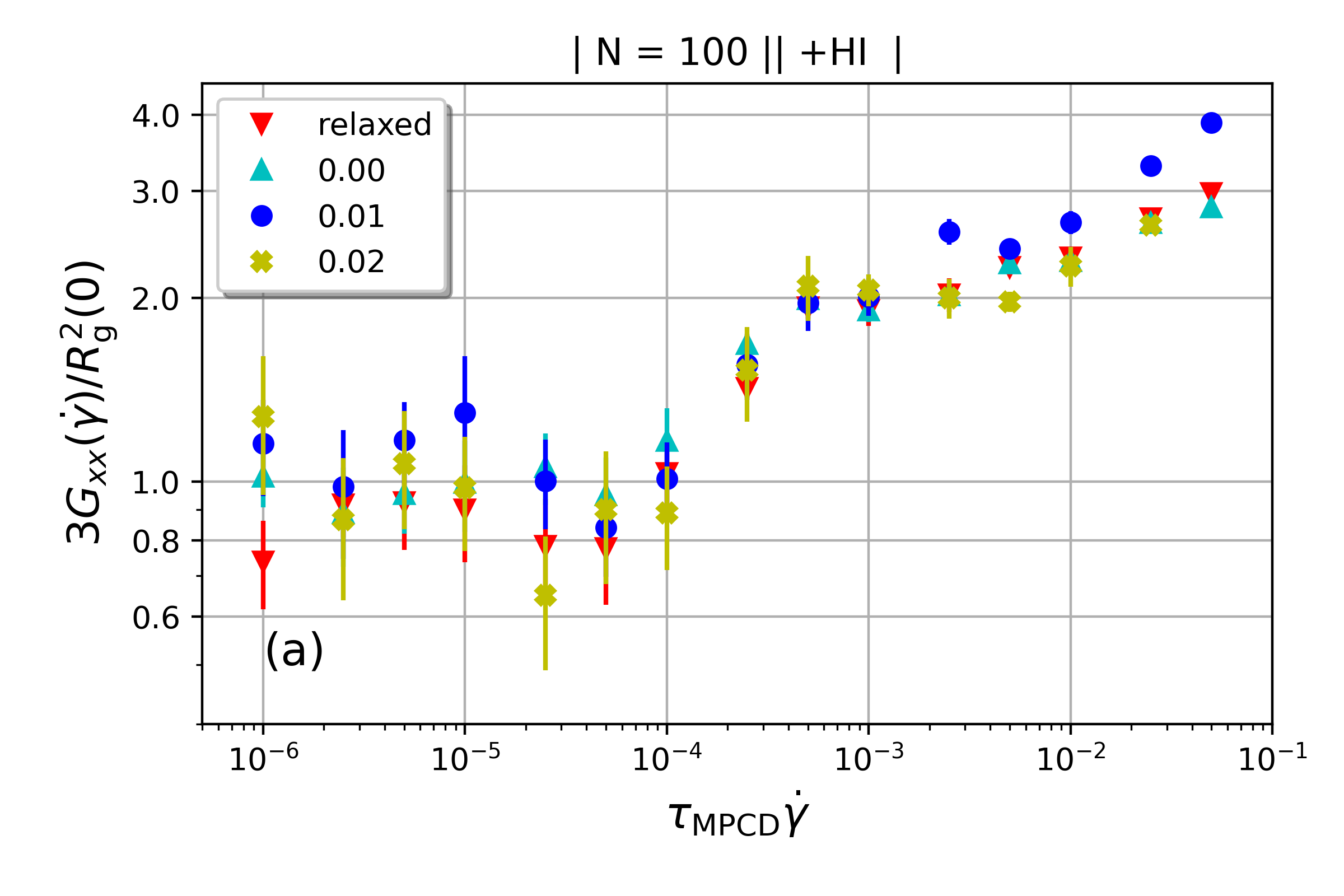}
  \includegraphics[width=0.4\textwidth]{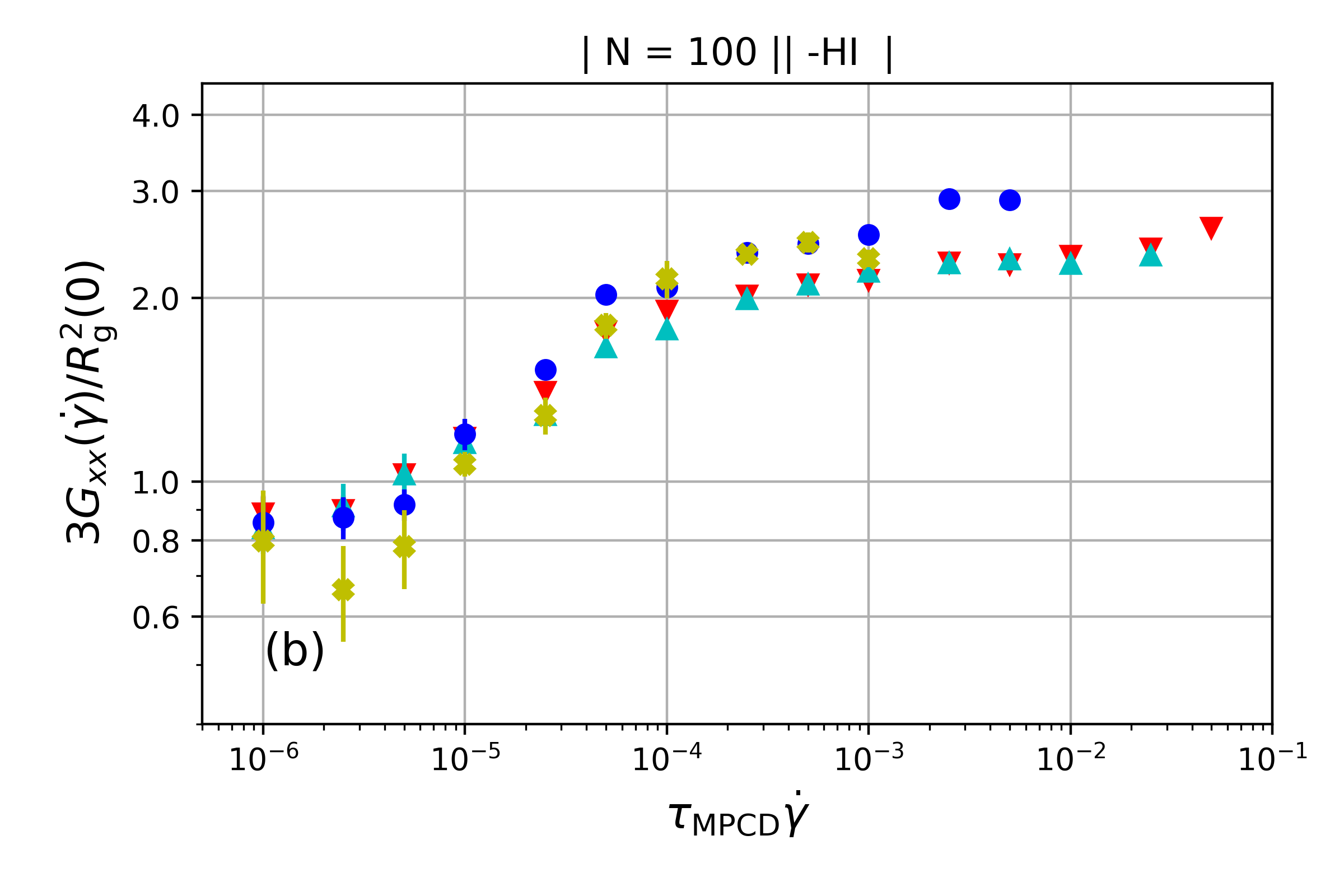}
  \includegraphics[width=0.4\textwidth]{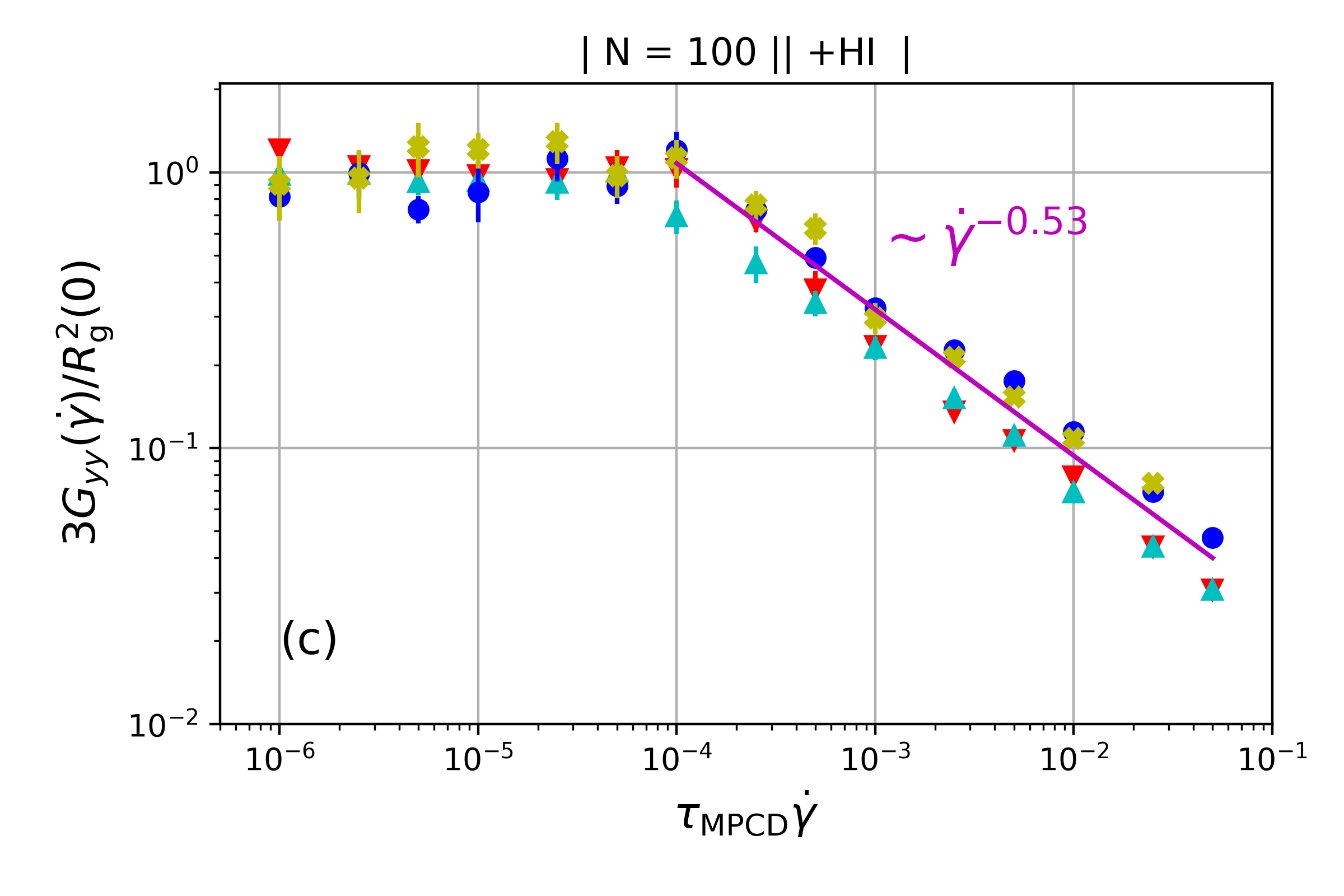}
  \includegraphics[width=0.4\textwidth]{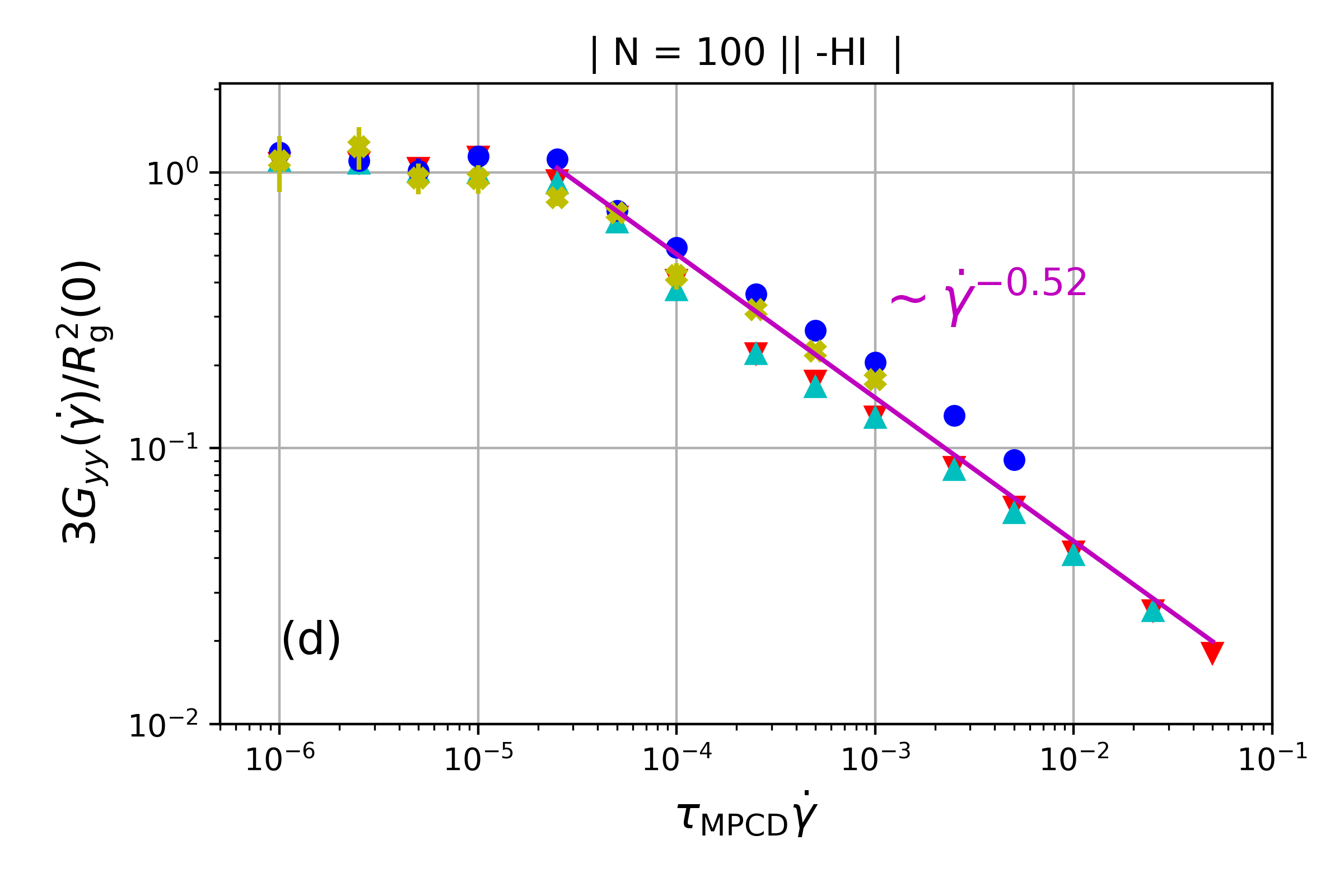}
  \includegraphics[width=0.4\textwidth]{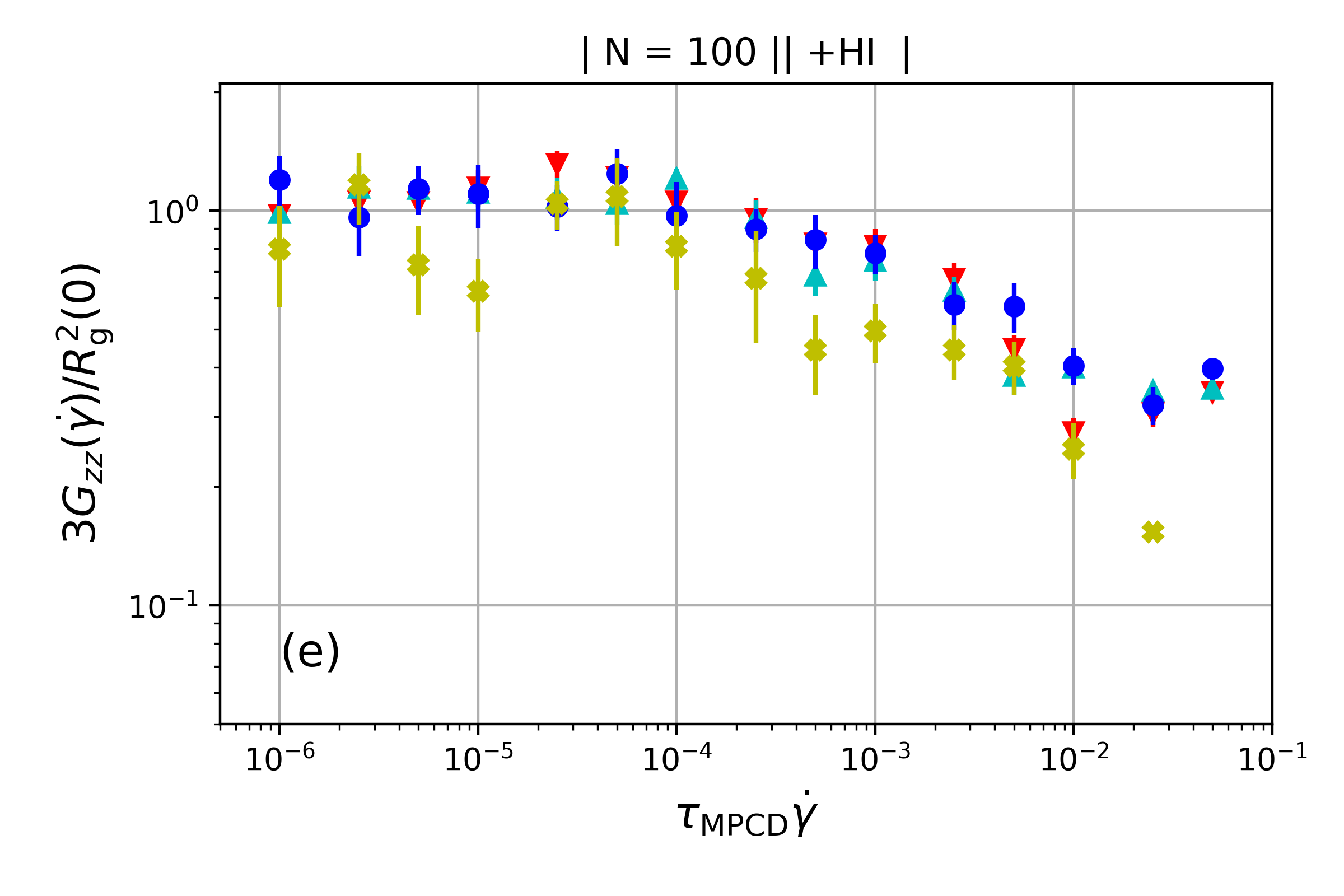}
  \includegraphics[width=0.4\textwidth]{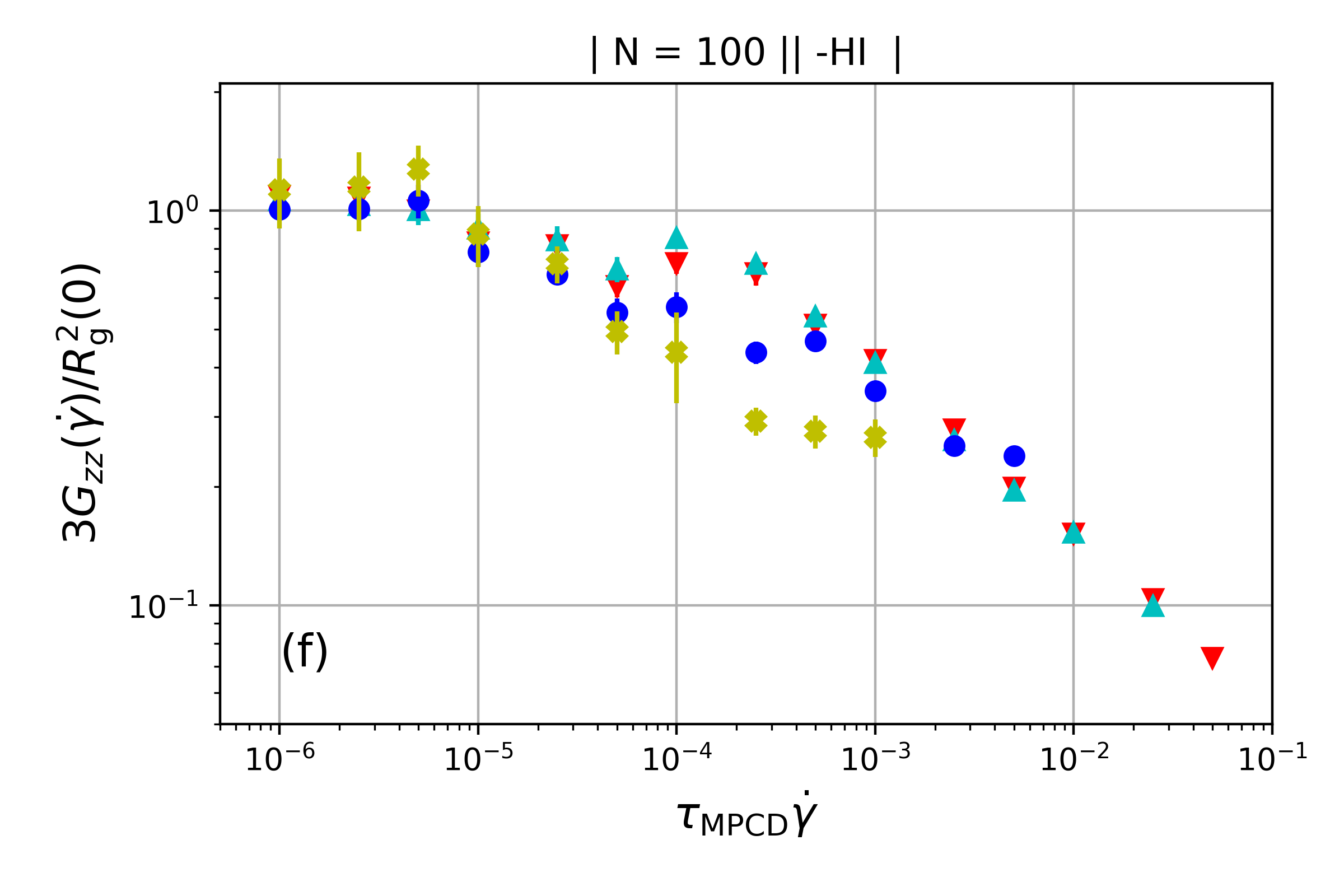}
  \caption{Time-averaged diagonal components of the gyration tensor, $G_{\alpha\alpha}(\dot\gamma)$,
  $\alpha = x,y,z$, as functions of the 
  shear rate for the cases of inclusion of HI [panels (a),(c),(e)] and exclusion of the same
  [panels (b),(d),(f)]. Error bars indicate the standard error of the mean.}
  \label{Fig:gyration_vs_shear}
\end{figure*}

In Fig. \ref{Fig:gyration_vs_shear} we show the diagonal elements of $G_{\alpha\beta}(\dot{\gamma})$ 
for both cases when hydrodynamic interactions are active ($+$HI) as well as the situation when the polymer is 
coupled to the random MPCD solvent ($-$HI). All ring polymers considered in this work
are not affected by the shear flow up to a certain threshold shear rate 
of $\tau_{\mathrm{MPCD}}\dot{\gamma}_{\times} \approx 10^{-4}$ for $+$HI and 
$\tau_{\mathrm{MPCD}}\dot{\gamma}_{\times} \approx 10^{-5}$ for $-$HI. As of the behaviour of polymers under 
shear flow starts deviating from its equilibrium pattern at Weissenberg number $\mathrm{Wi}_{\times} \cong 1$,
this suggests that the longest relaxation times should be of the order $\tau_{\mathrm{R}} \cong 10^4$ for $+$HI 
and $\tau_{\mathrm{R}} \cong 10^5$ for $-$HI, in agreement with previous findings for 
flexible ring polymers\cite{liebetreu:acsml:2018} or star polymers,\cite{Jaramillo2018} where 
the two relaxation times were indeed found to differ by about one order of magnitude for
polymers of $N \cong 100$.\footnote{We show in the Appendix that these estimates agree well
with the characteristic decay times of the orientational correlation functions of the rings at
equilibrium.}

%%%%%%%%%%%%%%%%%%%%%%%%%%%%%%%%%%%%%%%%%%%%%%%%%%%%%%%%%%%%%%%%%%%%%%%%
%%%%%%%%%%%%%%%%%%%%%%%%%%%%%%%%%%%%%%%%%%%%%%%%%%%%%%%%%%%%%%%%%%%%%%%%

%%%%%%%%%%%%%%%%%%%%%%%%%%%%%%%%%%%%%%%%%%%%%%%%%%%%%%%%%%%%%%%%%%%%%%%%
\begin{figure*}[htb]
\centering
  \includegraphics[width=0.4\textwidth]{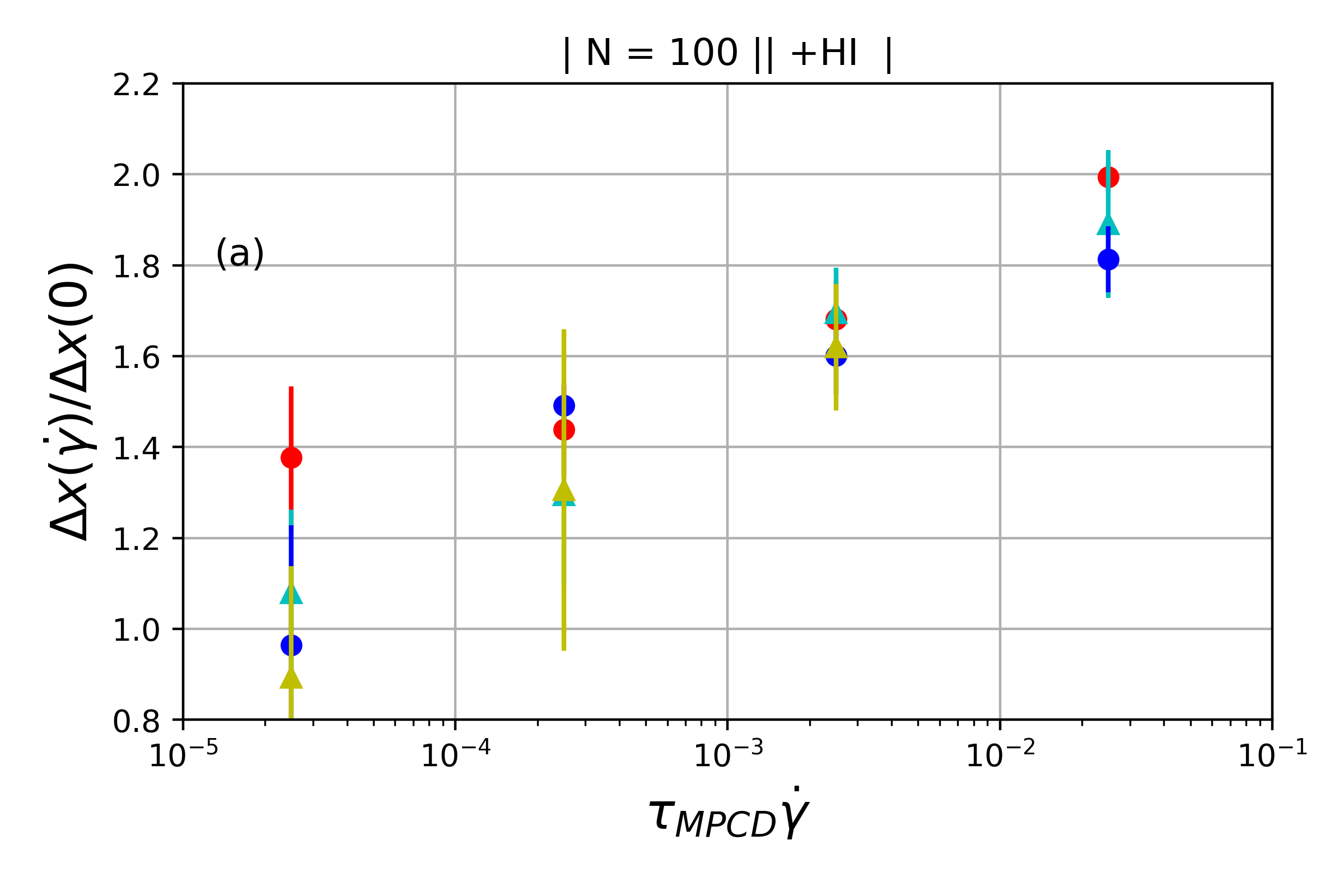}
  \includegraphics[width=0.4\textwidth]{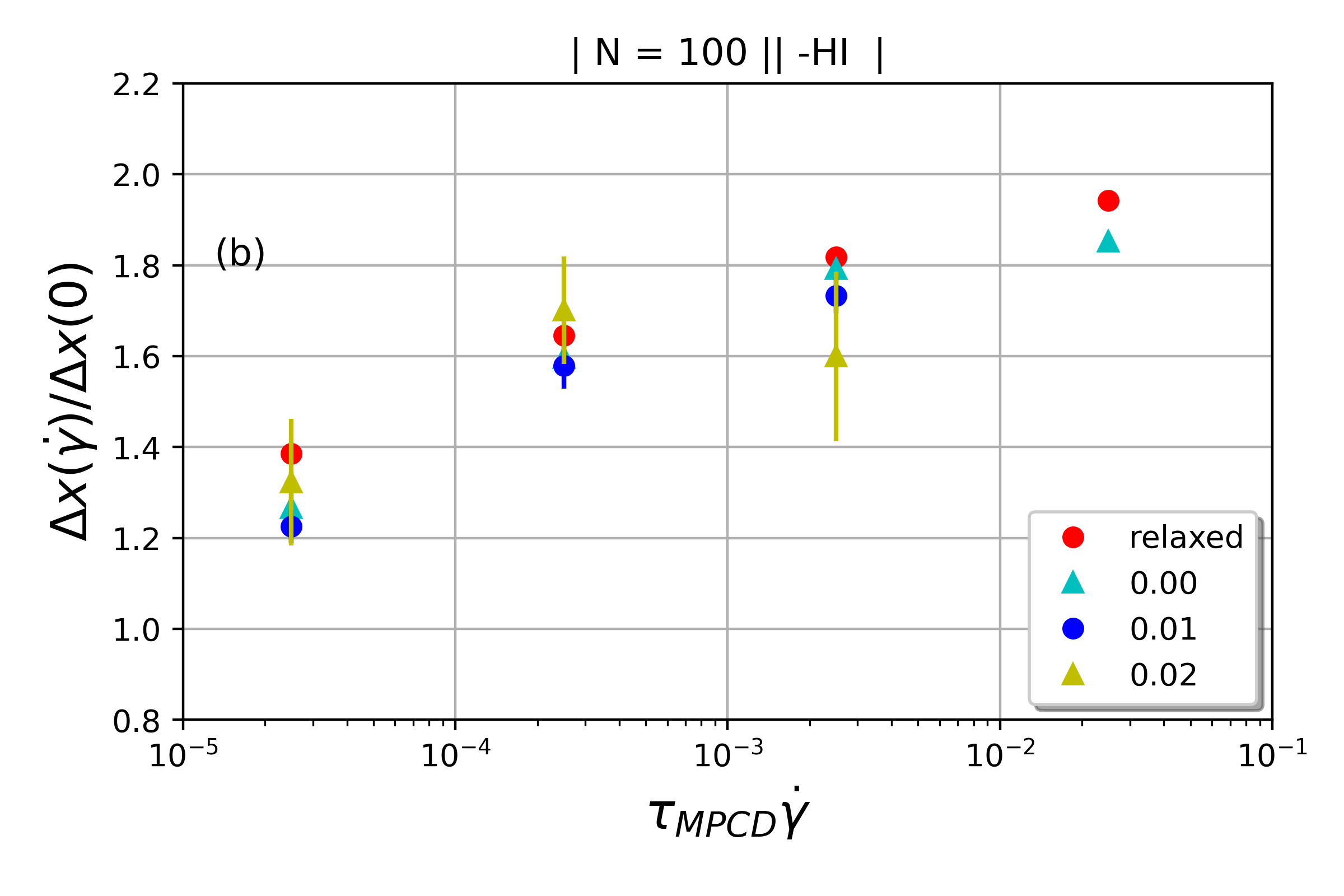}
  \caption{The extent $\Delta x(\dot\gamma)$ of the polymer along the flow direction normalized over its equilibrium value, $\Delta x(0)$,
  as a function of the shear rate $\dot\gamma$. (a) $+$HI-case; (b) $-$HI-case.}
  \label{Fig:deltax}
\end{figure*}
%%%%%%%%%%%%%%%%%%%%%%%%%%%%%%%%%%%%%%%%%%%%%%%%%%%%%%%%%%%%%%%%%%%%%%%%

For all degrees of supercoiling as well as for the relaxed rings the component of the gyration tensor in the flow direction grows with 
increasing shear rates while the components in vorticity and gradient directions both get smaller.
As the curves do not differ between relaxed and supercoiled rings, we can attribute the trends
as well as their deviations from those of flexible rings to the non-vanishing bending rigidity 
of the polymers.
Interestingly, the growth of $G_{xx}(\dot{\gamma})$ (i.e. size extension in the flow direction) with shear rate,
shown in Fig.~\ref{Fig:gyration_vs_shear}(a), is moderate 
limited to a factor of $\approx 2 - 3$ with respect to its equilibrium value for all shear rates covered, 
which is small compared to the growth factor of the fully 
flexible rings,\cite{liebetreu:acsml:2018,liebetreu:commats:2020}
which reach values up to 
$\approx 15-20$. The rings at hand do not 
stretch as much in the flow direction as the flexible ones, due to the stiffness of the bonds
that prevents elongated conformations, which would necessarily result into sharp turns and thus
very high bending penalties at the tips of the rings. For the $-$HI case, 
Fig.~\ref{Fig:gyration_vs_shear}(b) there is even a saturation of $G_{xx}(\dot\gamma)$ at the 
highest shear rates, corresponding to $10^2 \lesssim {\mathrm{Wi}} \lesssim 10^3$.
In Fig.~\ref{Fig:deltax}, we show a quantity closely related to $G_{xx}(\dot\gamma)$, namely the 
extent $\Delta x(\dot\gamma)$. The growth of the extent over its equilibrium value is by about
a factor 2 for the whole range of shear rates applied, much less than values experimentally
measured for fully flexible rings under similar conditions in, e.g., extensional flow.~\cite{hsiao:mm:2016} 

In the gradient and vorticity directions ($y$ and $z$ directions respectively), the diagonal elements of the gyration tensor
decrease with shear rate but again in ways that differ from those of the flexible cyclic polymers.
The contraction in gradient direction is more pronounced than in vorticity direction,
as is also the case for flexible rings, but here it decreases with a stronger power-law,
$G_{yy}(\dot{\gamma}) \sim \dot{\gamma}^{-0.52}$ as opposed to the power-law
$G_{yy}(\dot{\gamma}) \sim \dot{\gamma}^{-0.43}$, valid for flexible rings,\cite{liebetreu:acsml:2018}
independently of the inclusion or exclusion of HI, see 
figs.~\ref{Fig:gyration_vs_shear}(c) and \ref{Fig:gyration_vs_shear}(d).
One would naively expect that this is due to a stronger alignment of the rigid rings with the
flow direction but, as we will establish in what follows, this is not the case. The rigid and
supercoiled rings at hand align, in fact, much less with the flow direction than flexible ones
but at the same time they also tumble much less. Accordingly, time intervals of tumbling,
in which the (flexible) rings extend stronger into the gradient direction contribute much 
less to the time-averages and the contraction of the gradient-direction extent with the 
shear rate is less strong for the flexible rings, resulting into new exponents of the semiflexible ones.

The vorticity-direction dependence of the gyration tensor,
shown in figs.~\ref{Fig:gyration_vs_shear}(e) and \ref{Fig:gyration_vs_shear}(f), 
is also quite unique for the rigid rings.
The phenomenon of vorticity swelling of the 
flexible rings\cite{liebetreu:acsml:2018,liebetreu:commats:2020} is absent here, because
the solvent back-flow is not strong enough to stretch the stiff bonds of these rings.
One sees, nevertheless, the solvent backflow effect in comparing the $+$HI-case,
Fig.~\ref{Fig:gyration_vs_shear}(e), in which it is present, with the 
$-$HI-case, Fig.~\ref{Fig:gyration_vs_shear}(f), in which it is absent. Indeed, in the 
latter the ring contracts much more in the vorticity direction than in the former,
where solvent backflow creates additional swelling pressure in the ring's 
interior.\cite{liebetreu:acsml:2018} In this respect, it is important to notice
that the conversion of writhe to twist \textit{and} the reduction in vorticity swelling
are both effects that take place solely in the presence of HI. Accordingly, we claim that
the solvent backflow is crucial for both, i.e., only solvent backflow generates the 
necessary local stresses along the polymer backbone to bring about the topological 
conversion while at the same time keeping the ring sufficiently open in the vorticity
direction, consistently with the unwrithed conformation seen, e.g., in 
Fig.~\ref{Fig:topological_vs_shear}(f).

\begin{figure*}[htb]
\centering

  \includegraphics[width=0.4\textwidth]{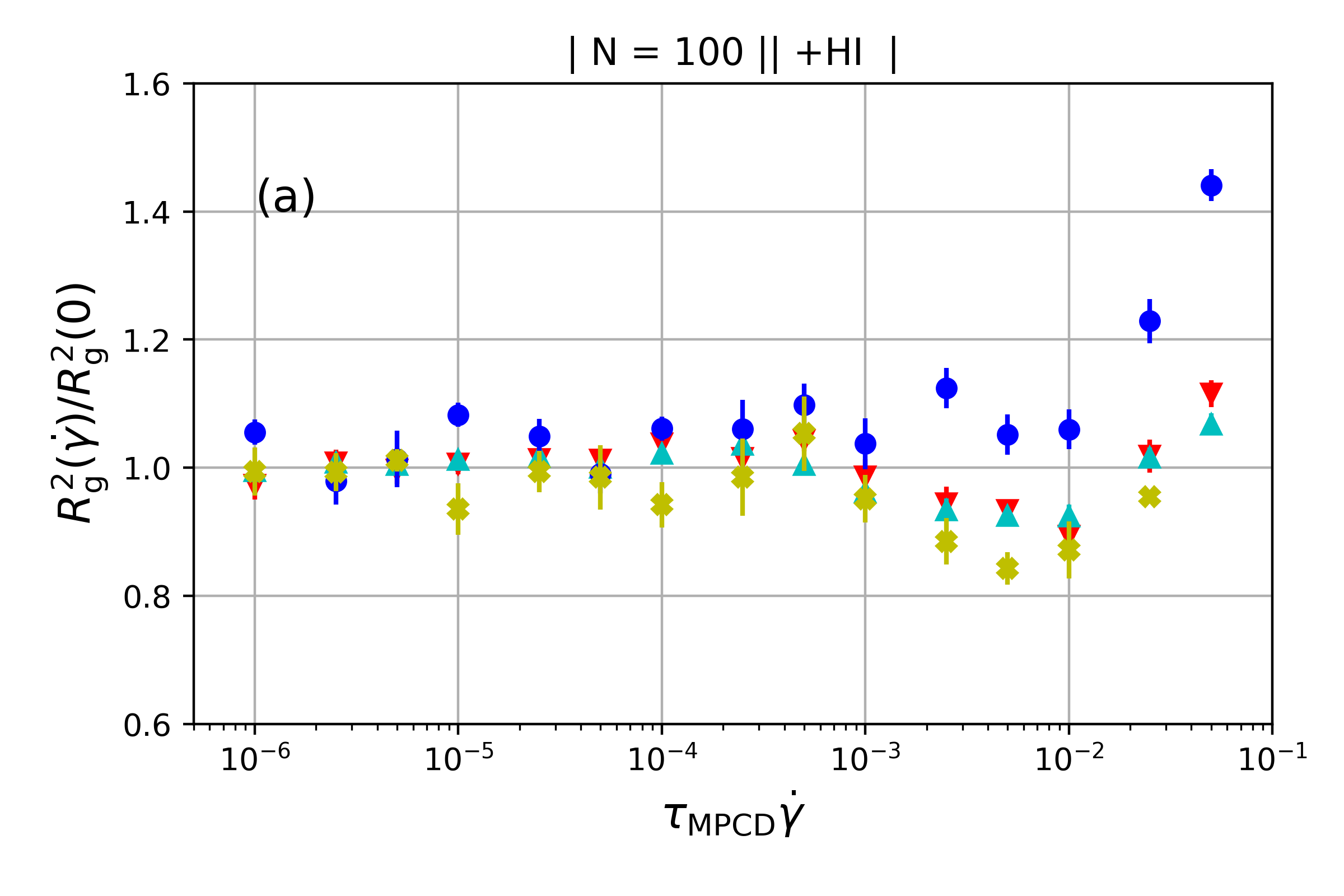}
  \includegraphics[width=0.4\textwidth]{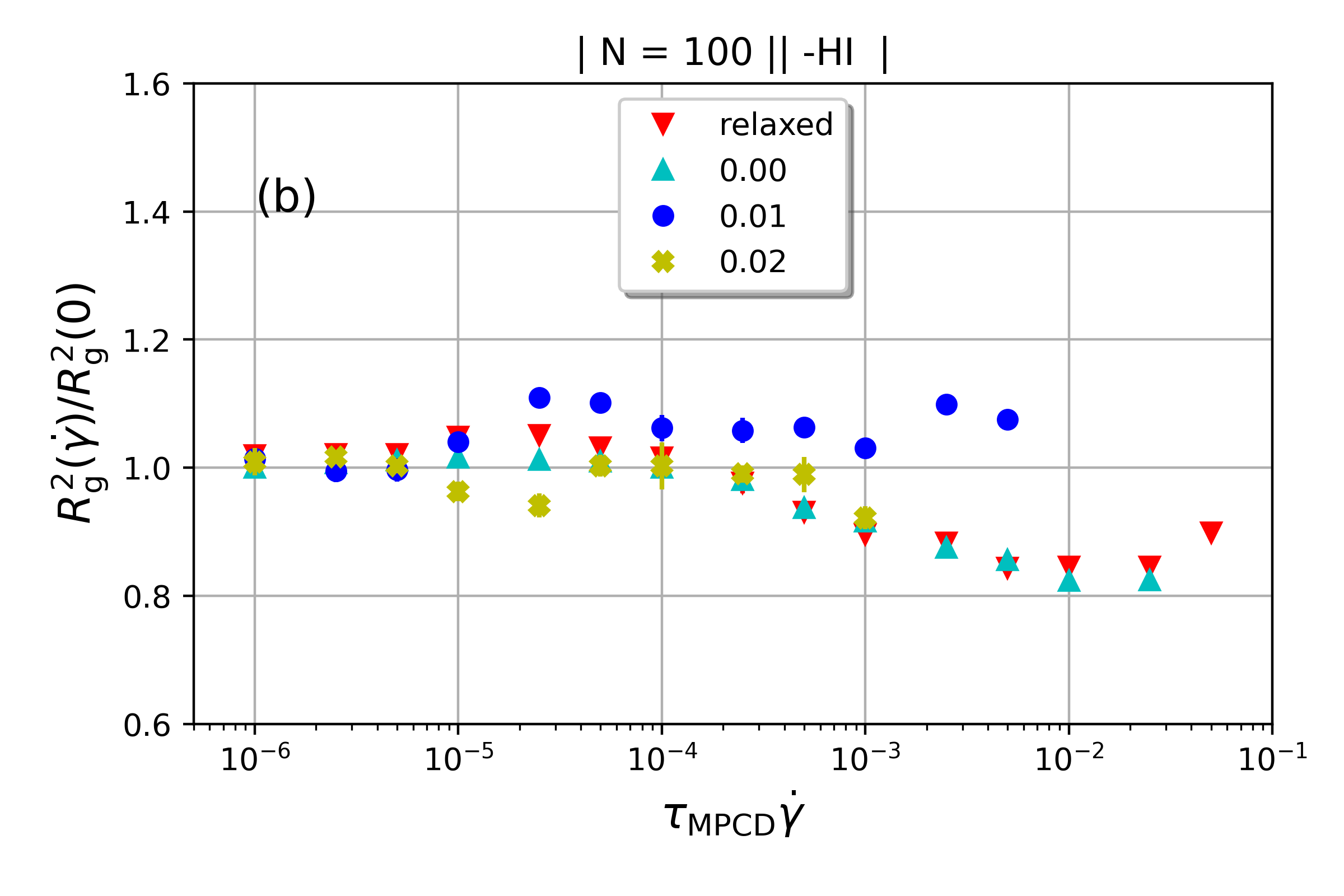}
  \includegraphics[width=0.4\textwidth]{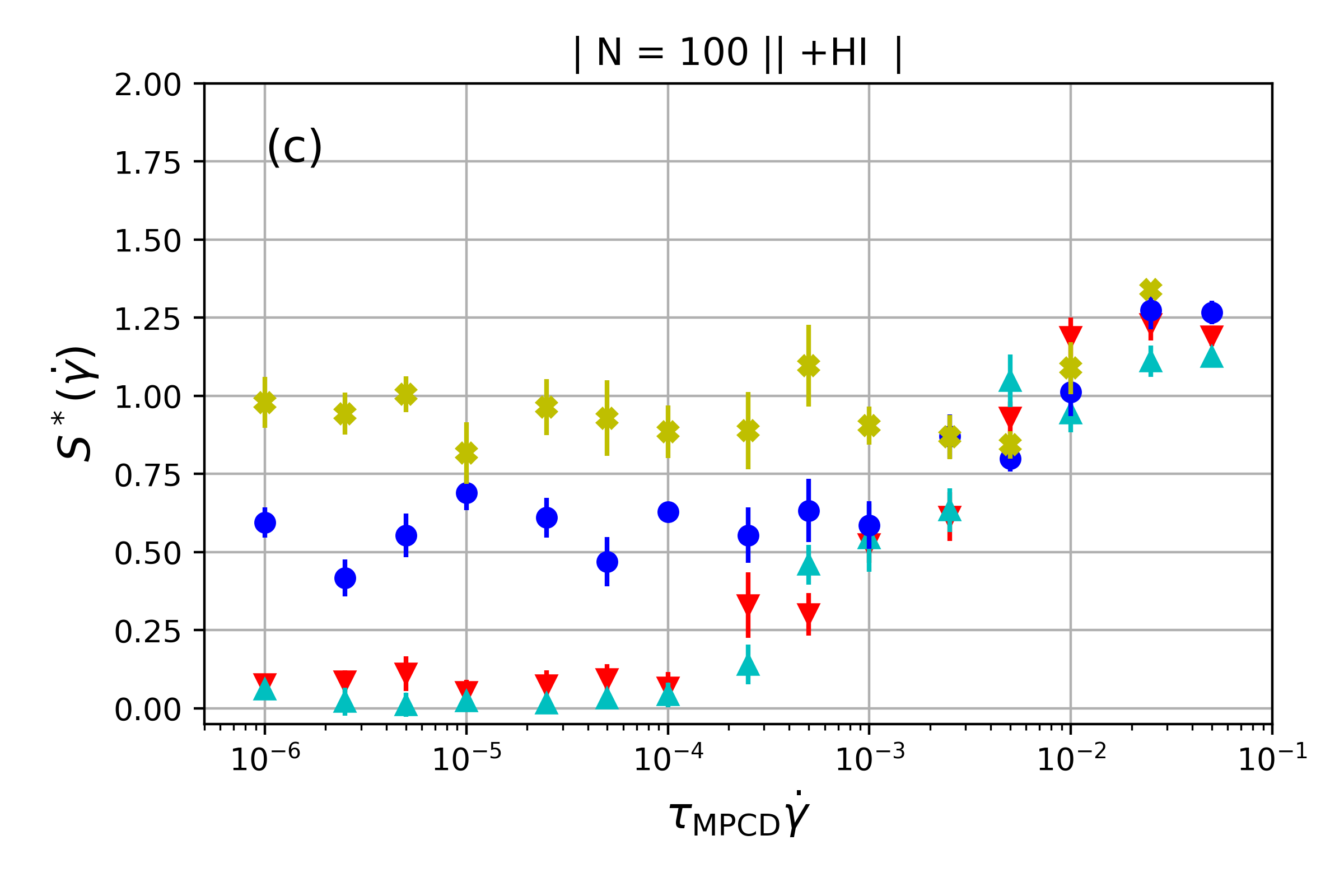}
  \includegraphics[width=0.4\textwidth]{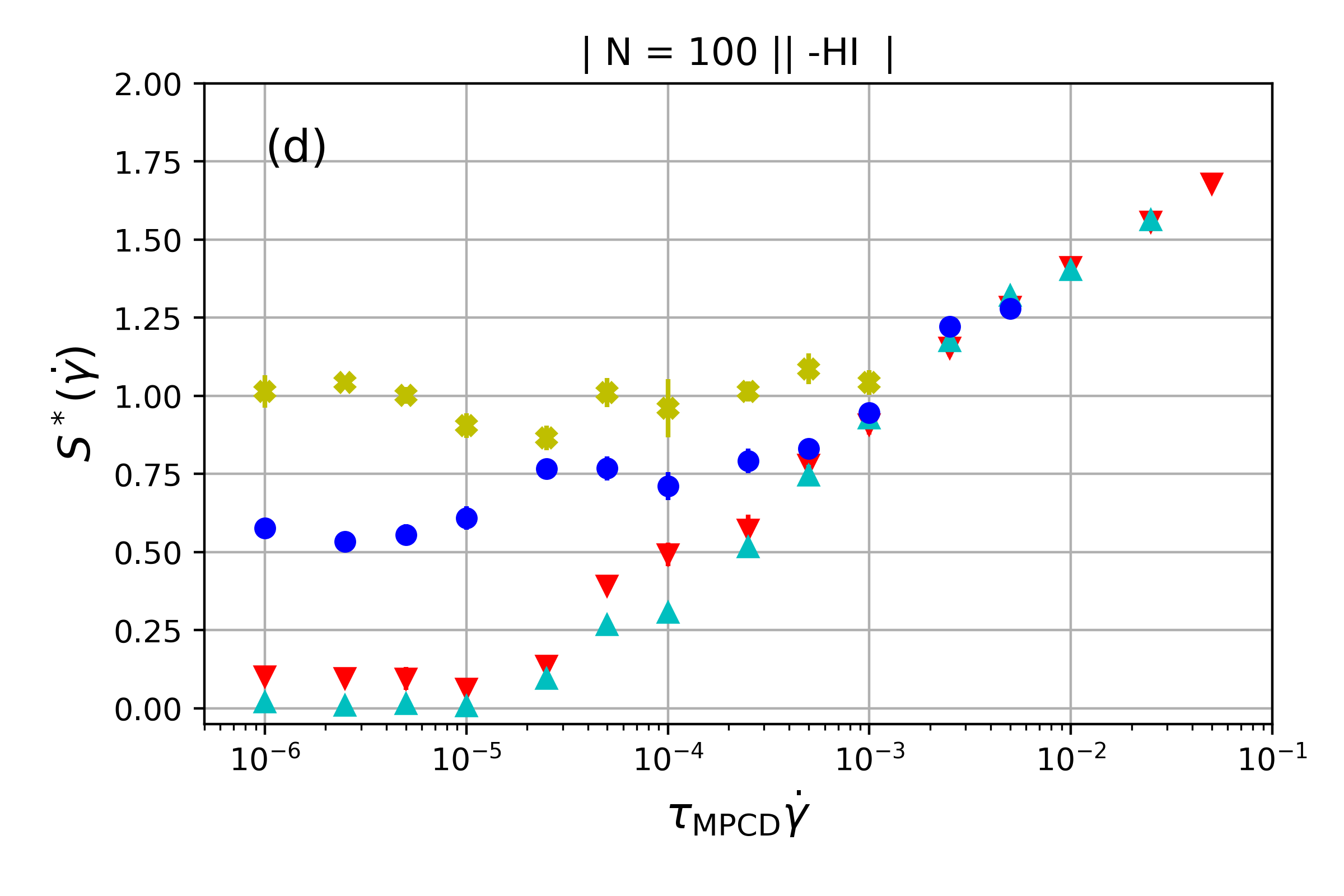}
  \includegraphics[width=0.4\textwidth]{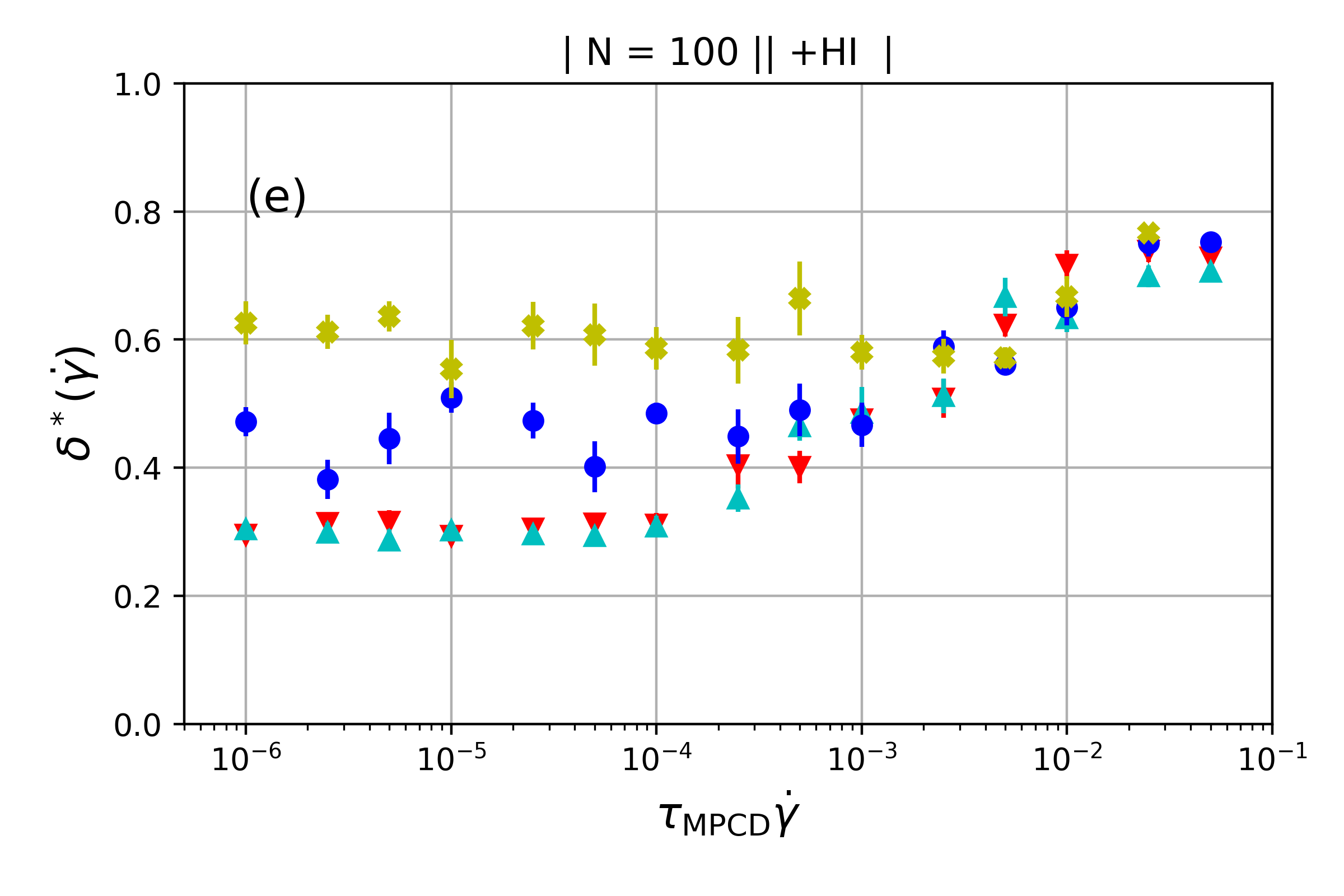}
  \includegraphics[width=0.4\textwidth]{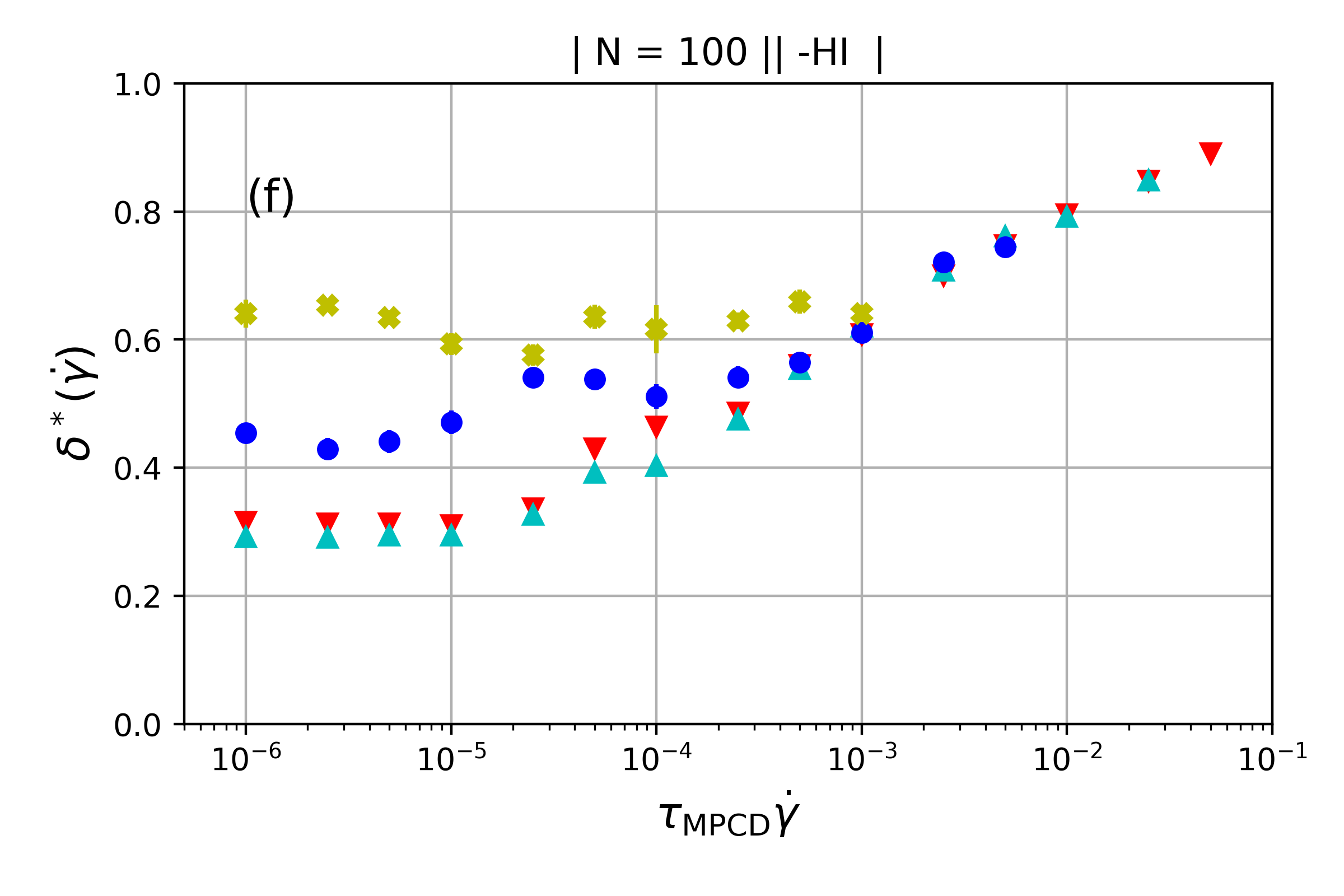}
  \caption{Time-averages of the squared radius of gyration $R_\mathrm{g}^2(\dot\gamma)$ [panels (a), (b)], 
  the prolateness $S^*(\dot{\gamma})$ [panels (c), (d)] and the shape anisotropy $\delta^*(\dot{\gamma})$ [panels (e),(f)] as functions of the shear rate. Panels (a), (c), (e) are for
  the $+$HI-case, and panels (b), (d), (f) for the $-$HI-case. Error bars indicate the standard error of the mean.}
  \label{Fig:shape_parameters_vs_shear}
\end{figure*}

The dependence of the shape parameters on the shear rate is summarized 
in Fig.~\ref{Fig:shape_parameters_vs_shear}.
The decrease of $G_{yy}(\dot{\gamma})$ and $G_{zz}(\dot{\gamma})$ is compensated by the increase in the 
flow direction, so that $R_\mathrm{g}^2(\dot{\gamma})=\mathrm{tr}[G_{\alpha\beta}(\dot{\gamma})]$ is more or less 
constant over most of the range of $\dot{\gamma}$, see 
figs.~\ref{Fig:shape_parameters_vs_shear}(a) and \ref{Fig:shape_parameters_vs_shear}(b). 
This is again in stark contrast with flexible rings, for which $R_\mathrm{g}(\dot\gamma)$ is dominated
by $G_{xx}(\dot\gamma)$ and it grows rapidly with shear 
rate.\cite{liebetreu:acsml:2018,liebetreu:commats:2020} Here, in fact,
the radius of gyration of all ring types decreases by a small amount before experiencing growth again for 
very high shear rates. The effect is observed more 
clearly in the case of $-$HI because the sample size covers longer time periods. 

%%%%%%%%%%%%%%%%%%%%%%%%%%%%%%%%%%%%%%%%%%%%%%%%%%%%%%%%%%%%%%%%%%%%%%%%
%%%%%%%%%%%%%%%%%%%%%%%%%%%%%%%%%%%%%%%%%%%%%%%%%%%%%%%%%%%%%%%%%%%%%%%%

For the overall polymer size, as expressed by the gyration radius, neither the degree of
supercoiling nor the hydrodynamics seem to significantly affect its dependence on shear rate.
This changes if one looks at more detailed shape characteristics, such as the 
prolateness, Figs.~\ref{Fig:shape_parameters_vs_shear}(c) and \ref{Fig:shape_parameters_vs_shear}(d), 
and anisotropy,
Figs.~\ref{Fig:shape_parameters_vs_shear}(e) and \ref{Fig:shape_parameters_vs_shear}(f). 
As they share common characteristics,
we discuss $S^*(\dot\gamma)$ and $\delta^*(\dot\gamma)$ together. First of all, for the lower
shear rates, there is now a clear splitting of the curves depending on the degree of supercoiling.
Whereas the relaxed ring as well as the $\sigma_{\rm sc} = 0$-ring start with $S^*(\dot\gamma) \cong 0$
and low $\delta^*(\dot\gamma)$ values at low shear rates, which grow as $\dot\gamma$ increases,
the supercoiled rings are already prolate and very anisotropic at equilibrium, due to the 
writhed conformations they assume, see Fig.~\ref{Fig:topological_zero_shear}. Accordingly, the 
prolateness and shape anisotropy are affected very weakly by shear for the rings with high supercoiling degree, 
especially for the $+$HI-case, whereas in the $-$HI case prolateness and anisotropy
become monotonically increasing at high $\dot\gamma$ and all rings follow the same
curves there. As mentioned above, solvent backflow for the $+$HI-case reduces prolateness
and anisotropy by temporarily stabilizing more open, oblate conformations of the rings
and by converting writhe into twist for the supercoiled ones.

%%%%%%%%%%%%%%%%%%%%%%%%%%%%%%%%%%%%%%%%%%%%%%%%%%%%%%%%%%%%%%%%%%%%%%%%
\begin{figure}[htb]
\centering
  
  \includegraphics[width=0.9\columnwidth]{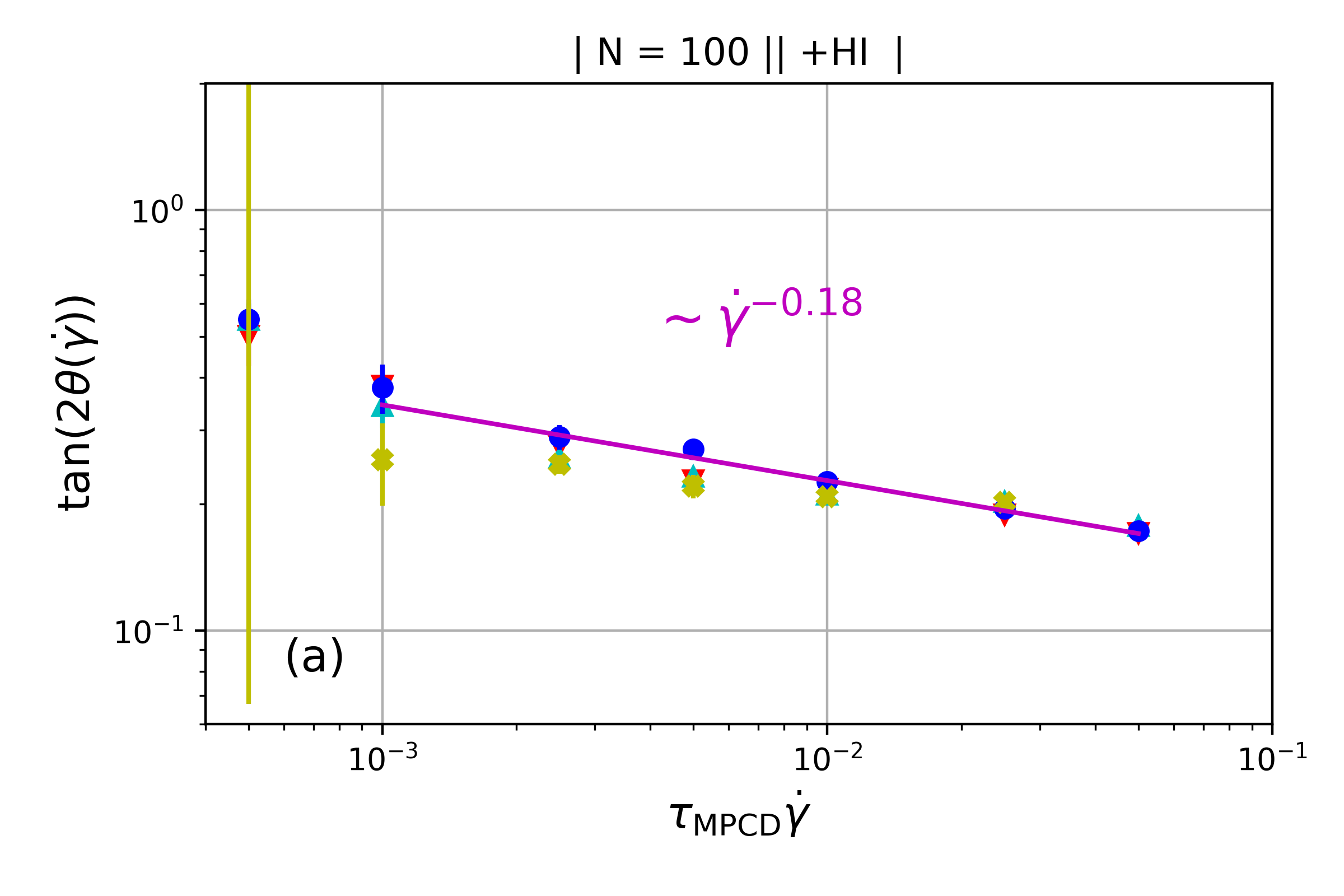}
  \includegraphics[width=0.9\columnwidth]{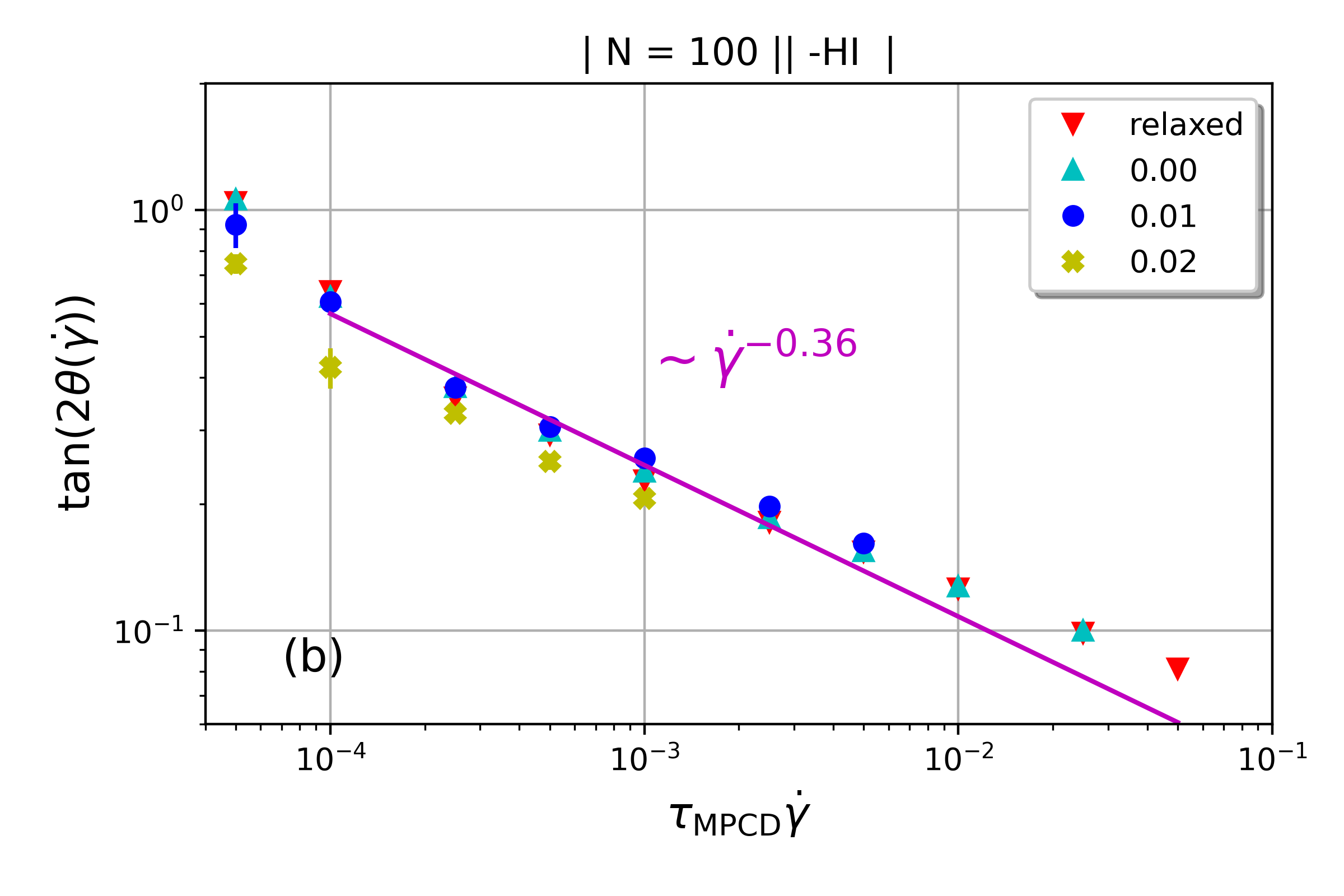}
  \caption{The tangent of twice the alignment angle, $\tan(2\theta(\dot{\gamma}))$, 
  of the rings
  for $+$HI [panel (a)], and $-$HI [panel (b)]. 
  Data points for lower shear rates are cut off as they are very noisy and fluctuate without a visible trend. The power-law fits were computed according to the combined data points of all four ring types. Error bars indicate the standard error of the mean.}
  \label{Fig:alignment_angle_vs_shear}
\end{figure}
%%%%%%%%%%%%%%%%%%%%%%%%%%%%%%%%%%%%%%%%%%%%%%%%%%%%%%%%%%%%%%%%%%%%%%%%
 
The average alignment of a polymer with the shear flow direction and its dependence on shear rate
can be quantified by the orientational resistance $m_G(\dot\gamma)$, 
a dimensionless quantity 
that can be constructed from the gyration tensor elements as:\cite{orientational_resistance}
\begin{equation}
    m_G(\dot{\gamma})  = \tau_{\mathrm{R}} \dot{\gamma} \; \frac{2 G_{xy}(\dot{\gamma}) }{G_{xx}(\dot{\gamma})  - G_{yy}(\dot{\gamma})  }  = \tau_{\mathrm{R}} \dot{\gamma} \; \mathrm{tan}(2\theta(\dot{\gamma})),
    \label{defres}
\end{equation}
\newline
where the angle $\theta(\dot{\gamma})$ is subtended
between the flow direction and the axis of largest extension, i.e., the eigenvector corresponding to 
the largest eigenvalue, $\lambda_1(\dot{\gamma})$, of the 
gyration tensor. We show in Fig.~\ref{Fig:alignment_angle_vs_shear} the dependence of 
$\tan(2\theta(\dot\gamma))$ on the shear rate for both the $+$HI-case, 
Fig.~\ref{Fig:alignment_angle_vs_shear}(a) and the $-$HI-case, 
Fig.~\ref{Fig:alignment_angle_vs_shear}(b). Here, the effect of hydrodynamics is quite
pronounced, as we obtain quite different power-law dependencies for each case. 
As expected, the long axis of the rings tends to align with flow direction as 
$\dot\gamma$ grows but much weaker in the $+$HI-case than in the $-$HI-case. Indeed,
the orientational resistance $m_G \propto \dot\gamma\tan(2\theta(\dot\gamma))$ is very strong
for the $+$HI-case, scaling as $m_G \sim \dot\gamma^{0.82}$,
to be compared with the scaling $m_G \sim \dot\gamma^{0.6}$ 
for flexible rings,\cite{liebetreu:commats:2020}
whereas $m_G \sim \dot\gamma^{0.64}$ in the $-$HI-case. Rigid and supercoiled
rings are therefore orientationally stiff under shear, since the solvent backflow
hinders strong alignment with the flow axis. The forces of the solvent on the rings
are `used up' in bringing about occasional opening of the ring as well as the associated
conversion of writhe into twist for the supercoiled ones, and not for aligning the 
rings with the flow, whereas in the $-$HI-case, the absence of coupling with the 
solvent allows the undisturbed velocity profile of the latter to align the polymer
more strongly with the $x$-axis.

%%%%%%%%%%%%%%%%%%%%%%%%%%%%%%%%%%%%%%%%%%%%%%%%%%%%%%%%%%%%%%%%%%%%%%%%
\begin{figure}[htb]
\centering
  
  \includegraphics[width=0.49\columnwidth]{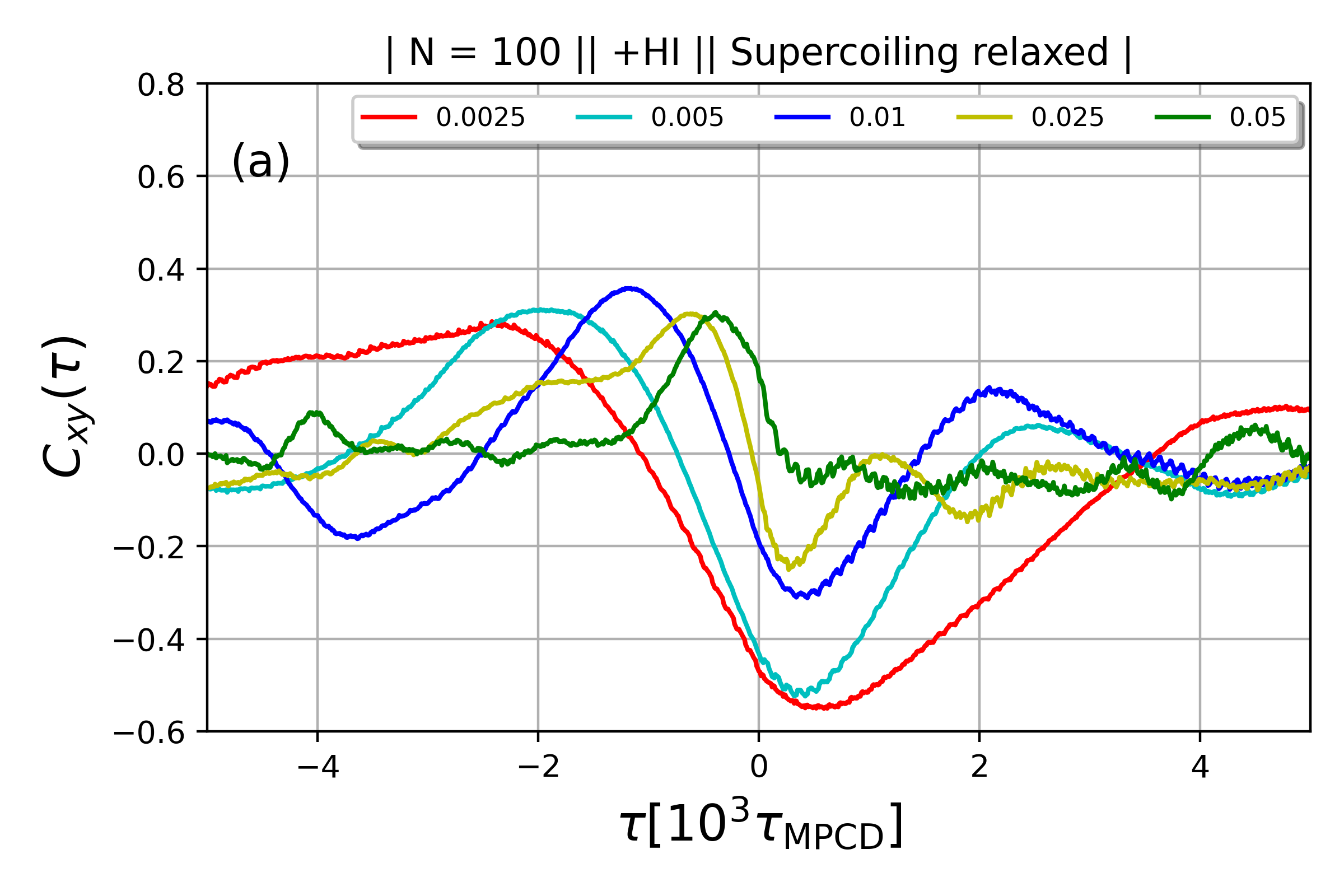}
  \includegraphics[width=0.49\columnwidth]{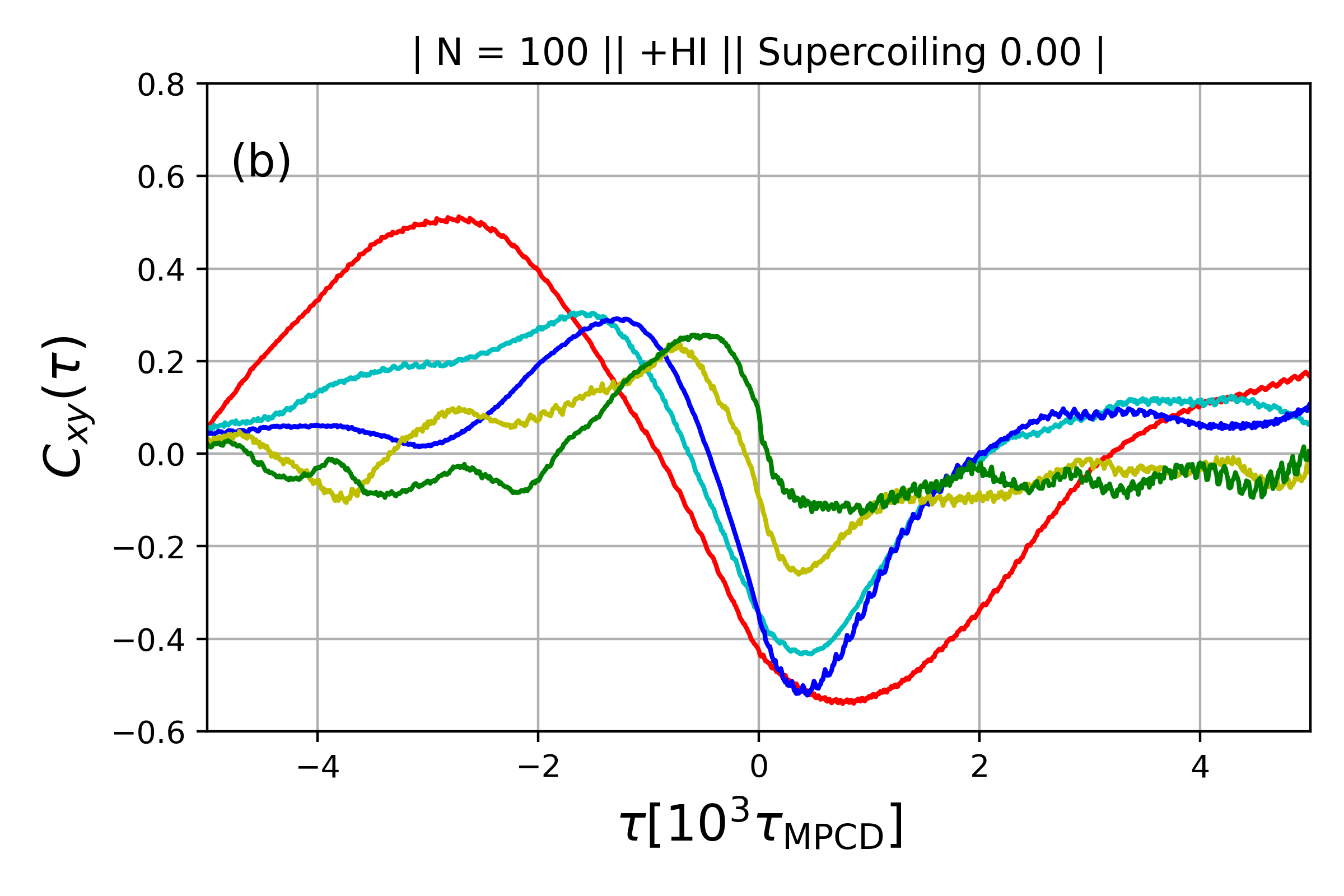}
  \includegraphics[width=0.49\columnwidth]{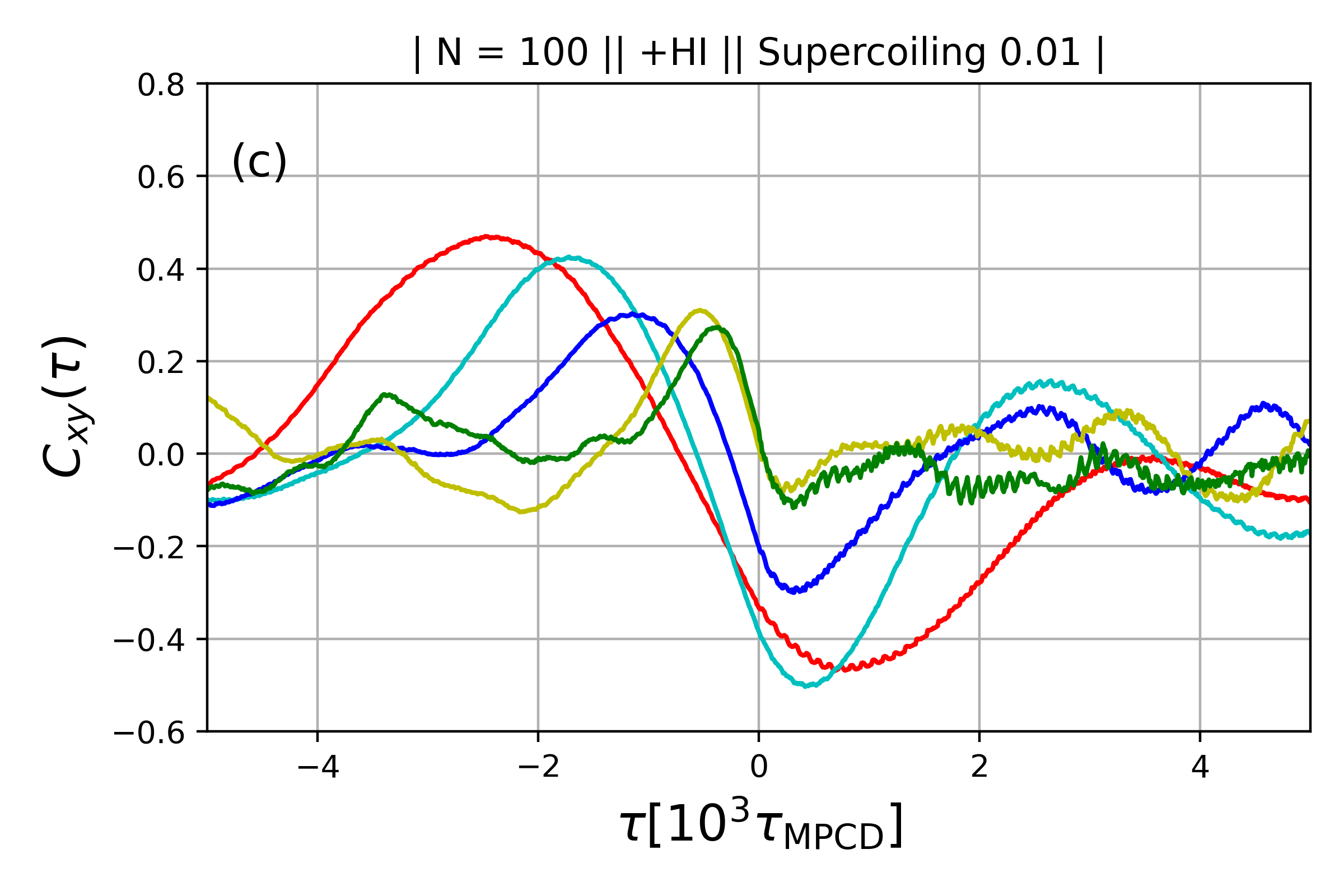}
  \includegraphics[width=0.49\columnwidth]{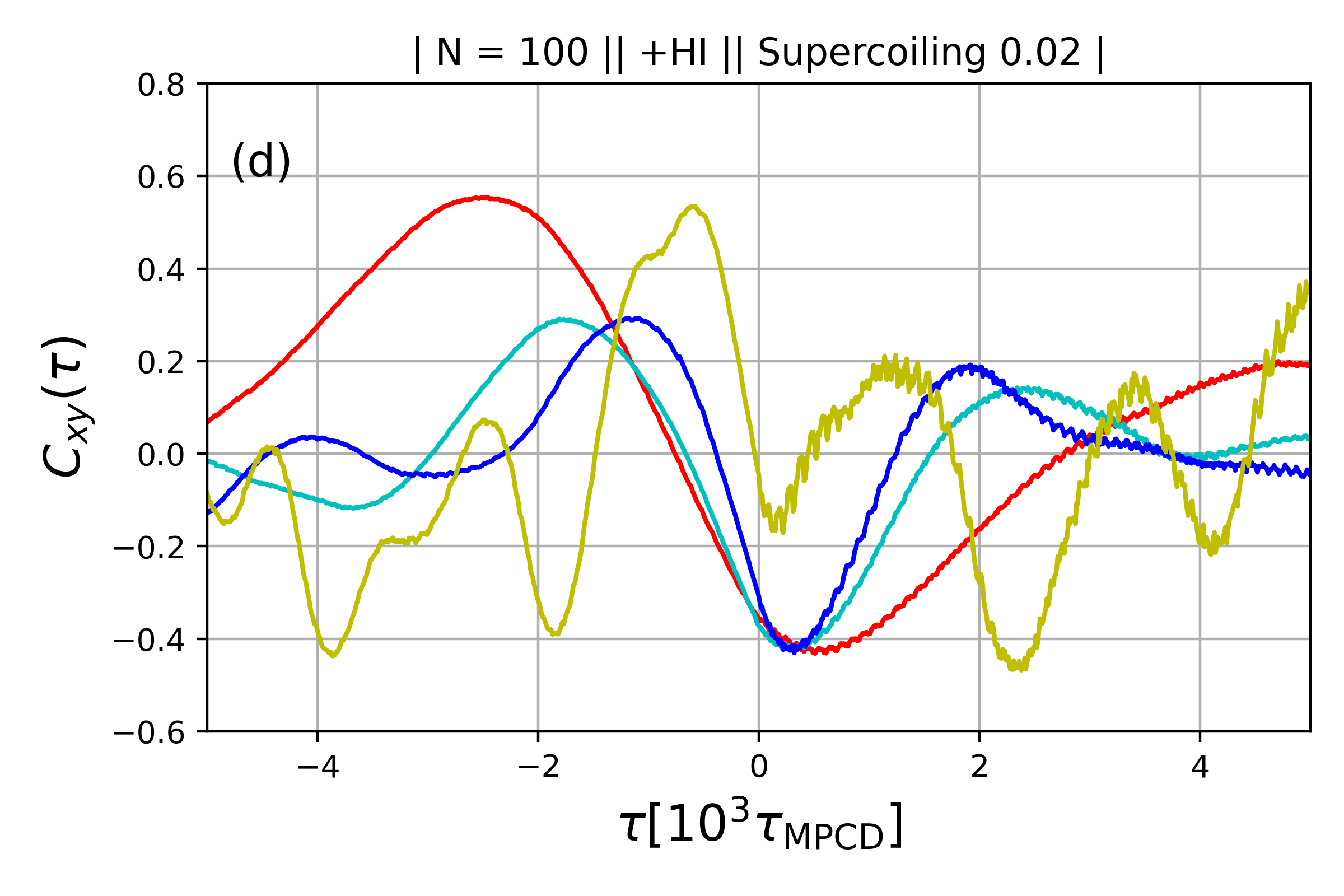}
  \caption{Tumbling cross correlation function $C_{xy}(\tau;\dot{\gamma})$ at high shear rates, as indicated in the legend of panel (a), in presence of HI. (a) Relaxed ring without torsion ($k_{\mathrm{torsion}}=0$), (b) $\sigma_{\mathrm{sc}}=0.00$, (c) $\sigma_{\mathrm{sc}}=0.01$, (d) $\sigma_{\mathrm{sc}}=0.02$.}
  \label{Fig:tumbling}
\end{figure}
%%%%%%%%%%%%%%%%%%%%%%%%%%%%%%%%%%%%%%%%%%%%%%%%%%%%%%%%%%%%%%%%%%%%%%%%

The strong orientational resistance of the supercoiled polymers, in conjunction with the
previous results on the dependence of $G_{\alpha\alpha}(\dot\gamma)$ on the shear rate,
shown in Fig.~\ref{Fig:gyration_vs_shear}, creates an apparent contradiction. Indeed, if we visualize
the polymer as an object oriented on a stable configuration that forms an angle 
$\theta(\dot\gamma)$ with the $x$-axis, and making the small-angle approximation
$\tan(\theta(\dot\gamma)) \cong \theta(\dot\gamma)$, valid only
for the $-$HI-case, we can write:
\begin{equation}
\sqrt{G_{yy}(\dot\gamma)} \cong \sqrt{G_{xx}(\dot\gamma)}\tan(\theta(\dot\gamma)) \cong 
\sqrt{G_{xx}(\dot\gamma)} \theta(\dot\gamma).
\label{eq:trigonometry}
\end{equation}
Ignoring the weak dependence of $G_{xx}(\dot\gamma)$ on the shear rate [Fig.~\ref{Fig:gyration_vs_shear}(b)] (fitted exponent $0.03$, for $\tau_{\mathrm{MPCD}}\dot{\gamma}>10^{-4}$, not shown)
and writing $\theta(\dot\gamma) \sim \dot\gamma^{-0.36}$, eq.~(\ref{eq:trigonometry}) yields the 
power-law $G_{yy}(\dot\gamma) \sim \dot\gamma^{-0.72}$,  stronger than the measured power-law
$G_{yy}(\dot\gamma) \sim \dot\gamma^{-0.53}$ in Fig.~\ref{Fig:gyration_vs_shear}(c). Moreover,
the corresponding power-laws for flexible polymers read $\theta(\dot\gamma) \sim \dot\gamma^{-0.60}$
and $G_{yy}(\dot\gamma) \sim \dot\gamma^{-0.43}$, i.e., in that case the gradient-direction
contraction is \textit{weaker} than that of the rigid rings although their alignment with 
the flow direction is stronger. This can be partially explained by the fact that for the
flexible ring case, $G_{xx}(\dot\gamma) \sim \dot\gamma^{0.60}$; even so, it is evident
that the picture of an effective ellipsoid orienting itself at a stable angle 
$\theta(\dot\gamma)$ with the $x$-axis during shear flow is too simplistic, as the 
polymer dynamics is richer in patterns developed by the combination of thermal and
hydrodynamic forces on the monomers.

\begin{figure}[htb]
\centering
 
  \includegraphics[width=0.9\columnwidth]{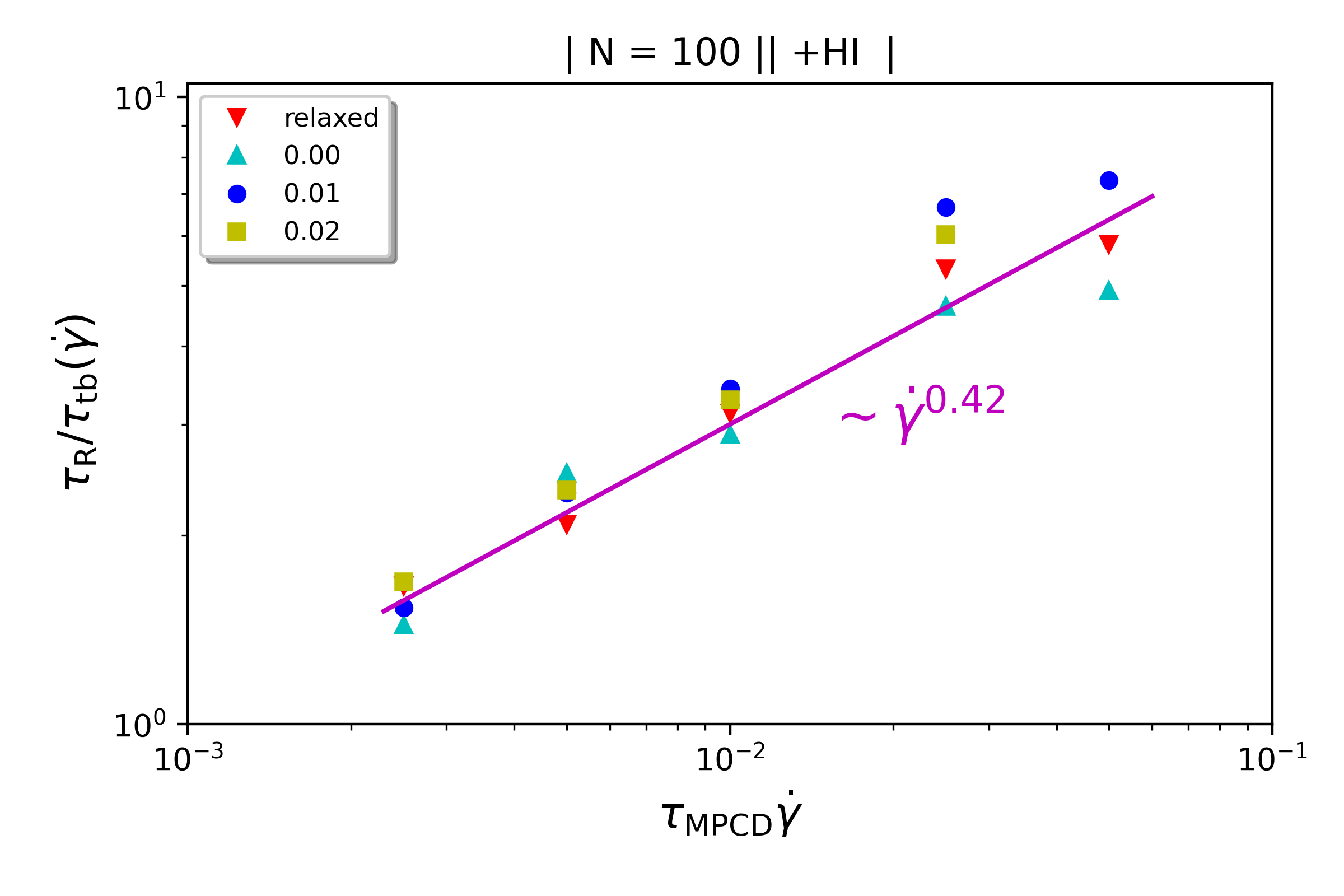}
  \caption{Tumbling frequency $f_{\mathrm{tb}}(\dot{\gamma})=\tau_{R}/\tau_{\mathrm{tb}}(\dot{\gamma})$. The longest relaxation time $\tau_{R}$ is approximated by $\tau_{R} \approx 10^{4}\tau_{\mathrm{MPCD}}$.}
  \label{Fig:tumbling_vs_sr}
\end{figure}

Polymers have been shown to undergo tumbling motion 
under shear flow both in experiment\cite{teixeira:mm:2005,schroeder:prl:2005} 
and simulations.\cite{liebetreu:acsml:2018, Chen2013}
The tumbling dynamics are characterized by a rapid collapse in one direction 
($x$-axis) followed by the expansion in another direction
($y$-axis). Such events are cyclical yet not periodic, as the molecule undergoes
intermittent phases between stable orientations and vivid tumbling motions. 
During the latter, a tumbling frequency $f_{\mathrm{tb}}(\dot{\gamma})$ can be
deducted from the cross correlation function between fluctuations of 
$G_{xx}(t;\dot{\gamma})$ and $G_{yy}(t;\dot{\gamma})$ around their mean values 
separated by a time difference $\tau$:\cite{teixeira:mm:2005,schroeder:prl:2005,liebetreu:acsml:2018}
\begin{equation}
    \centering
C_{xy}(\tau;\dot{\gamma}) = \frac{ \langle \delta G_{xx}(t;\dot{\gamma}) \delta G_{yy}(t + \tau;\dot{\gamma}) \rangle _{t}}{\sigma_{xx}(\dot{\gamma}) \sigma_{yy}(\dot{\gamma})},
\label{cxy}
\end{equation}
where 
$\delta G_{\alpha\alpha}(t;\dot{\gamma}) = G_{\alpha\alpha}(t;\dot{\gamma}) - G_{\alpha\alpha}(\dot{\gamma}) $ and 
$\sigma_{\alpha\alpha}(\dot{\gamma})=\sqrt{\langle G_{\alpha\alpha}^2(t;\dot{\gamma}) \rangle_t - \langle G_{\alpha\alpha} (t;\dot{\gamma})\rangle_t ^2}$.
If the polymer undergoes tumbling motions, the cross correlation function will 
have a damped oscillatory behaviour with a pronounced maximum at a 
time $t_+(\dot{\gamma})<0$ and a pronounced minimum at $t_-(\dot{\gamma})>0$. 
Similar to their flexible counterparts,
supercoiled rings feature tumbling dynamics at high shear rates, 
see Fig.~\ref{Fig:tumbling}, with clear minimum and maximum peaks. 
The oscillation period $\tau_{\mathrm{tb}}(\dot\gamma)$ of this motion 
can be extracted from the correlation function $C_{xy}(\tau;\dot\gamma)$ as:
\begin{equation}
\tau_{\mathrm{tb}}(\dot{\gamma})=2\left(t_-(\dot{\gamma}) - t_+(\dot{\gamma})\right).
\label{eq:tumbling_period}
\end{equation}
The resulting tumbling frequencies $f_{\mathrm{tb}}(\dot\gamma) = \tau_{\mathrm{tb}}^{-1}(\dot\gamma)$, 
normalized with the inverse of the
relaxation time, are shown in Fig.~\ref{Fig:tumbling_vs_sr}. As expected, the tumbling frequency
grows with the shear rate; however, the resulting growth 
$f_{\mathrm{tb}}(\dot\gamma) \sim \dot\gamma^{0.42}$ is much weaker than the 
$f_{\mathrm{tb}}(\dot\gamma) \sim \dot\gamma^{0.67}$-dependence of the flexible rings.\cite{liebetreu:acsml:2018}
Rigidity slows down the rotational motion of the rings under shear, so that a rigid ring
at a specific Weissenberg number tumbles and orients to the flow in a way akin to a flexible
ring at some lower Weissenberg number, where tumbling is less frequent. In this way, 
for the same Weissenberg number, the
rigid rings have fewer phases than their flexible counterparts in which they undergo 
tumbling motions that cause $G_{yy}(t;\dot\gamma)$ to grow, resulting thereby in a 
stronger gradient-direction contraction than for flexible cyclic macromolecules.

%%%%%%%%%%%%%%%%%%%%%%%%%%%%%%%%%%%%%%%%%%%%%%%%%%%%%%%%%%%%%%%%%%%%%%%%
%%%%%%%%%%%%%%%%%%%%%%%%%%%%%%%%%%%%%%%%%%%%%%%%%%%%%%%%%%%%%%%%%%%%%%%%

\begin{figure}[htb]
\centering
  
  \includegraphics[width=0.9\columnwidth]{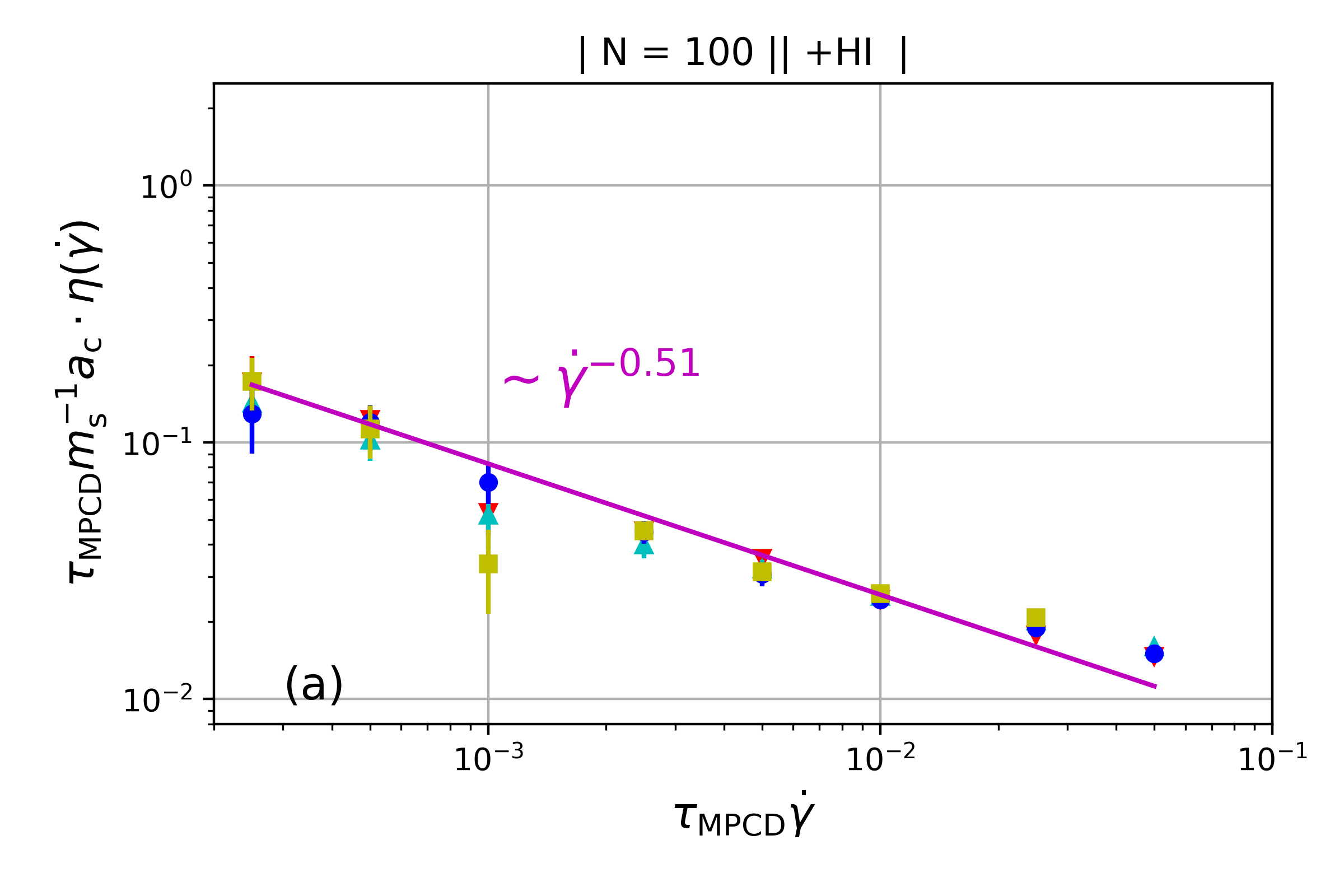}
  \includegraphics[width=0.9\columnwidth]{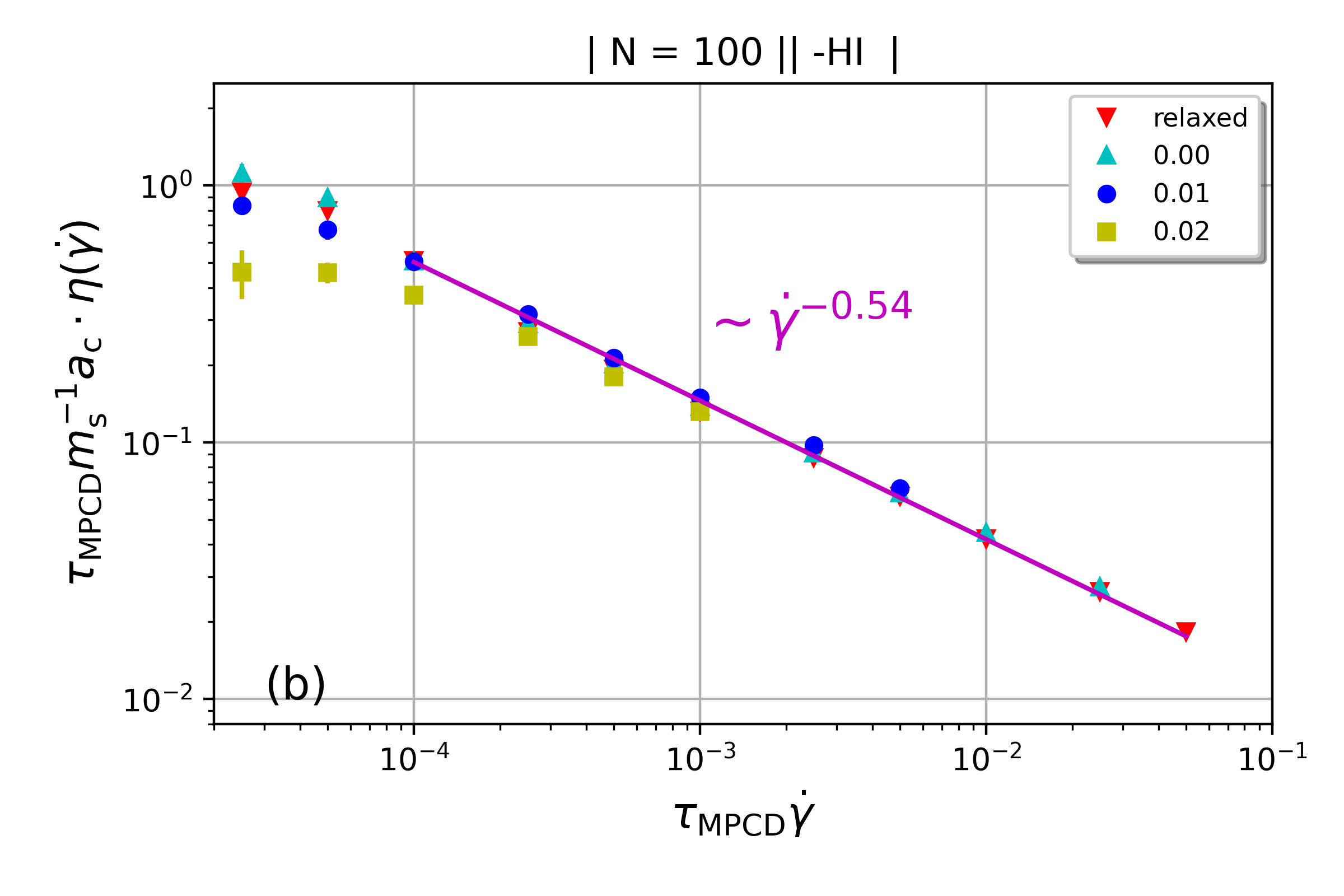}

  \caption{The intrinsic viscosity of the various types of rigid polymers considered in this work
  for the cases $+$HI [panel (a)], and $-$HI [panel (b)]. Data points for lower shear rates are cut off as they are very noisy and fluctuate without a visible trend. The power-law fit curves are computed according to the combined data points of all four rings tested. The values are normalized by the 
  simulation viscosity unit, $\tau_{\mathrm{MPCD}}^{-1} m_{\mathrm{s}} a_{\mathrm{c}}^{-1}$. Error bars indicate the standard error of the mean.}
  \label{Fig:intr_visc_vs_shear}
\end{figure}

We finally discuss the effect of the polymers' presence on the solvent's viscosity, 
by extracting their intrinsic viscosity $\eta(\dot\gamma)$, 
defined as:
\begin{equation}
    \eta (\dot{\gamma})= -\frac{\tau_{xy}(\dot{\gamma})}{\dot{\gamma}}.
    \label{eq:intr_visc}
\end{equation}
In eq.~(\ref{eq:intr_visc}) above, the polymer stress tensor $\tau_{\alpha\beta}(\dot\gamma)$
is calculated in the simulation as a time average:\cite{Ryder2006}
\begin{equation}
    \tau_{\alpha\beta}(\dot{\gamma}) = \frac{1}{V} \sum^{N}_{i=1}
    \langle \overline{r}_{i,\alpha}(t;\dot\gamma){F}_{i,\beta}(t;\dot\gamma) \rangle_t,
    \label{eq:stress_tensor}
\end{equation}
where $V=L_xL_yL_y = 2\cdot10^5 \, a_{\mathrm{c}}^3$ is the volume of the simulation box,
the quantity $\overline{r}_{i,\alpha}(t;\dot\gamma)$ has been defined in 
eq.~(\ref{gyration_tensor}) and ${F}_{i,\beta}(t;\dot\gamma)$ is the $\beta$-component
of the total force from other beads acting on monomer $i$ at time $t$.
Results are shown in Fig.~\ref{Fig:intr_visc_vs_shear}.
As with other polymer architectures, upon a certain threshold value of $\dot{\gamma}$, 
shear thinning takes place and the shear-rate-dependent viscosity has a
power-law dependence on $\dot\gamma$
with an exponent matching that of the scaling of $G_{yy}(\dot{\gamma})$.
This is in agreement with predictions of the Giesekus approximation for the 
stress tensor,\cite{doyle:jfm:1997,schroeder:mm:2005,huang:mm:2010} 
according to which $\eta(\dot\gamma) \sim G_{yy}(\dot\gamma)$.
As the latter is based on a free-draining approximation (no HI) but the 
scaling holds here also for the $+$HI-case, we surmise that the intrinsic viscosity
is not affected by the presence of hydrodynamic interactions, despite the significant
effects these have on the molecular orientation in the shear cell. 

%%%%%%%%%%%%%%%%%%%%%%%%%%%%%%%%%%%%%%%%%%%%%%%%%%%%%%%%%%%%%%%%%%%%%%%%
%%%%%%%%%%%%%%%%%%%%%%%%%%%%%%%%%%%%%%%%%%%%%%%%%%%%%%%%%%%%%%%%%%%%%%%%

\section{Conclusions}
\label{sec:conc}

By means of computer simulations that take into account hydrodynamic interactions,
we have examined in detail the behavior of supercoiled ring polymers in steady shear,
comparing the results to those obtained for flexible polymers and therefore identifying
the key factors that influence the conformation and dynamics under steady flow.
We found that the combination of bending rigidity and supercoiling renders 
such rings much more robust in their sizes, shapes and orientations than their
flexible counterparts and that it also reduces the frequency of their tumbling motions.
Despite the fact that they display a much higher orientational resistance to flow 
than flexible rings, supercoiled rings are indeed affected by shear in nontrivial ways,
and mostly when hydrodynamic interactions are taken into account. Particularly
relevant are the effects of the solvent on the vorticity extension of the ring,
where back-flow significantly reduces the vorticity contraction seen in simulations
in the absence of hydrodynamics. The topological properties of the rings, namely
their writhe and twist, are quite robust for a wide range of shear rates, with
the polymers maintaining mostly writhed configurations with very little twist.
However, at sufficiently strong shear rates, writhe is rapidly reduced in favour
of the twist in the case in which hydrodynamics is active.
We have not been able to test whether the same topological
transformation of writhe to twist would also take place
in the absence of hydrodynamic interactions due to current
simulation limitations outlined before.

Our results highlight the interplay between
topology, hydrodynamics and thermal fluctuations in dilute
polymer solutions. Experimentally, they can in principle be tested
in suitably constructed microfluidic devices with single-molecule techniques.\cite{tu:mm:2020, tanyeri:nl:2013, Schroeder_Single_molecule_review_JR18} Indeed, properties of ring polymers in flows are being elucidated with these techniques\cite{Schroeder_hysteresis_Science03,Schroeder_RL_NATCOM19,hsiao:mm:2016,Soh_PRL19,Schroeder_Single_molecule_review_JR18} and even viscoelasticity of supercoiled plasmids has been considered, however only in higher concentration blends so far.\cite{Schroeder_supercoiled_SM20,janspaper} Our results include experimentally accessible chain extent and other characteristics, to be addressed experimentally, but the DNA plasmids would have to be first sorted according to their supercoiling degree. As our simulations concerned only relatively short rings, the gyration radius depends only weakly on the supercoiling, hence a method other than gel electrophoresis would have to be used.
Future directions include the study of the dynamical behaviour in higher 
concentrations or for topologically linked rigid rings, such as catenanes and 
the kinetoplast.
As a further highly interesting avenue of exploration, particularly in connection to experiments, we see the investigation of longer and/or more flexible rings with torsional stiffness in shear flow. Such rings with high supercoiling degrees display branched structure in equilibrium. The orientation of the branches as well as their restructuring and hydrodynamic swelling connected to the conversion of writhe to twist in shear flow is highly desired, not only due to the biological context, but also as a future perspective tool to tune mechanical properties of polymer solutions using topology in nonequilibrium conditions. The present work presents the first step on the way to achieve these goals. 

\section*{Author Contributions}
CS   wrote the simulation code, performed the calculations, analysed the results and wrote the paper. 
JS and AZ designed research, advised on the calculations, 
analysed the results and wrote the paper.
CNL designed research, analysed the results and wrote the paper.

\section*{Conflicts of interest}
There are no conflicts to declare.

\section*{Acknowledgements}
We acknowledge support from the European Union (Horizon-MSCA-Doctoral  Networks) through the project QLUSTER  (HORIZON-MSCA-2021-DN-01-GA101072964).
The computations leading to the results of this work have been carried
out in part at the Vienna Scientific Cluster (VSC).

%\newpage

\section*{Appendix}
\subsection*{Software details}
The code for the simulations is written in Fortran 90, visualisations of polymers are created with VMD. \cite{HUMP96} 
%Random numbers are generated during the MPCD algorithm (axis of randomly orientated rotation matrices, placing into cells and setting starting velocities during initialisation, Poisson distributed particle numbers and effective momenta of random solvent) using the ziggurat library. \cite{ziggurat} 

\subsection*{Velocity Verlet algorithm with rotational degrees of freedom} 
The rotational degrees of freedom have to be considered within the velocity Verlet algorithm which is extended by the rotational dynamics consisting of the following three steps:
\begin{equation}
\begin{split}
\label{velocity verlet rotation}
    \vec{\omega}_i\left(t + \frac{\delta t}{2}\right) & = \vec{\omega}_i(t) + \frac{\delta t}{2} \dot{\vec{\omega}}_i(t)
    \\
    \chi_i(t + \delta t) & = R_{\chi} \biggl[ \delta t \vec{\omega}_i\left(t + \frac{\delta t}{2}\right) \biggr]  \chi_i(t) 
    \\
    \vec{\omega}_i(t + \delta t) & = \vec{\omega}_i\left(t + \frac{\delta t}{2}\right) + \frac{\delta t}{2} \dot{\vec{\omega}}_i(t + \delta t),
\end{split}
\end{equation}
where $\dot{\vec{\omega}}_i(t + \delta t) \equiv \dot{\vec{\omega}}_i(\{\vec{r}_j(t + \delta t)\},\{\chi_j(t + \delta t)\})$, i.e.~the angular acceleration at time $t+\delta t$ is computed from the positions $\vec{r}_{j}$ and orientations $\chi_{j}$ at time $t+\delta t$. The steps are performed simultaneously to the translational counterparts where the monomer is moved as a whole according to the bead velocity $\vec{v}_i$. In the rotational part, first, the angular velocity is boosted by a half-kick of the current angular acceleration. A rotation $R_{\chi} \bigl[ \vec{\omega}_i\left(t + \frac{\delta t}{2}\right) \delta t \bigr]$ of the monomer orientation $\chi_i(t)$ by an angle $\bigl|\vec{\omega}_i\left(t + \frac{\delta t}{2} \right)\bigr| \delta t$ around rotation axis $\vec{\omega}_i/\left|\vec{\omega}_i\right|$ follows. Next, $\dot{\vec{\omega}}_i(t)$ is calculated as $ \dot{\vec{\omega}}_i(t) = \vec{T}_i(t)/I_{\mathrm{sphere}}$. The torque and hence also the angular acceleration depend on both, the location (all potentials) and the orientation (torsion and alignment) of the monomers. $\vec{T}_i$ and $\dot{\vec{\omega}}_i$ are always related by the linear equation $\vec{T}_i= \sum_{k=1}^3 \vec{r}_{\mathrm{pat},i,k} \times \vec{F}_{\mathrm{pat},i,k}=I_{\mathrm{sphere}}\,\dot{\vec{\omega}}_i$. The forces are calculated as a sum over the gradients of the potentials (\ref{WCA}), (\ref{FENE}), (\ref{bending_potential}), (\ref{torsion_potential}), and (\ref{align_potential}). Finally, in step 4 a second half-kick updates the angular velocity with the new $\dot{\vec{\omega}}_i(t+\delta t)$. $\vec{r}_{\mathrm{cm},i}(t+\delta t)$ is computed during the calculation of the torque in step 3 of (\ref{velocity verlet rotation}) at time $t$ and is therefore stored for the next rotation at $t + \delta t$. 

\subsection*{Computation of forces on patches}

The potentials for bending (\ref{bending_potential}) and alignment (\ref{align_potential}) depend on angles formed by two segments connecting three positions, respectively, so that their gradients are the same function with different arguments. In contrast, the torsion potential (\ref{torsion_potential}) involves angles formed by three segments connecting four positions and has an offset $\psi_{0}$ in the argument of the cosine $\cos(\psi_{\mathrm{b/r}}-\psi_0)$. The cosine and sine in (\ref{torsion_potential}) are computed as
\begin{equation}
    \centering
    \begin{aligned}
        &\cos(\psi_{\mathrm{b/r}}) = \frac{(\vec{r}_{i,i+1} \times \vec{r}_{\mathrm{pat},i,\mathrm{b/r}}) \cdot (\vec{r}_{i,i+1} \times \vec{r}_{\mathrm{pat},i+1,\mathrm{b/r}})}{|\vec{r}_{i,i+1} \times \vec{r}_{\mathrm{pat},i,\mathrm{b/r}}||\vec{r}_{i,i+1} \times \vec{r}_{\mathrm{pat},i+1,\mathrm{b/r}})|}
        \\
        & = \frac{|\vec{r}_{i,i+1}|^2}{|\vec{r}_{i,i+1} \times \vec{r}_{\mathrm{pat},i,\mathrm{b/r}}||\vec{r}_{i,i+1} \times \vec{r}_{\mathrm{pat},i+1,\mathrm{b/r}}|} \cdot
        \\
        & \biggr[(\vec{r}_{\mathrm{pat},i,\mathrm{b/r}} \cdot \vec{r}_{\mathrm{pat},i+1,\mathrm{b/r}})
         - (\frac{\vec{r}_{i,i+1}}{|\vec{r}_{i,i+1}|}\cdot \vec{r}_{\mathrm{pat},i,\mathrm{b/r}})(\frac{\vec{r}_{i,i+1}}{|\vec{r}_{i,i+1}|} \cdot \vec{r}_{\mathrm{pat},i+1,\mathrm{b/r}})\biggr], 
    \end{aligned}
    \label{torsion_potential_cos}
\end{equation}
\newline
\begin{equation}
    \centering
    \begin{aligned}
        \sin(\psi_{\mathrm{b/r}}) & = \frac{(\vec{r}_{i,i+1} \times \vec{r}_{\mathrm{pat},i,\mathrm{b/r}}) \times (\vec{r}_{i,i+1} \times \vec{r}_{\mathrm{pat},i+1,\mathrm{b/r}})}{|\vec{r}_{i,i+1} \times \vec{r}_{\mathrm{pat},i,\mathrm{b/r}}||\vec{r}_{i,i+1} \times \vec{r}_{\mathrm{pat},i+1,\mathrm{b/r}})|} \cdot \frac{\vec{r}_{i,i+1}}{|\vec{r}_{i,i+1}|}
        \\
        & = \frac{|\vec{r}_{i,i+1}| \vec{r}_{i,i+1} \cdot (\vec{r}_{\mathrm{pat},i,\mathrm{b/r}} \times \vec{r}_{\mathrm{pat},i+1,\mathrm{b/r}}) } {|\vec{r}_{i,i+1} \times \vec{r}_{\mathrm{pat},i,\mathrm{b/r}}||\vec{r}_{i,i+1} \times \vec{r}_{\mathrm{pat},i+1,\mathrm{b/r}})|} .
    \end{aligned}
    \label{torsion_potential_sin}
\end{equation}
\newline
The forces on beads and patches are derived from their gradients with respect to the position $\vec{r}_a$ where $a$ stands for either a bead or a patch. The force on particle $a$ resulting from the three potentials amounts to:
\begin{equation}
    \centering
    \begin{aligned}
    \vec{F}_a & = -\sum_{c}\vec{\nabla}_a k_{\mathrm{bend}}(1-\cos(\theta_{\mathrm{b}}^c)) - \sum_{c}\vec{\nabla}_a k_{\mathrm{align}}(1-\cos(\theta^c_{\mathrm{a}})) 
    \\
    & \; - \sum_{c}\vec{\nabla}_a k_{\mathrm{torsion}}(1-\cos(\psi_0)\cos(\psi^c_{\mathrm{b/r}})-\sin(\psi_0)\sin(\psi^c_{\mathrm{b/r}}))
    \\
    & = k_{\mathrm{bend}}\sum_{c}\vec{\nabla}_a \cos(\theta_{\mathrm{b}}^c)) + k_{\mathrm{align}}\sum_{c}\vec{\nabla}_a \cos(\theta^c_{\mathrm{align}})
    \\
    & \; + k_{\mathrm{torsion}}\sum_{c}(\cos(\psi_0)\vec{\nabla}_a\cos(\psi^c_{\mathrm{b/r}}) + \sin(\psi_0)\vec{\nabla}_a\sin(\psi^c_{\mathrm{b/r}})),
    \end{aligned}
    \label{potential_forces}
\end{equation}
where the index $c$ indicates summation over all the bonds that particle $a$ is a part of. The position of every bead particle appears in three bending terms, two torsion terms and two alignment terms, whereas the blue and red patches appear only in one torsion bond each and the green patches only in one alignment bond. The computation of $\vec{\nabla}_a \cos(\theta_{\mathrm{b}}^c))$, $\vec{\nabla}_a \cos(\theta^c_{\mathrm{a}})$, and $\vec{\nabla}_a\cos(\psi^c_{\mathrm{b/r}})$ is performed following the calculations of Allen and Tildesley \cite{compsimliq} (App. C.2, p. 491-494). The authors omit the gradient terms of the sine appearing in (\ref{potential_forces}). The missing derivations are performed in an analogous way by rewriting the sine of the torsional angle (\ref{torsion_potential_sin}) formed by beads $\vec{r}_i$, $\vec{r}_{i+1}$ and patches $\vec{r}_{\mathrm{pat},i,\mathrm{b/r}}$, $\vec{r}_{\mathrm{pat},i+1,\mathrm{b/r}}$:
\begin{equation}
    \centering
        \sin(\psi_{\mathrm{b/r}}) = \frac{|\vec{r}_{i,i+1}|\vec{r}_{i,i+1} \cdot (\vec{r}_{\mathrm{pat},i,\mathrm{b/r}} \times \vec{r}_{\mathrm{pat},i+1,\mathrm{b/r}})}{|\vec{r}_{i,i+1} \times \vec{r}_{\mathrm{pat},i,\mathrm{b/r}}||\vec{r}_{i,i+1} \times \vec{r}_{\mathrm{pat},i+1,\mathrm{b/r}})|} 
        = \frac{r_{i,i+1}}{D_1 D_2}V_{\mathrm{triple}},
    \label{torsional_sine}
\end{equation}
\newline
where $V_{\mathrm{triple}}=\vec{r}_{i,i+1} \cdot (\vec{r}_{\mathrm{pat},i,\mathrm{b/r}} \times \vec{r}_{\mathrm{pat},i+1,\mathrm{b/r}})$ and $D_1=|\vec{r}_{i,i+1} \times \vec{r}_{\mathrm{pat},i,\mathrm{b/r}}|$ and $D_2=|\vec{r}_{i,i+1} \times \vec{r}_{\mathrm{pat},i+1,\mathrm{b/r}})|$ are defined for convenience. The gradients with respect to bead position $i$ and $i+1$ are computed by the following derivatives with $C_1 = \vec{r}_{i,i+1}\cdot\vec{r}_{\mathrm{pat},i,\mathrm{b/r}}$ and $C_2 = \vec{r}_{i,i+1}\cdot\vec{r}_{\mathrm{pat},i+1,\mathrm{b/r}}$:
\begin{equation}
    \centering
    \begin{aligned}
        &\vec{\nabla}_{\vec{r}_{i}}\sin(\psi_{\mathrm{b/r}}) =  \frac{r_{i,i+1}}{D_1 D_2} \vec{r}_{\mathrm{pat},i+1,\mathrm{b/r}}\times(\vec{r}_{\mathrm{pat},i,\mathrm{b/r}}-\vec{r}_{i,i+1})  
        \\
        & - \frac{V_{\mathrm{triple}}}{r_{i,i+1}D_1D_2} \Bigg[ \Bigg. \left(1-\frac{r_{i,i+1}^2(l_{\mathrm{pat}}^2+C_1)}{D_1^2} - \frac{r_{i,i+1}^2 l_{\mathrm{pat}}^2}{D_2^2}\right)\vec{r}_{i,i+1} 
        \\
        \\
        & - \frac{r_{i,i+1}^2(r_{i,i+1}^2+C_1)}{D_1^2}\vec{r}_{\mathrm{pat},i,\mathrm{b/r}} + \frac{r_{i,i+1}^2 C_2}{D_2^2}\vec{r}_{\mathrm{pat},i+1,\mathrm{b/r}} \Bigg. \Bigg],
    \end{aligned}
    \label{grad_sin_bead1}
\end{equation}
\newline
\begin{equation}
    \centering
    \begin{aligned}
        &\vec{\nabla}_{\vec{r}_{i+1}}\sin(\psi_{\mathrm{b/r}}) = \frac{r_{i,i+1}}{D_1 D_2} \vec{r}_{\mathrm{pat},i,\mathrm{b/r}}\times(\vec{r}_{\mathrm{pat},i+1,\mathrm{b/r}} + \vec{r}_{i,i+1}) 
        \\
        & + \frac{V_{\mathrm{triple}}}{r_{i,i+1}D_1D_2} \Bigg[ \Bigg. \left(1-\frac{r_{i,i+1}^2(l_{\mathrm{pat}}^2+C_2)}{D_2^2} - \frac{r_{i,i+1}^2 l_{\mathrm{pat}}^2}{D_1^2}\right)\vec{r}_{i,i+1} 
        \\
        & + \frac{r_{i,i+1}^2(r_{i,i+1}^2+C_2)}{D_2^2}\vec{r}_{\mathrm{pat},i+1,\mathrm{b/r}} - \frac{r_{i,i+1}^2 C_1}{D_1^2}\vec{r}_{\mathrm{pat},i,\mathrm{b/r}} \Bigg. \Bigg],
    \end{aligned}
    \label{grad_sin_bead2}
\end{equation}
\newline
while the gradients with respect to the patches $k=$ blue, red of particles $i$ and $i+1$ are given by:

\begin{equation}
    \centering
    \begin{aligned}
        &\vec{\nabla}_{\vec{r}_{\mathrm{pat},i,\mathrm{b/r}}}\sin(\psi_{\mathrm{b/r}}) =
        \\
        &  \frac{r_{i,i+1}}{D_1 D_2} \Bigg[ \Bigg. \vec{r}_{\mathrm{pat},i+1,\mathrm{b/r}} \times \vec{r}_{i,i+1} 
         - \frac{V_{\mathrm{triple}}}{D_1^2}\left( C_1 \vec{r}_{i,i+1} + r_{i,i+1}^2 \vec{r}_{\mathrm{pat},i,\mathrm{b/r}} \right) \Bigg. \Bigg],
    \end{aligned}
    \label{grad_sin_patch1}
\end{equation}

\begin{equation}
    \centering
    \begin{aligned}
        & \vec{\nabla}_{\vec{r}_{\mathrm{pat},i+1,\mathrm{b/r}}}\sin(\psi_{\mathrm{b/r}}) = 
        \\
        & \frac{r_{i,i+1}}{D_1 D_2} \Bigg[ \Bigg. - \vec{r}_{\mathrm{pat},i,\mathrm{b/r}} \times \vec{r}_{i,i+1} +
         \frac{V_{\mathrm{triple}}}{D_2^2}\left(  C_2 \vec{r}_{i,i+1} - r_{i,i+1}^2 \vec{r}_{\mathrm{pat},i+1,\mathrm{b/r}} \right) \Bigg. \Bigg].
    \end{aligned}
    \label{grad_sin_patch2}
\end{equation}

\subsection*{Quaternion formalism}
Quaternions are used to quantify the monomer orientation and the corresponding rotational dynamics for the implementation of the $\chi_i$ into the simulations. This reduces the rotational dynamics to quaternion multiplications and offers a way of regaining the patch positions relative to their beads if they are needed for calculations. A quaternion can be written as the 4-tuple $q=(q_0,q_1,q_2,q_3) = (q_0, \vec{q})$ where $q_0 \in \mathbb{R}$ is called the scalar part and $\vec{q} = (q_1,q_2,q_3) \in \mathbb{R}^3$ the vector part. $q=(0,\vec{q})$ with zero scalar part is called a vector quaternion. $q=(q_0,\vec{0})$ is called a scalar quaternion for $\vec{q}=\vec{0}$ in which case it is equivalent to a scalar $q_0$. The multiplication of two quaternions $q_a$ and $q_b$ is defined in the following way:
\begin{equation}
    \centering
    q_a \, q_b = (q_{a0},\vec{q}_a)(q_{b0},\vec{q}_b) = (q_{a0} q_{b0} - \vec{q}_a \cdot \vec{q}_b, \, q_{a0}\vec{q}_b + q_{b0}\vec{q}_a + \vec{q}_a \times \vec{q}_b).
    \label{quaternion_mult}
\end{equation}

The quaternions form a non-abelian group with respect to their multiplication which is associative. The conjugate of a quaternion $q=(q_{0},\vec{q})$ is defined as $q^*=(q_{0},-\vec{q})$ which brings about the quaternion norm $|q|=\sqrt{qq^*}=\sqrt{q_{0}^2+q_{1}^2+q_{2}^2+q_{3}^2}$ and the inverse quaternion $q^{-1}=q^*/|q|^2$, \,$qq^{-1}=q^{-1}q=1$. A rotation $\mathbb{R}^3 \rightarrow \mathbb{R}^3, \vec{p} \mapsto \vec{p}\,'$ by an angle $\alpha$ around axis $\vec{u}$ can be expressed by quaternion multiplication. A 3D vector $\vec{p}$ is projected onto the vector quaternion $p_q=(0,\vec{p})$ and the rotation is performed with a unit quaternion $q(t)=\left(\cos(\frac{\alpha}{2}),\sin(\frac{\alpha}{2})\vec{u}\right)$ with $|q|=|\vec{u}| = 1$, so that\cite{quaternions} $q^*=q^{-1}$:
\begin{equation}
    \centering
    p_q' = q \, p_q \, q^* = (0,\vec{p}\,').
    \label{quaternion_rotation}
\end{equation}

The rotated vector $\vec{p}\,'$ is read off the resulting vector quaternion $p_q'$. 
The monomer orientations $\chi_i(t)$ are represented by unit quaternions $q_i(t)=(q_{i,0},q_{i,1},q_{i,2},q_{i,3})=(q_{i,0},\vec{q}_i)=(\cos(\frac{\alpha_i}{2}),\sin(\frac{\alpha_i}{2})\vec{u}_i)$ with $|\vec{u}_i| = 1$. The $q_i(t)$ carry the information about a rotation axis $\vec{u}_i$ and an angle $\alpha_i$ corresponding to the orientation of monomer $i$ at time $t$. The patch positions $k=1,2,3\equiv$ blue, red, green are recovered by scaling the standard basis vectors to $l_{\mathrm{pat}}\hat{e}_k$ and rotating them with $q_i(t)$:
\begin{equation}
    \centering
    \begin{aligned}
        (0,l_{\mathrm{pat}} \hat{e}_k) \mapsto (0,\vec{r}_{\mathrm{pat},i,k}(t)) & = q_i(t) (0,l_{\mathrm{pat}}\hat{e}_k) q_i^*(t).
    \end{aligned}
    \label{patch_rotation}
\end{equation}

In this way, the patch vectors are always guaranteed to both have the correct distance to the bead and be orthogonal to each other so that they span an orthogonal reference system. At the beginning of a simulation run the monomers are given a starting orientation $q_i(0)$ that sets the initial patch positions $(0,\vec{r}_{\mathrm{pat},i,k}(0))= l_{\mathrm{pat}} q_i(0) (0,\hat{e}_k) q_i^*(0)$. The second step of the rotational velocity Verlet algorithm (\ref{velocity verlet rotation}) comprises a rotation of the orientation $q_i(t)$ around axis $\vec{\omega}_i/|\vec{\omega}_i|$ and by an angle $\vec{\omega}_i \delta t$ to $q_i(t+\delta t)$. In terms of quaternions it is computed in the following way with $\vec{\omega}_{i} = \vec{\omega}_{i}(t+\frac{\delta t}{2})$:
\begin{equation}
    \centering
    \begin{aligned}
    q_i(t+\delta t) & = q_{\vec{\omega}_{i}\delta t} \left(t + \frac{\delta t}{2} \right) \, q_i(t) 
    \\
    & = \left(\cos\left(\frac{|\vec{\omega}_i|\delta t}{2}\right),\sin\left(\frac{|\vec{\omega}_i|\delta t}{2}\right)\frac{\vec{\omega}_i}{|\vec{\omega}_i|}\right) \, q_i(t).
    \end{aligned}
    \label{verlet_quaternion_rotation}
\end{equation}

Because of associativity the updated $q_i(t+\delta t)$ imparts the same 3D rotation of a vector $\vec{p}$ to $\vec{p}\,'(t+\delta t)$ as $q_i(t)$ followed by a rotation according to $\vec{\omega}_i \delta t$:
 \begin{equation}
    \centering
    \begin{aligned}
    (0,\vec{p}\,'(t+\delta t)) & = p_q'(t+\delta t) = q_i(t+\delta t) \, p_q \, q_i^*(t+\delta t)
    \\
    & = \biggl( q_{\vec{\omega}_{i}\delta t} \left(t + \frac{\delta t}{2} \right) \, q_i(t) \biggr) \, p_q \, \biggl( q_{\vec{\omega}_{i}\delta t} \left(t + \frac{\delta t}{2} \right) \, q_i(t) \biggr)^*
    \\
    & = \biggl( q_{\vec{\omega}_{i}\delta t} \left(t + \frac{\delta t}{2} \right) \, q_i(t) \biggr) \, p_q \, \biggl( q_i^*(t) \, q_{\vec{\omega}_{i}\delta t} ^*\left(t + \frac{\delta t}{2} \right) \biggr)
    \\
    & = q_{\vec{\omega}_{i}\delta t} \left(t + \frac{\delta t}{2} \right) \, \biggl( q_i(t) \, p_q \, q_i^*(t) \biggr) \, q_{\vec{\omega}_{i}\delta t} ^*\left(t + \frac{\delta t}{2} \right)
    \\
    & = q_{\vec{\omega}_{i}\delta t} \left(t + \frac{\delta t}{2} \right) \, p_q'(t) \, q_{\vec{\omega}_{i}\delta t} ^*\left(t + \frac{\delta t}{2} \right)
    \\
    & = q_{\vec{\omega}_{i}\delta t} \left(t + \frac{\delta t}{2} \right) \, (0,\vec{p}\,'(t)) \, q_{\vec{\omega}_{i}\delta t} ^*\left(t + \frac{\delta t}{2} \right).
    \end{aligned}
    \label{verlet_quaternion_associative}
\end{equation}

At time $t$, 
the orientation is given by the time-ordered product of all rotations to the left of $q_i(0)$:
\begin{equation}
    \centering
    \begin{aligned}
     q_i(t) & = 
     q_{\vec{\omega}_{i}\delta t} \left(t - \frac{\delta t}{2}  \right) \, q_{\vec{\omega}_{i}\delta t} \left(t - \frac{3\delta t}{2} \right)\times \, ... \, \\
     & \times q_{\vec{\omega}_{i}\delta t} \left(\frac{3\delta t}{2} \right) \, q_{\vec{\omega}_{i}\delta t} \left(\frac{\delta t}{2} \right) \, q_i(0),
     \end{aligned}
    \label{quaternion_prod}
\end{equation}
\newline
and the step (\ref{verlet_quaternion_associative}) can be repeated $t/\delta t + 1$ many times:
\begin{equation}
    \centering
    \begin{aligned}
     p_q'(t) & = q_i(t) \; p_q \; q_i^*(t)
      \\
      & = q_{\vec{\omega}_{i}\delta t} \biggl(t - \frac{\delta t}{2} \biggr) \, ... \, q_{\vec{\omega}_{i}\delta t} \left(\frac{\delta t}{2} \right) 
      q_{i}\left( 0 \right) \;  p_q \times
      \\
      & \times \; q_{i}^*\left( 0 \right) q_{\vec{\omega}_{i}\delta t}^* \left(\frac{\delta t}{2} \right) \, ... \, q_{\vec{\omega}_{i}\delta t}^*\left(t - \frac{\delta t}{2} \right).
    \end{aligned}
    \label{verlet_quaternion_associative_2}
\end{equation}

Inserting $p_q=(0,\vec{p})=(0,l_{\mathrm{pat}}\hat{e}_k$) and $p_q\,'(t)=(0,\vec{p}\,'(t)) = (0,\vec{r}_{\mathrm{pat},i,k}(t))$ into (\ref{verlet_quaternion_associative}) and (\ref{verlet_quaternion_associative_2}) makes evident that (\ref{patch_rotation}) and (\ref{verlet_quaternion_rotation}) reproduce the monomer orientations at all time steps. The quaternions must always have unity norm, i.e., $|q_i(t)|=1$ so that they represent proper rotations. It holds that $\bigl| q_{\vec{\omega}_{i}\delta t}\left(t + \frac{\delta t}{2} \right) \bigr| = 1$, i.e., normalization to unity is guaranteed for all time steps $t$, if the initial $q_i(0)$ have unity norm. The quaternions are normalized $q_i(t) \mapsto q_i(t)/|q_i(t)|$ every 100th MD step to counteract numerical errors that can possibly occur along the way.

\subsection*{Relaxation time of supercoiled rings}

%%%%%%%%%% PUT TO SI

%%%%%%%%%%%%%%%%%%%%%%%%%%%%%%%%%%%%%%%%%%%%%%%%%%%%%%%%%%%%%%%%%%%%%%%%
\begin{figure}[htb]
\centering
  \includegraphics[width=0.9\columnwidth]{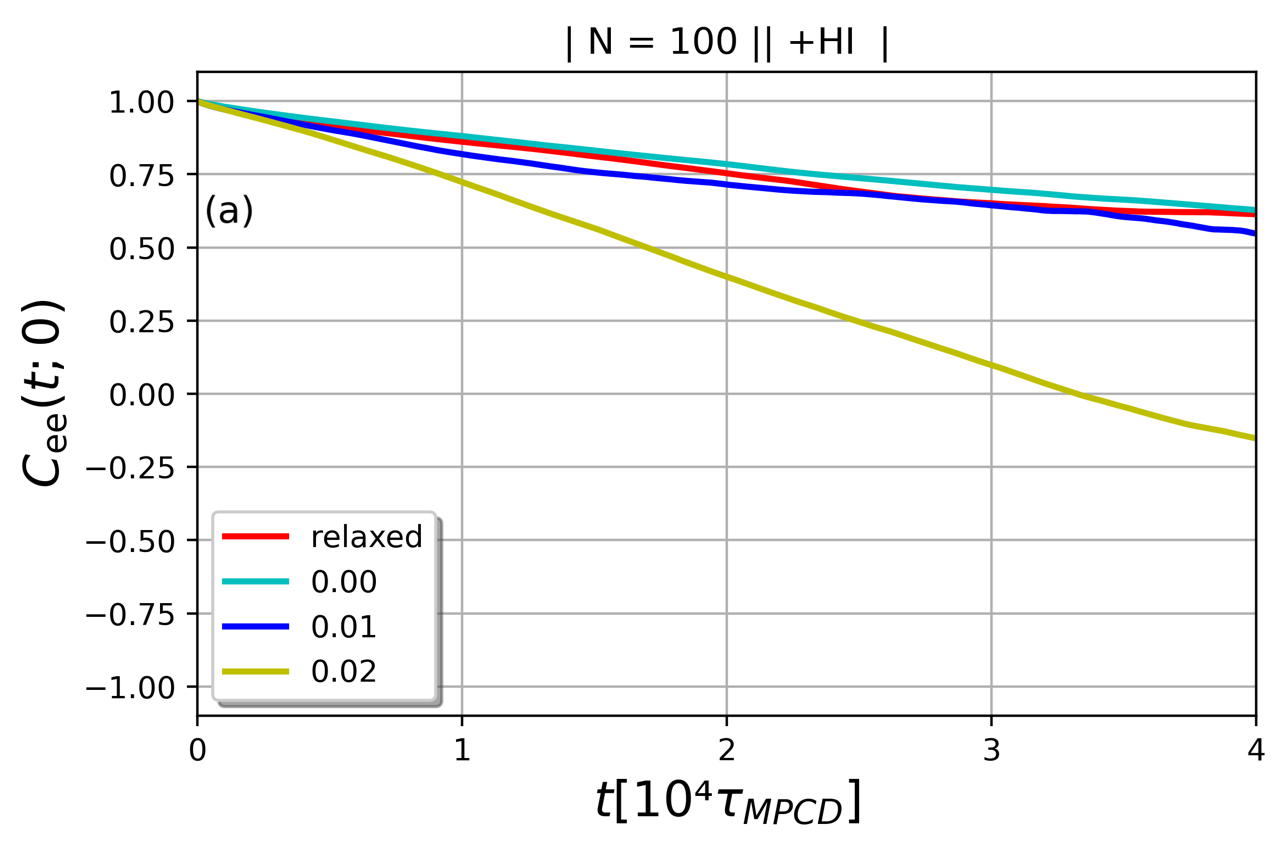}
  \includegraphics[width=0.9\columnwidth]{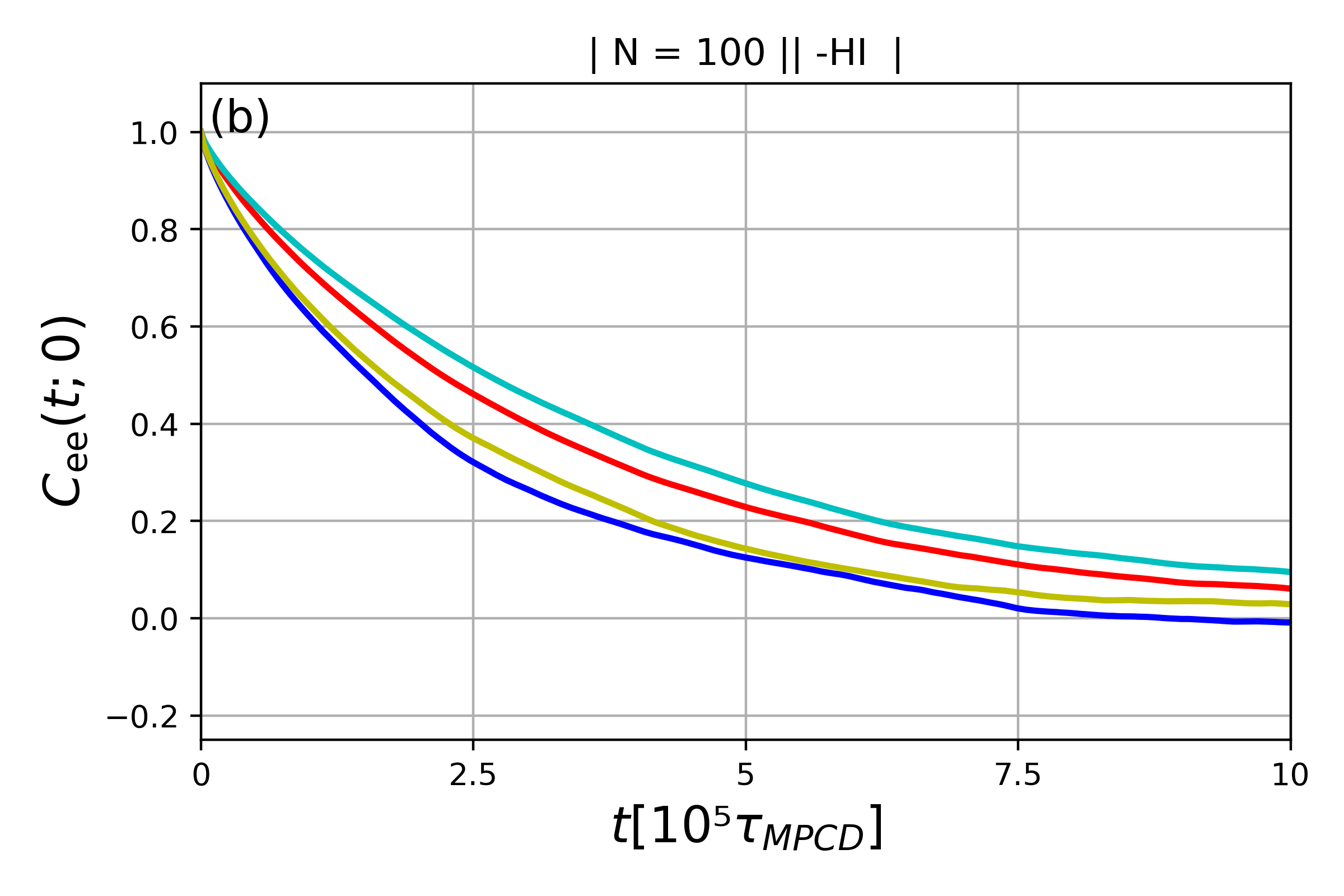}
  \caption{The equilibrium end-to-end correlation function of supercoiled rings of length $N=100$ as a function of time starting from the initialization states for $+$HI, panel (a), and $-$HI, panel (b).}
 \label{Fig:corr_ee_zero_shear}

\end{figure}
%%%%%%%%%%%%%%%%%%%%%%%%%%%%%%%%%%%%%%%%%%%%%%%%%%%%%%%%%%%%%%%%%%%%%%%%
The relaxation times of supercoiled rings can be extracted by means of the ``end-to-end'' correlation function,\cite{liebetreu:commats:2020,corr_ee}
\begin{equation}
\label{end_to_end_correlation}
    C_{\mathrm{ee}}(t;0) = \frac{{\langle\vec{R}_{\mathrm{ee}}(t;0)\cdot\vec{R}_{\mathrm{ee}}(0;0)\rangle}_m}
    {{\langle\vec{R}_{\mathrm{ee}}^2(0;0)\rangle}_m},
\end{equation}
\newline
where $\vec{R}_{\mathrm{ee}}(t;\dot\gamma)$ defines a vector joining two monomers of the chain that are separated by $N/2$ monomers and $\langle ... \rangle_m$ represents an average over the $N/2$ end-to-end connections; results are shown in  
Fig.~\ref{Fig:corr_ee_zero_shear}. For the $+$HI-case, shown in Fig.~\ref{Fig:corr_ee_zero_shear}(a), we extended the simulation up to times $t \cong 4\cdot 10^4\,\tau_{\rm MPCD}$, whereas for the $-$HI-case, shown in Fig.~\ref{Fig:corr_ee_zero_shear}(b),
which is computationally cheaper, we were able to reach times $t \cong 10^6\,\tau_{\rm MPCD}$. The latter case clearly shows the characteristic exponential decay of the correlation functions, establishing a relaxation time $\tau_{\rm R} \cong 3\cdot 10^5\,\tau_{\rm MPCD}$,
in very good agreement with the estimate presented in the main text. On the other hand, the curves in Fig.~\ref{Fig:corr_ee_zero_shear}(a) demonstrate that the correlation functions for the $+$HI-case decay much faster (by about one order of magnitude), again confirming 
the estimate of $\tau_{\rm R}$ in the main text.

%%%END OF MAIN TEXT%%%


\balance


%%%REFERENCES%%%
%\bibliography{topology} %You need to replace "rsc" on this line with the name of your .bib file
\bibliographystyle{rsc} %the RSC's .bst file

\providecommand*{\mcitethebibliography}{\thebibliography}
\csname @ifundefined\endcsname{endmcitethebibliography}
{\let\endmcitethebibliography\endthebibliography}{}

\end{document}